%% file: main.tex
\documentclass[12pt]{article}
\pdfoutput=1 

\usepackage{jheppub} 

\usepackage{float, array, xspace, amscd, amsthm}
\usepackage{fancyhdr}
\usepackage{longtable}

\makeatletter
\g@addto@macro\@floatboxreset\centering
\makeatother

\RequirePackage{xcolor}
\definecolor{darkblue}{rgb}{0.0,0.0,0.4} 
\hypersetup{pdftitle={Topological G2 and Spin(7) strings at 1-loop from double complexes},linkcolor=darkblue, urlcolor=darkblue, citecolor=darkblue, filecolor=darkblue,anchorcolor=darkblue,menucolor=darkblue}

\usepackage{setspace}
\allowdisplaybreaks

\usepackage{amsmath}
\usepackage{slashed}
\usepackage{microtype}

\makeatletter
\g@addto@macro\bfseries{\boldmath}
\makeatother

\let\originalleft\left
\let\originalright\right
\renewcommand{\left}{\mathopen{}\mathclose\bgroup\originalleft}
\renewcommand{\right}{\aftergroup\egroup\originalright}

\usepackage{amssymb,euscript,bbold}
\usepackage{amsfonts}
\usepackage{mathrsfs}
\usepackage{graphicx}
\usepackage{mathtools}
\usepackage{slashed}
\usepackage{amsthm}
\usepackage{amscd}

\usepackage{epsfig}                    
\usepackage[matrix,arrow]{xy}          
\usepackage{xspace}                    
\usepackage{stmaryrd}                  
\usepackage{slashed}
\usepackage{appendix,graphicx}
\usepackage{enumitem}

\usepackage{units} 
\usepackage{tikz}
\usepackage{tikz-cd}
\usepackage{color}

\usepackage{mathdots}

\PassOptionsToPackage{linecolor=blue,backgroundcolor=blue!25,bordercolor=blue,textsize=scriptsize}{todonotes}
\usepackage{todonotes}

\usepackage{tikz-cd}
\usepackage{subfigure}

\DeclareSymbolFont{bbold}{U}{bbold}{m}{n}
\DeclareSymbolFontAlphabet{\mathbbold}{bbold}

\newcommand{\dd}{{\rm d}}

\newcommand{\Dorf}{L}

\DeclareMathOperator{\tr}{tr}

\def\bea#1\eea{\begin{align}#1\end{align}}

\usepackage{color}

\theoremstyle{definition}

\global\long\def\ext{\Lambda}%
\global\long\def\op#1{\operatorname{#1}}%

\newcommand{\ii}{\mathrm{i}}
\newcommand{\ee}{\mathrm{e}}

\newcommand{\der}{\partial}
\newcommand{\del}{\partial}
\newcommand{\delb}{\bar{\partial}}

\newcommand{\bbR}{\mathbb{R}}
\newcommand{\bbC}{\mathbb{C}}

\global\long\def\Orth#1{\mathit{O}(#1)}%
\global\long\def\Gx#1{\mathit{G}_#1}%
\global\long\def\eqspace{\mathrel{\phantom{{=}}{}}}%

\DeclareMathOperator{\SU}{\mathit{SU}}
\DeclareMathOperator{\Uni}{\mathit{U}}

\DeclareMathOperator{\SL}{\mathit{SL}}
\DeclareMathOperator{\GL}{\mathit{GL}}

\DeclareMathOperator{\Spin}{\mathit{Spin}}
\DeclareMathOperator{\G}{\mathit{G}}

\DeclareMathOperator{\Cliff}{Cliff}

\newcommand{\rep}[1]{\mathbf{#1}}

\newcommand{\repp}[2]{(\rep{#1}, \rep{#2})}
\newcommand{\id}{\mathbbold{1}}

\DeclareMathOperator{\re}{Re}

\DeclareMathOperator{\vol}{vol}
\DeclareMathOperator{\Vol}{V}

\newcommand{\Gs}[1]{\Gamma(#1)}

\newcommand{\Lgen}{L}
\newcommand{\LHgen}{L^{H}}

\newcommand{\anchor}{a}

\renewcommand{\Pr}[2]{\mathcal{P}^{#1}_{\rep{#2}}}

\newcommand{\bgen}[2]{\left\llbracket#1,#2\right\rrbracket}

\newcommand{\Dgen}{{D}}

\newcommand{\LC}{\nabla}

\newcommand{\dc}{\check{\dd}} 
\newcommand{\Deltac}{\check{\Delta}}

\newcommand{\Riem}{\mathcal{R}}
\newcommand{\Ric}{\mathcal{R}}
\newcommand{\Scalar}{\mathcal{R}}

\newcommand{\GenRic}{R^{\scriptscriptstyle 0}}

\newcommand{\GenS}{R}

\DeclareMathOperator{\Diff}{Diff}
\DeclareMathOperator{\GDiff}{GDiff}

\newcommand{\ba}{\bar{a}}
\newcommand{\bb}{\bar{b}}
\newcommand{\bc}{\bar{c}}
\newcommand{\bd}{\bar{d}}
\newcommand{\be}{\bar{e}}

\newcommand{\fspin}{\mathfrak{spin}_7}

\newcommand{\gxg}{G_2 \times G_2}
\newcommand{\sxs}{Spin(7)\times Spin(7)}

\newcommand{\dil}{\phi}
\newcommand{\gphi}{\varphi}

\newcommand{\A}{\mathcal{A}}

\newcommand{\QR}{Q_R}
\newcommand{\QL}{Q_L}

\newcommand{\QA}{Q_A}
\newcommand{\QB}{Q_B}

\global\long\def\DeAB{\overset{\bullet}{\Delta}}%
\global\long\def\DeCD{\underset{\bullet}{\Delta}}%
\global\long\def\DeAC{\raisebox{1pt}{$\scriptstyle\bullet$}\Delta}%
\global\long\def\DeBD{\Delta\raisebox{1pt}{$\scriptstyle\bullet$}}%

\global\long\def\AomAB{\overset{\bullet}{\mathcal{A}}}%
\global\long\def\AomCD{\underset{ \bullet}{\mathcal{A}}}%
\global\long\def\AomAC{\raisebox{1pt}{$\scriptstyle\bullet$}\mathcal{A}}%
\global\long\def\AomBD{\mathcal{A}\raisebox{1pt}{$\scriptstyle\bullet$}}%

\title{Topological G$_{2}$ and Spin(7) strings at 1-loop from double complexes}

\author[a,b]{Anthony Ashmore,}
\emailAdd{ashmore@uchicago.edu}
\author[c]{Andr\'e Coimbra,}
\emailAdd{andre.coimbra@aei.mpg.de}
\author[d]{Charles Strickland-Constable,}
\emailAdd{c.strickland-constable@herts.ac.uk}
\author[e]{Eirik Eik Svanes}
\emailAdd{eirik.e.svanes@uis.no}
\author[f]{and David Tennyson}
\emailAdd{d.tennyson16@imperial.ac.uk}

\affiliation[a]{Enrico Fermi Institute \& Kadanoff Center for Theoretical Physics,\\ University of Chicago, Chicago, IL 60637, USA} 
\affiliation[b]{Sorbonne Universit\'e, CNRS, LPTHE, 75005 Paris, France}
\affiliation[c]{Max Planck Institute for Gravitational Physics (Albert Einstein Institute),\\
	Am Muhlenberg 1, D-14476 Potsdam, Germany}
\affiliation[d]{School of Physics, Astronomy and Mathematics, University of Hertfordshire,\\
	College Lane, Hatfield, AL10 9AB, UK}
\affiliation[e]{Department of Mathematics and Physics, University of Stavanger,\\ Kristine Bonnevies vei 22, 4021, Stavanger, Norway}
\affiliation[f]{Department of Physics, Imperial College London,	Prince Consort Road, \\London, SW7 2AZ, UK}

\abstract{We study the topological $G_2$ and $Spin(7)$ strings at 1-loop. We define new double complexes for supersymmetric NSNS backgrounds of string theory using generalised geometry. The 1-loop partition function then has a target-space interpretation as a particular alternating product of determinants of Laplacians, which we have dubbed the analytic torsion.  In the case without flux where these backgrounds have special holonomy, we reproduce the worldsheet calculation of the $G_2$ string and give a new prediction for the $Spin(7)$ string. We also comment on connections with topological strings on Calabi--Yau and K3 backgrounds.}

\begin{document} 

\maketitle


\section{Introduction}

Topological string models with Calabi--Yau target spaces provide us with subsectors of string theory in which certain quantities can be computed exactly. The key to this is that $(2,2)$ $\sigma$-models with Calabi--Yau targets admit certain topological twists. To be precise, there are two distinct twists which give the A- and B-models~\cite{Witten:1991zz}. In both of these models, the metric is not a fundamental degree of freedom -- the A- and B-models are theories of K\"ahler and complex structures respectively -- which suggests that the resulting theories may be topological. At the quantum level, the A-model can be defined on any K\"ahler manifold, while the B-model can be defined consistently only on Calabi--Yau targets. With these assumptions, one can show that observables indeed do not depend on the metric and so deserve the name topological.

The connection between topological strings and geometries captured by invariant functionals was first discussed in \cite{Dijkgraaf:2004te,Gerasimov:2004yx,Nekrasov:2004vv}, where the partition function of the topological B-model and its conjugate \cite{Witten:1991zz,Witten:1988xj,Witten:1988ze,Bershadsky:1993cx} on a six-dimensional target space $M$ was argued to be encoded in the Hitchin functional for an $SL(3,\mathbb{C})$ structure on the same target space~\cite{Hitchin:2000jd}. Pestun and Witten later observed that there is a discrepancy between the two at 1-loop, and showed that the 1-loop partition function of the B-model is actually given by the partition function of an extended Hitchin functional~\cite{Pestun:2005rp}. Generalising the real three-form that characterises an $SL(3,\mathbb{C})$ structure, this extended functional is written in terms of a polyform which defines a generalised Calabi--Yau structure. A key insight here was that although the critical points of the two functionals agree, at the quantum level the fluctuating degrees of freedom of the two structures are different. Thus, it was essential to view the target space as a background in generalised geometry in order to match the topological B-model calculation.

One can view the above calculation by first starting with a conventional $\sigma$-model with a Calabi--Yau target space. In the large-volume limit, the worldsheet theory is captured by an effective theory on the target space. The topological twist that leads to the B-model corresponds to a subsector of the target-space theory described by a generalised Calabi--Yau structure. There is a similar construction for $\sigma$-models with a $G_2$ holonomy target space. One starts with the $G_2$ worldsheet algebra, which contains an extended $(1,1)$ supersymmetry algebra~\cite{Shatashvili:1994zw}. Importantly, this algebra has a $c=7/10$ subalgebra, known as a tri-critical Ising model, which can be used to define a topological twist of the $\sigma$-model~\cite{deBoer:2005pt}. One might expect that there is a subsector of the target-space theory that captures this twisted sector. An attempt at constructing this theory was made in \cite{deBoer:2005pt}, where a target-space action was proposed by starting from a Hitchin functional for a generalised $\gxg$ structure. In this case, the 1-loop partition function of the topological $G_2$ string disagreed with the target-space calculation, differing by a factor of the Ray--Singer torsion of the background $G_2$ manifold.

The goal of this paper is to resolve this discrepancy and give a target-space interpretation of the topological $G_2$ string calculation. Rather than starting from a functional on the target space, we follow a different path and suggest that these results can be obtained by considering a certain double complex that arises naturally in generalised geometry. 

To be precise, we will give a double complex for $\gxg$ that realises the BRST complex of the topological $G_2$ string. The degrees of freedom of the worldsheet theory break into right- and left-moving sectors, each with their own BRST operators, with the string states given by tensor products of these sectors. Physical states are then cohomology classes of the total BRST operator. We will give a target-space interpretation for each of these ingredients, culminating in an expression for the 1-loop partition function that agrees with the worldsheet calculation of de Boer et al.~\cite{deBoer:2005pt} and a target-space action whose BV quantisation reproduces this answer. Given the conjectured existence of theories in seven and eight dimensions that unify the A- and B-models~\cite{Gerasimov:2004yx,Dijkgraaf:2004te,Nekrasov:2004vv,Grassi:2004xr,Anguelova:2004wy}, it seems sensible to consider how topological $Spin(7)$ strings might also be captured by generalised geometry. Following the same logic as for the $G_2$ string, we make a conjecture for its 1-loop partition function.

Our construction does not require that the target space has special holonomy, but instead requires only the weaker condition of being a purely NSNS Minkowski background (with metric, dilaton and $B$ field) preserving at least $N=1$ supersymmetry. In outline, starting with an $O(d,d)\times\mathbb{R}^+$ generalised geometry description of the target space, we show that supersymmetry implies the existence of a torsion-free $G\times G$ structures. From these, one can construct a compatible, torsion-free generalised connection which, together with certain projectors onto representations of $G\times G$, can be used to define a pair of differentials $(\dd_+, \dd_-)$. These differentials give the maps in a double complex for $G\times G \subset O(d)\times O(d)\subset O(d,d)$. After working out the Hodge theory and the analogue of K\"ahler identities for these differentials, we conjecture that a certain alternating sum of determinants of the Laplacian defined by $\hat\dd=\dd_+ + \dd_-$ determines the 1-loop partition function of the corresponding worldsheet theory. Upon restricting to honest special holonomy backgrounds with vanishing $H$ flux, our expression reduces to the known result for the $G_2$ string and the A- and B-models, and gives a prediction for the $Spin(7)$ string.

Though our work gives a target-space description of the 1-loop partition function for these topological strings, we have not been able to find target-space actions that reproduce these calculations upon quantisation in all cases. The central result of \cite{Pestun:2005rp} was the construction of a target-space theory based on an extended Hitchin functional for $SL(3,\mathbb{C})$ whose BV quantisation gives precisely the 1-loop partition function of the B-model on a Calabi--Yau target. For the $G_2$ string, a similar calculation was attempted in \cite{deBoer08c} with less success -- the quantisation of neither the conventional nor the extended $G_2$ Hitchin functionals reproduced the 1-loop partition function of the $G_2$ string. To the authors' knowledge, there has been no attempt to repeat this for the $Spin(7)$ string. In this paper, we give a target-space action whose quantisation does agree with the $G_2$ string, but it is not based on an invariant functional that we are aware of. For the $Spin(7)$ string, we have not been able to write down a target-space action.

A summary of our results follows:
\begin{itemize}
    \item In Section \ref{sec:G2G2_double_complex}, we introduce a new double complex for $G_2 \times G_2$ structures on seven-dimensional manifolds within generalised geometry. The differential operators that appear in this double complex are defined using a generalised connection that is compatible with the $G_2\times G_2$ structure. We show that the operators are nilpotent and commute in the correct manner (so that they define a complex) if the generalised connection is torsion-free, which implies that the underlying string background is an NSNS Minkowski solution preserving at least $N=1$ supersymmetry. We define Laplacians for these operators and their Hodge theory. 
    \item We conjecture that the 1-loop partition function of the corresponding topological string is given by a certain alternating product of determinants of the Laplace operators acting on the double complex. In Section \ref{sec:relation_to_G2_string} we restrict to the case of a $G_2$ holonomy background, and show that our general expression for the 1-loop partition function agrees with the worldsheet calculation of de Boer et al.~\cite{deBoer:2005pt}.
    \item We give a target-space action whose BV quantisation reproduces our expression for the 1-loop partition function in Section \ref{sec:G2-quad-action}. It does not seem that this action comes from considering variations of an invariant functional, unlike the case of the B-model~\cite{Pestun:2005rp}.
    \item In Section \ref{spin7xspin7} we repeat the above analysis for $Spin(7)\times Spin(7)$ structures on eight-dimensional manifolds. We again show that one can define a double complex provided there exists a torsion-free connection compatible with the generalised structure, equivalent to the corresponding NSNS Minkowski background preserving some supersymmetry. In Section \ref{sec:top_spin7_string} we compute the alternating product of determinants, and in the case of a $Spin(7)$ holonomy manifold conjecture that this gives the 1-loop partition function of the $Spin(7)$ string. We find it to be
    \begin{equation}
    Z_{1} = (\det{}'\Delta_{\rep{1}})^{-1}(\det{}'\Delta_{\rep{7}})(\det{}'\Delta_{\rep{21}})^{-1/2}(\det{}'\Delta_{\rep{27}})^{-1/2},
\end{equation}
    where $\Delta_{\rep{r}}$ is the Laplacian acting on the $\rep r$ representation of $Spin(7)$ and $\det{}'$ is the $\zeta$-regularised determinant.
    \item We outline in Section \ref{sec:other-examples} how our formalism applies to the A- and B-models with flux and show that the relevant Laplacian is associated to the Lie algebroid defined by the corresponding generalised complex structure, agreeing with~\cite{Kapustin:2004gv}. We also comment on topological strings on K3 surfaces where one finds that the 1-loop contribution is trivial.
\end{itemize}
We begin with an overview of the worldsheet theories for the A- and B-models, $G_2$ and $Spin(7)$ in Section \ref{sec:review_top_strings}. We then review the complexes that one can define on manifolds with $G$-structures and outline their Hodge theory in Section \ref{sec:G_complexes}, before moving onto the results outlined above. The appendices contain our conventions and useful identities, a discussion of determinants and partition functions, and a quick review of $O(d,d)\times\mathbb{R}^+$ generalised geometry.


\section{Review of topological strings}\label{sec:review_top_strings}

Since we will be proposing a target-space interpretation of various topological theories, we will first spend some time reviewing topological strings from the worldsheet, starting with the well-known A/B-models \cite{Witten:1988xj,Bershadsky:1993cx,Witten:1991zz,WITTEN1990281} and then moving on to the topological $\G_{2}$ \cite{Shatashvili:1994zw,deBoer:2005pt,deBoer08b,deBoer08c} and $\Spin(7)$ strings \cite{Shatashvili:1994zw}. 

In each of these cases, special holonomy of the target space implies the existence of an extended worldsheet symmetry which allows a twisting procedure that renders the theory topological. In brief, one looks for an operator $\rho$, often related to the extended symmetry, with which to `twist' the energy-momentum tensor
\begin{equation}\label{eq:generic_twist}
    T \longrightarrow T_{\text{twist}} \sim T + \del^{2}\rho ,
\end{equation}
such that the central charge $c$ of the twisted theory vanishes. This twisted energy-momentum tensor endows operators of the theory with new charges under Lorentz transformations. Interestingly, the twisting operator $\rho$ is intimately related to the spectral flow operator, or analogues thereof, which is used to generate target-space supersymmetry. 

Next, one looks for a nilpotent scalar\footnote{This is a scalar with respect to the twisted Lorentz symmetry.} operator $Q$:
\begin{equation}
    Q^{2} = 0.
\end{equation}
Typically, $Q$ is built from the supersymmetry generators and then identified as the relevant BRST operator. One then requires that $T_{\text{twist}}$ is a $Q$-trivial operator.\footnote{A necessary condition for this is that the central charge vanishes, hence the need to look for a $T_{\text{twist}}$ with vanishing central charge.} This is usually done by requiring the action to be written as a $Q$-exact piece, plus  terms independent of the target-space metric. If this is the case, one can use localisation techniques to obtain exact results for correlators by evaluating them on fixed points of the BRST symmetry \cite{Witten:1991zz}. We further require that physical operators fall into $Q$-cohomology classes. These physical fields form a closed ring under the OPE, called the chiral ring, which is often related to certain cohomological data of the target space. The chiral ring generates certain highest-weight states in the NS sector, which can be related to the R sector ground states through spectral flow.

For \emph{closed} topological strings, one has independent left- and right-moving sectors. States of the theory are built from tensor products of left- and right-moving states, and the BRST operator can be split into a left- and a right-moving operator
\begin{equation}
    Q = \QL + \QR, \qquad \QL^{2} = \QR^{2} = \{\QL,\QR\} = 0.
\end{equation}
Grading the states by left- and right-moving fermion number $(p,q)$, we find that observables fit into a double complex
\begin{equation}
\label{eq:BRST_double_complex}
    \begin{tikzcd}[column sep = normal, row sep = normal]
    & \vdots & \vdots & \vdots & \\
    \dots \arrow{r} & \mathcal{O}^{p-1,q+1} \arrow{u} \arrow {r} & \mathcal{O}^{p,q+1} \arrow{u} \arrow{r} & \mathcal{O}^{p+1,q+1} \arrow{u} \arrow{r} & \dots \\
    \dots \arrow{r} & \mathcal{O}^{p-1,q} \arrow{u}\arrow{r}{Q_{L}} & \mathcal{O}^{p,q} \arrow{u}{Q_{R}} \arrow{r}{Q_{L}} & \mathcal{O}^{p,q+1} \arrow{u} \arrow{r} & \dots \\
    \dots \arrow{r} & \mathcal{O}^{p-1,q-1} \arrow{u} \arrow{r} & \mathcal{O}^{p,q-1} \arrow{u}{Q_{R}} \arrow{r} & \mathcal{O}^{p+1,q-1} \arrow{u} \arrow{r} & \dots \\
    & \vdots \arrow{u} & \vdots \arrow{u} & \vdots \arrow{u} &
    \end{tikzcd}
\end{equation}
This double complex is understood for the A/B-model \cite{Kapustin:2004gv}. It is the main goal of this paper to show that there is a nice target-space interpretation of \eqref{eq:BRST_double_complex} for \emph{any} topological string with a special holonomy target space, at least in the infinite-volume limit, and that the 1-loop partition function of the worldsheet theory calculates a particular quantity of the double complex that we have dubbed the \emph{analytic torsion}.

We will now go into more detail about the twisting procedure and the cohomological structure of the topological string for A/B, $\G_{2}$ and $\Spin(7)$ strings.

\subsection{The A- and B-models}\label{sec:review_A_B_models}

When the target space $M$ is K\"ahler and the $H$-flux vanishes, the worldsheet supersymmetry is enhanced to $N = (2,2)$. This symmetry is built from a left-moving and a right-moving sector which each have an energy-momentum tensor $T$, two supercurrents $G^{\pm}$, and a $\Uni(1)$ current $J$. The $\pm$ on the supercharges correspond to their charge under the $\Uni(1)$. We will denote fields in the right-moving sector with a bar.

After the usual mode expansion, the relevant commutators for the left-moving sector are
\begin{align}
    [L_{m},L_{n}] &= (m-n)L_{m+n} + \frac{c}{12}(m^{3}-m)\delta_{m+n,0} ,\\
    [L_{m},G_{n\pm a}^{\pm}] &= \left(\tfrac12 m-n \mp a\right) G^{\pm}_{m+n\pm a} ,\\
    [L_{m},J_{n}] &= -nJ_{m+n} ,\\
    [J_{m},J_{n}] &= \frac{c}{3}m \,\delta_{m+n,0} ,\\
    [J_{m},G^{\pm}_{n\pm a}] &= \pm G^{\pm}_{m+n\pm a},
\end{align}
and similarly for the right-movers. Strikingly, if one defines a twisted energy-momentum operator via
\begin{equation}\label{eq:N=2_twist}
    T_{\text{twist}} = T + \tfrac12 \del J,
\end{equation}
then the new modes $\tilde{L}_{m} = L_{m} - \tfrac{1}{2}(m+1)J_{m}$ satisfy
\begin{equation}
    [\tilde{L}_{m},\tilde{L}_{n}] = (m-n)\tilde{L}_{m+n}, \qquad [\tilde{L}_{0},G^{+}_{-1/2}] = 0.
\end{equation}
Hence we see that the central charge of the twisted algebra \emph{vanishes}, and the supercharge $G^{+}_{-1/2}$ is a scalar with respect to the new Lorentz symmetry. We can therefore use it as the left-moving BRST operator.

Given this twist, there are two inequivalent twists of the right-moving sector given by~\cite{Lerche:1989uy}
\begin{equation}
    \bar{T}_{\text{twist}} = \bar{T} \pm \tfrac12 \del \bar{J},
\end{equation}
where upper sign corresponds to the B-model and the lower leads to the A-model. Both twists result in algebras with vanishing central charge, but with different nilpotent scalar operators. One then finds that the relevant right-moving BRST operators are
\begin{equation}
    [\bar{\tilde{L}}_{m},\bar{G}^{\pm}_{-1/2}] = 0.
\end{equation}
The total BRST operators for each model are then
\begin{equation}
    \QA = G^{+}_{-1/2} + \bar{G}^{-}_{-1/2} , \qquad \QB = G^{+}_{-1/2} + \bar{G}^{+}_{-1/2}.
\end{equation}
One can then examine the cohomology of local observables in each case.

Let us also briefly note how this twist is related to spectral flow. This is a symmetry of the algebras given by
\begin{align}
    L^{\eta}_{n} &= L_{n} + \eta J_{n} + \frac{c}{6}\eta^{2}\delta_{n,0}, \\
    G^{\eta\pm}_{n\pm a} &= G^{\pm}_{n\pm (a+\eta)} \label{eq:spectral_flow}, \\
    J^{\eta}_{n} &= J_{n} + \frac{c}{3}\eta\,\delta_{n,0}.
\end{align}
In particular, we see from \eqref{eq:spectral_flow} that for $\eta = \frac{1}{2}$, we have a map from the NS to the R sector. This is generated by an operator which one can bosonise to $\ee^{\ii\rho/2}$. One then finds that the $\Uni(1)$ generator can be written in terms of $\rho$ as
\begin{equation}
    J = \del\rho.
\end{equation}
Inserting this into \eqref{eq:N=2_twist}, one finds a formula for the twisted energy-momentum tensor  more like that of  \eqref{eq:generic_twist}.

\subsubsection*{The A-model}

The A-model action can be written as
\begin{equation}\label{eq:A-model_action}
    S = \left\{ \QA, \int_{\Sigma}V \right\} + \int_{x(\Sigma)}\omega,
\end{equation}
for some $V$, where $\omega$ is the K\"ahler form on $M$, $\Sigma$ is the worldsheet and $x\colon\Sigma \rightarrow M$ is a map from the worldsheet to the target space. The action is thus $\QA$-exact, up to a term that depends only on the homology class of $x(\Sigma)\subset M$ and so is independent of the worldsheet metric. This is sufficient to show that the energy-momentum tensor is $\QA$-exact.\footnote{In fact, to write $S$ as in \eqref{eq:A-model_action}, one has to use the equations of motion. It is possible to show that there exists an operator $\tilde{Q}_{A}$, which is related to $\QA$ by the equations of motion, such that \eqref{eq:A-model_action} also holds off-shell.} Note further that the second term is also independent of the target-space complex structure and hence all correlators will depend only on the K\"ahler moduli. This topological string is thus quasi-topological, depending on some but not all the target-space moduli.

Localisation techniques allow us to evaluate correlators exactly by restricting the calculation to solutions of the equations of motion for $V$. It turns out that, for the bosonic sector, these are
\begin{equation}
    \delb x = \del\bar{x} = 0.
\end{equation}
The theory therefore localises on holomorphic maps $x\colon\Sigma \rightarrow M$. Coupling this theory to gravity, one must then integrate correlators over the moduli space of complex structures on $\Sigma$. This ensures the result is indeed independent of the target-space complex structure.

To study the $\QA$ cohomology ring, and hence the physical local operators, it is useful to go to the infinite-volume limit in which contributions from non-trivial homology classes $x(\Sigma)$ drop out of all correlation functions. In this limit, it is possible to show that the chiral operators take the form
\begin{equation}
    \mathcal{O}_{\alpha} = \alpha(x)_{\mu_{1}\dots\mu_{p}\bar{\nu}_{1}\dots\bar{\nu}_{q}}\chi^{\mu_{1}}\ldots\chi^{\mu_{p}} \bar{\chi}^{\bar{\nu}_{1}}\ldots\bar{\chi}^{\bar{\nu}_{q}},
\end{equation}
where the $\chi^{\mu}$ and $\bar{\chi}^{\bar{\nu}}$ are the left- and right-moving worldsheet fermions with $U(1)$ charge $+1$ and $-1$ respectively. Note that under the twisted Lorentz symmetry, these are scalars and thus dimension-zero operators, as is required for a topological observable. Moreover, one can identify $\chi^{\mu} \in x^{*}(T^{1,0})$ and $\bar{\chi}^{\bar{\nu}}\in x^{*}(T^{0,1})$.\footnote{Here, and throughout the paper, we will use the shorthand $T=TM$, $T^{1,0} = T^{1,0}M$, and so on.} Hence, the space of operators is identified with standard $(p,q)$-forms on $M$. Under this identification, one finds that
\begin{equation}
    G^{+}_{-1/2} \sim \del, \qquad \bar{G}^{-}_{-1/2} \sim \delb, \qquad \QA \sim \del+\delb = \dd.
\end{equation}
Therefore, the chiral ring is isomorphic to the de Rham cohomology ring of $M$ in the infinite-volume limit. Furthermore, the double complex given in \eqref{eq:BRST_double_complex} maps onto the Dolbeault complex of $M$.

At finite volume, the chiral ring is deformed by worldsheet instantons coming from the second term in \eqref{eq:A-model_action}. While the operators can still be identified with $(p,q)$-forms, $\QA$ no longer matches the de Rham operator, and the ring structure does not match the de Rham cohomology. Instead, one finds what is called the quantum-deformed cohomology of $M$ which takes into account holomorphic multiwrappings of the worldsheet on Riemann surfaces in $M$~\cite{Candelas:1990rm,Aspinwall:1991ce}.

The correlators are interesting in their own right as they compute Gromov--Witten invariants~\cite{gromov1985pseudo}. Unfortunately, contributions from worldsheet instantons make them difficult to calculate directly. In practice, one uses either the holomorphic anomaly equations to relate correlators at genus $g$ to lower-genus correlators, or mirror symmetry to relate correlators in the A-model to those in the B-model.

\subsubsection*{The B-model}

Unlike the A-model, which exists for any K\"ahler target space, the axial R-symmetry generated by the pair $(J,-\bar{J})$ is anomalous unless $c_{1}(M)=0$. Hence, for the B-model twist to be well defined at the quantum level, one must restrict to Calabi--Yau target spaces.

With this restriction, it is possible to write the action as a $\QB$-exact piece, plus a term that is independent of both the worldsheet metric and the target-space K\"ahler form.\footnote{We do not give the action explicitly as we did for the A-model as it is not enlightening in this case. It can be found in e.g.~\cite{Vonk:2005yv}} Thus one finds that correlators depend only on the complex structure moduli, and again one has a quasi-topological theory. The theory localises on solutions to
\begin{equation}
    \dd x = \dd\bar{x} = 0,
\end{equation}
and hence we can calculate exact results by restricting to constant maps $x\colon\Sigma \rightarrow M$. This observation often makes B-model correlators easier to calculate (though perhaps less interesting from a mathematical point of view).

To study the $\QB$ cohomology ring, it is once again useful to go to the infinite-volume limit. There one finds the dimension-zero operators take the form
\begin{equation}
    \mathcal{O}_{\beta} = \beta(x)^{\mu_{1}\ldots\mu_{p}}{}_{\bar{\nu}_{1}\ldots\bar{\nu}_{q}} \theta_{\mu_{1}}\ldots\theta_{\mu_{p}}\bar{\eta}^{\bar{\nu}_{1}}\ldots\bar{\eta}^{\bar{\nu}_{q}},
\end{equation}
where the $\theta_{\mu}$ and $\bar{\eta}^{\bar{\nu}}$ are left- and right-moving fermions (though scalars under the twisted Lorentz symmetry), both with $\Uni(1)$ charge $+1$. We can identify $\theta_{\mu}\in x^{*}(T^{*1,0})$ and $\bar{\eta}^{\bar{\nu}}\in x^{*}(T^{0,1})$, and so the space of operators corresponds to sections of $\ext^{q}T^{*0,1}\otimes \ext^{p}T^{1,0}$. Using this identification, one finds
\begin{equation}\label{eq:B-model_BRST}
    G^{+}_{-1/2} \sim \tfrac{1}{2}(\delb + \del^{\dagger}), \qquad \bar{G}^{+}_{-1/2}\sim \tfrac{1}{2} (\delb-\del^{\dagger}), \qquad \QB \sim \delb.
\end{equation}
The chiral ring of physical operators is therefore isomorphic to the bundle-valued Dolbeault cohomology groups
\begin{equation}
    H_{\delb}^{\bullet}(M,\ext^{\bullet}T^{1,0}) = \bigoplus_{p,q} H^{q}_{\delb}(M,\ext^{p}T^{1,0}).
\end{equation}
The holomorphic $(n,0)$-form of the Calabi--Yau target space then gives an isomorphism between this and the usual Dolbeault complex on $(n-p,q)$-forms.

Looking at \eqref{eq:B-model_BRST}, we see that the left and right BRST operators do not correspond to the Dolbeault operators, but instead raise the antiholomorphic degree while lowering the holomorphic degree of forms. This means that the left and right fermion numbers cannot be matched to the holomorphic and antiholomorphic degree of the form respectively, and the BRST double complex \eqref{eq:BRST_double_complex} cannot be identified with a Dolbeault complex on the target space. However, the \emph{total} cohomology can still be identified with the Dolbeault cohomology as above \cite{Bershadsky:1993cx}. Moreover, given that the B-model is independent of the K\"ahler moduli, this chiral ring is exact at finite volume even though it was derived at infinite volume.

\subsubsection*{The 1-loop partition function}

Finally, we will briefly review the 1-loop partition functions of the A- and B-models. One can calculate the 1-loop partition function from the free energy which is given by~\cite{hep-th/9209085,Bershadsky:1993ta,Bershadsky:1993cx}
\begin{equation}\label{eq:1-loop_free_energy}
    \mathcal{F}_{1} = \frac{1}{2}\int\frac{\dd\tau\dd\bar{\tau}}{\tau_{2}} (-1)^{F}F_{L}F_{R}\,\ee^{2\pi \ii \tau H_{L}}\ee^{-2\pi\ii \bar{\tau}H_{R}},
\end{equation}
where $F_{L}$ and $F_{R}$ are the left- and right-moving fermion number operators respectively, $F = F_{L}+F_{R}$ is the total fermion number operator, $H_{L}=\{Q_{L},Q_{L}^{\dagger}\}$ is the left-moving Hamiltonian, and similarly for $H_{R}$. The BRST operators are given by $Q_{L} = G^{+}_{-1/2}$ and $Q_{R} = \bar{G}^{\pm}_{-1/2}$ depending on whether we are in the A- or B-model. Integrating over the upper half-plane, this can be shown to be equal to~\cite{Bershadsky:1993cx,deBoer08c}
\begin{equation}\label{eq:free_energy_2}
    \mathcal{F}_{1} = \delta(H_{L}-H_{R})\frac{1}{2}\log \left[\prod_{F_{L},F_{R}} (\det{}'(H_{L}+H_{R}))^{(-1)^{F}F_{L}F_{R}} \right],
\end{equation}
with the partition function then given by $\ee^{-\mathcal{F}_{1}}$.

From a worldsheet perspective, this is simply an alternating product of determinants of Hamiltonians acting on the double BRST complex \eqref{eq:BRST_double_complex}. To understand what this calculates on the target space, one needs to use the target-space identification of \eqref{eq:BRST_double_complex}. For the A-model at infinite volume ($\omega\to\infty$), this identification is clear and \eqref{eq:free_energy_2} becomes an alternating product of Laplacians acting on the Dolbeault complex:
\begin{equation}\label{eq:A-model_1-loop}
    Z^{\text{A}}_{1} \overset{\omega\rightarrow \infty}{=} \left[ \prod_{p,q}(\det{}'\Delta^{p,q})^{(-1)^{p+q}pq} \right]^{-1/2} ,
\end{equation}
which can be written in terms of holomorphic Ray--Singer torsions:
\begin{equation}
   Z^{\text{A}}_{1} \overset{\omega\rightarrow \infty}{=} \frac{I_{1}}{I_{0}^{3}} .
\end{equation}
For finite volume, the answer will receive contributions from strings wrapping cycles in $M$~\cite{Bershadsky:1993ta}.

For the B-model, understanding the 1-loop calculation in terms of the double BRST complex is more opaque as the target-space BRST complex is more complicated. Despite this, one can show that, up to moduli-independent terms (i.e.~a multiplicative constant), one obtains the same answer as for the A-model~\cite{Bershadsky:1993cx}:
\begin{equation}
    Z^{\text{B}}_{1} = \frac{I_{1}}{I_{0}^{3}}.
\end{equation}
This holds for arbitrary Calabi--Yau target spaces, even for finite volume.

We emphasise the form of the 1-loop partition function given by the right-hand side of \eqref{eq:A-model_1-loop}, as it will appear again when we look at the 1-loop partition function of the $\G_{2}$ and $\Spin(7)$ strings. The 1-loop partition function calculates a quantity related to the target-space BRST double complex, given by a particular product of determinants of Laplacians on that complex as shown. Given the similarity to the analytic torsion of one-dimensional complexes \cite{RaySinger1,Cheeger2651}, we shall refer to the quantity \eqref{eq:A-model_1-loop}, when applied to arbitrary complexes, as the \emph{analytic torsion} of the double complex.

In \cite{Pestun:2005rp}, Pestun and Witten showed that this result for $Z^{\text{B}}_{1}$ could be obtained by BV quantising the target-space theory defined by
\begin{equation}\label{eq:pestun_action}
    S = \int_{M} b_{00}\wedge \del\delb b_{22} + b_{11}\wedge \del\delb b_{11},
\end{equation}
where the subscripts denote the $(p,q)$-form degree. Furthermore, they showed that this action has a natural interpretation as the quadratic variation of the Hitchin functional for a generalised Calabi--Yau structure, where the variation is taken within a fixed cohomology class~\cite{hitchin2003}. We review the generalised Hitchin functional in appendix \ref{app:gcy}. This provides a link between topological strings at 1-loop and geometric structures in the $\Orth{d,d}$ geometry of Hitchin that we will explore further in this paper. 

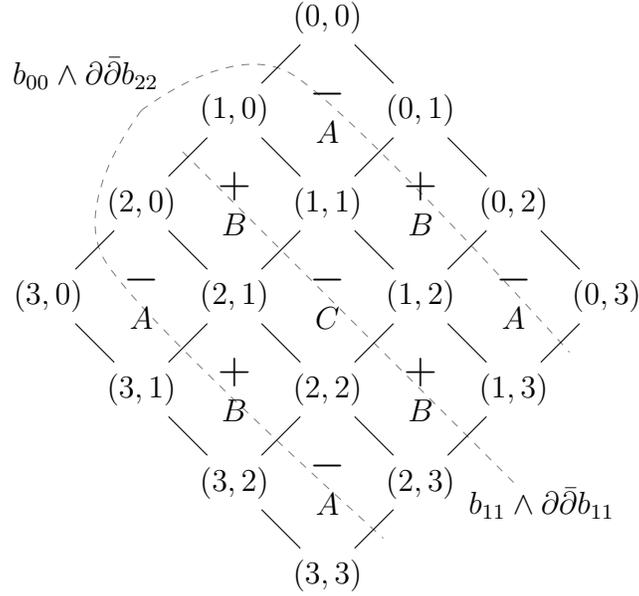
\begin{figure}
\centering
\begin{tikzpicture}[scale=1.75]
\draw[black] (0,0) grid [rotate=45] (3,3);
\node[circle,fill=white,draw=white,align=center,text=black,inner sep=0pt,minimum size=0pt] at (0*1.414,3*1.414) {$(0,0)$};
\node[circle,fill=white,draw=white,align=center,text=black,inner sep=0pt,minimum size=0pt] at (-.5*1.414,2.5*1.414) {$(1,0)$};
\node[circle,fill=white,draw=white,align=center,text=black,inner sep=0pt,minimum size=0pt] at (.5*1.414,2.5*1.414) {$(0,1)$};
\node[circle,fill=white,draw=white,align=center,text=black,inner sep=0pt,minimum size=0pt] at (1*1.414,2*1.414) {$(0,2)$};
\node[circle,fill=white,draw=white,align=center,text=black,inner sep=0pt,minimum size=0pt] at (0*1.414,2*1.414) {$(1,1)$};
\node[circle,fill=white,draw=white,align=center,text=black,inner sep=0pt,minimum size=0pt] at (-1*1.414,2*1.414) {$(2,0)$};
\node[circle,fill=white,draw=white,align=center,text=black,inner sep=0pt,minimum size=0pt] at (-1.5*1.414,1.5*1.414) {$(3,0)$};
\node[circle,fill=white,draw=white,align=center,text=black,inner sep=0pt,minimum size=0pt] at (-0.5*1.414,1.5*1.414) {$(2,1)$};
\node[circle,fill=white,draw=white,align=center,text=black,inner sep=0pt,minimum size=0pt] at (0.5*1.414,1.5*1.414) {$(1,2)$};
\node[circle,fill=white,draw=white,align=center,text=black,inner sep=0pt,minimum size=0pt] at (1.5*1.414,1.5*1.414) {$(0,3)$};
\node[circle,fill=white,draw=white,align=center,text=black,inner sep=0pt,minimum size=0pt] at (1*1.414,1*1.414) {$(1,3)$};
\node[circle,fill=white,draw=white,align=center,text=black,inner sep=0pt,minimum size=0pt] at (0*1.414,1*1.414) {$(2,2)$};
\node[circle,fill=white,draw=white,align=center,text=black,inner sep=0pt,minimum size=0pt] at (-1*1.414,1*1.414) {$(3,1)$};
\node[circle,fill=white,draw=white,align=center,text=black,inner sep=0pt,minimum size=0pt] at (-.5*1.414,.5*1.414) {$(3,2)$};
\node[circle,fill=white,draw=white,align=center,text=black,inner sep=0pt,minimum size=0pt] at (.5*1.414,.5*1.414) {$(2,3)$};
\node[circle,fill=white,draw=white,align=center,text=black,inner sep=0pt,minimum size=0pt] at (0*1.414,0*1.414) {$(3,3)$};
\node[align=center] at (0*1.414,2.5*1.414) {\mbox{\Large$-$}\\$A$};
\node[align=center] at (0.5*1.414,2*1.414) {\mbox{\Large$+$}\\$B$};
\node[align=center] at (-0.5*1.414,2*1.414) {\mbox{\Large$+$}\\$B$};
\node[align=center] at (0*1.414,1.5*1.414) {\mbox{\Large$-$}\\$C$};
\node[align=center] at (1*1.414,1.5*1.414) {\mbox{\Large$-$}\\$A$};
\node[align=center] at (-1*1.414,1.5*1.414) {\mbox{\Large$-$}\\$A$};
\node[align=center] at (0.5*1.414,1*1.414) {\mbox{\Large$+$}\\$B$};
\node[align=center] at (-0.5*1.414,1*1.414) {\mbox{\Large$+$}\\$B$};
\node[align=center] at (0*1.414,0.5*1.414) {\mbox{\Large$-$}\\$A$};
%
%
\draw[dashed,draw opacity=0.5] (1*1.414,0.5*1.414) -- (-.8*1.414,2.3*1.414);
\node[align=center] at (1.15*1.414,0.39*1.414) {$b_{11}\wedge \partial\bar\partial b_{11}$};
\draw[dashed,draw opacity=0.5] plot [smooth] coordinates {(-1.*1.414,2.5*1.414) (-0.2*1.414,2.7*1.414) (1.3*1.414,1.2*1.414)};
\draw[dashed,draw opacity=0.5] plot [smooth] coordinates {(-1*1.414,2.5*1.414) (-1.2*1.414,1.7*1.414) (.3*1.414,0.2*1.414)};
\node[align=center] at (-1.3*1.414,2.70*1.414) {$b_{00}\wedge \del\delb b_{22}$};
\end{tikzpicture}
\caption{Figure adapted from \cite{Pestun:2005rp}. Complex conjugation, Hodge duality and contraction with the holomorphic 3-form $\Omega$ leave only three independent determinants which all $\det{}' \Delta^{p,q}$ can be expressed in terms of. For example, $\det{}' \Delta^{0,0}=A$ and $\det{}' \Delta^{1,1}=AB^2C$. The analytic torsion (the 1-loop partition function) is then given by $(A^{-4}B^4 C^{-1})^{1/2}$, in agreement with \eqref{eq:A-model_1-loop}. Note that upon BV quantising \eqref{eq:pestun_action}, the first and second terms contribute $A^{-4}B^2$ and $B^2 C^{-1}$ respectively, corresponding to the products of the determinants along the dotted lines in the figure.}
\label{fig:SU_diagram}
\end{figure}

Note that the 1-loop partition function has a nice pictorial interpretation in terms of the Dolbeault complex, as we illustrate in Figure \ref{fig:SU_diagram} \cite{Pestun:2005rp}. Briefly, the determinant of $\Delta^{p,q}$ can be decomposed into a product of Laplacians acting on the subspaces appearing in the Hodge decomposition of $\Omega^{p,q}(M)$. These four spaces are represented by the four squares surrounding each vertex in the complex. By Hodge duality and complex conjugation, there are only three independent values these can take, represented by $A$, $B$ and $C$ in the diamond. It turns out that the 1-loop partition function can be read off from the Hodge diamond by multiplying these factors together with alternating powers of $\pm\tfrac{1}{2}$, in a ``checkerboard pattern'', as shown in the figure. We give a brief review of $\zeta$-regularised determinants of Laplacians in Appendix \ref{app:zeta-dets}.

\subsection{The \texorpdfstring{$\G_{2}$}{G2} string}

The existence of a topological string with $\G_{2}$ target space was conjectured in \cite{Shatashvili:1994zw} and further studied in \cite{deBoer:2005pt,deBoer08b,deBoer08c}, yet its properties are still not fully understood. Evidence for the twisting procedure comes from the extended worldsheet symmetry implied by $\G_{2}$ holonomy of the target space. Indeed, given a $\G_{2}$ structure $\gphi \in \Omega^{3}(M)$, one can define the operators
\begin{align}
    \Phi &= \frac{1}{3!}\gphi_{\mu\nu\rho}\psi^{\mu}\psi^{\nu}\psi^{\rho}, \\
    K &= \frac{1}{2}\gphi_{\mu\nu\rho}\psi^{\mu}\psi^{\nu}\del x^{\rho} ,\\
    X &= -\frac{1}{4!}(*\gphi)_{\mu\nu\rho\sigma} \psi^{\mu}\psi^{\nu}\psi^{\rho}\psi^{\sigma} - \frac{1}{2}g_{\mu\nu}\psi^{\mu}\del\psi^{\nu}, \\
    M &= -\frac{1}{3!}(*\gphi)_{\mu\nu\rho\sigma}\psi^{\mu}\psi^{\nu}\psi^{\rho}\del x^{\sigma} - \frac{1}{2}g_{\mu\nu}\del x^{\mu}\del\psi^{\nu} + \frac{1}{2}g_{\mu\nu}\psi^{\mu} \del^{2}x^{\nu}.
\end{align}
These operators along with the $N=1$ superconformal operators $(T,G)$ define a closed algebra denoted by $\mathcal{SW}_{[0,\tfrac{21}{2}]}(\tfrac{3}{2},\tfrac{3}{2},2)$, a particular supersymmetric $\mathcal{W}$-algebra.\footnote{This is true in the free-field or infinite-volume limit. More generally, properties like the Jacobi identity hold only modulo the ideal generated by the null field $N$ defined by \cite{Figueroa-OFarrill:1996tnk}
\begin{equation}
    N = 4GX - 2\Phi K - 4\del M - \del^{2} G.
\end{equation}}

As in the A/B-models, one can identify a spectral flow-like operator which implements the twisting. It turns out that, in this case, it is easier to understand the theory via the states, rather than the chiral ring. In the topological theory, the chiral ring is in one-to-one correspondence with the R ground states of the untwisted theory. These become the physical states in the twisted theory and hence one obtains an equivalent description of the theory.

To study the states of the theory, we introduce the operators $T_{I}=-\tfrac{1}{5}X$ and $G_{I} = \tfrac{\ii}{\sqrt{15}}\Phi$. One finds they define an $N=1$ superconformal algebra of central charge $c = \tfrac{7}{10}$. This is a minimal model known as the tri-critical Ising model. One can write the original energy-momentum tensor $T$ as
\begin{equation}
    T = T_{I} + T_{r}, \qquad T_{I}(z)T_{r}(w) = \text{regular},
\end{equation}
where $T_{r}$ defines a Virasoro algebra commuting with $T_{I}$ with central charge $c = \tfrac{98}{10}$. States are then labeled by two quantum numbers, $\vert\Delta_{I}, \Delta_{r}\rangle$, specifying their weights under $T_{I}$ and $T_{r}$. Since $T_{I}$ defines a minimal model, we know the weights of the conformal primaries of the theory. They split into an NS and an R sector:
\begin{equation}
        \text{NS}\colon  \quad 0, \quad \tfrac{1}{10}, \quad \tfrac{6}{10}, \quad \tfrac{3}{2}, \qquad
        \text{R}\colon \quad \tfrac{7}{16}, \quad \tfrac{3}{80}.
\end{equation}

We also know that the R ground states of the full theory must have total weights $\Delta = \Delta_{I}+\Delta_{r} = \tfrac{d}{16} = \tfrac{7}{16}$. Therefore, we find that the R ground states are
\begin{equation}
    \left|\tfrac{7}{16},0 \right>, \qquad \left|\tfrac{3}{80}, \tfrac{2}{5} \right>.
\end{equation}
One can then use the state with only non-zero tri-critical Ising weight to define a map between the R-sector ground states and certain special NS states. This is the analogue of the spectral flow operator of $N=(2,2)$ theories. Indeed, using the fusion rules of the tri-critical Ising model, we have the following NS states:
\begin{equation}\label{eq:g2_NS}
    \left| 0,0 \right>,  \qquad \left|\tfrac{1}{10},\tfrac{2}{5} \right>, \qquad \left| \tfrac{6}{10},\tfrac{2}{5} \right>, \qquad \left| \tfrac{3}{2},0 \right>.
\end{equation}
By examining the total weight of the states, we notice that these states are respectively generated by the operators\footnote{One could ask why we do not include fields of the form $\del x$ or $\del\psi$. The reason is that the derivative ensures that these have conformal weight $\geq1$ and hence cannot be scalars under a twisted Lorentz symmetry. They should therefore not be included in a set of local physical topological operators.}
\begin{equation}
    f(x), \qquad A_{\mu}(x)\psi^{\mu}, \qquad B_{\mu\nu}(x)\psi^{\mu}\psi^{\nu}, \qquad C_{\mu\nu\rho}(x)\psi^{\mu}\psi^{\nu}\psi^{\rho},
\end{equation}
and so they define target-space 0-, 1-, 2-, and 3-forms. To ensure they have the correct weight under $T_{I}$, the coefficients must be restricted to lie in particular $\G_{2}$ representations. In particular, they must fall into the irreducible representations that appear in the $\G_{2}$ instanton complex of \cite{Carrion98a}, which will be explored in more detail in the next section.

One can also use the R ground state to twist the model to produce an energy-momentum tensor $T_{\text{twist}}$ whose algebra has vanishing central charge. Bosonising the theory, one can write
\begin{align}
    \Phi &= \exp\left(\tfrac{3\ii}{\sqrt{5}}\rho\right), \\
    X &= (\del\rho)^{2} + \frac{1}{4\sqrt{5}}\del^{2}\rho, \\
    \left|\tfrac{7}{16},0\right> &= \exp\left(\tfrac{-5\ii}{4\sqrt{5}}\rho\right). \label{eq:bosonised_R_state}
\end{align}
Given that one can relate twisted and untwisted correlators of the A/B-models by $2g-2$ insertions of the spectral flow operator at genus $g$, one may guess that a twisted $\G_{2}$ string is obtained by inserting $2g-2$ copies of \eqref{eq:bosonised_R_state}. This has the effect of shifting the energy-momentum tensor induced by $X$ to
\begin{equation}
    X_{\text{twist}} = (\del\rho)^{2} + \frac{3}{2\sqrt{5}}\del^{2}\rho.
\end{equation}
Taking the total twist $T_{\text{twist}} = \tfrac{1}{5}X_{\text{twist}} + T_{r}$, one finds an algebra with vanishing central charge. The NS states of \eqref{eq:g2_NS} then have a shifted $\Delta_{I}$ weight and become
\begin{equation}\label{eq:G2_NS_states}
    \left| 0,0 \right>, \qquad \left|-\tfrac{2}{5},\tfrac{2}{5} \right>, \qquad \left| -\tfrac{2}{5}, \tfrac{2}{5} \right>, \qquad \left|0,0\right>.
\end{equation}
In particular, they have total weight zero -- a necessary condition for a physical state in a topological theory.

It remains to be seen whether there exists a nilpotent operator $Q$ that is a scalar with respect to the twisted Lorentz algebra such that $T_{\text{twist}}$ is $Q$-exact. In \cite{deBoer:2005pt}, it was argued that the correct operator is a particular conformal block of the supersymmetry generator $G_{-1/2}$, which was denoted by $G_{-1/2}^{\downarrow}$.\footnote{Note that Fiset and Gaberdiel~\cite{Fiset:2021azq} show that the cohomology of $G_{-1/2}^{\downarrow}$ is not restricted to the chiral ring and so it cannot be the exact BRST operator which captures the geometry of the target space (though the honest BRST operator is likely related to $G_{-1/2}^{\downarrow}$). For our purposes, we need only the identification of the complex of special NS states as later we will identify the analogue of the correct BRST operator in the target space.} While it was not shown that $T_{\text{twist}}$ is exact with respect to this operator, it was argued that $G_{-1/2}^{\downarrow}$ is indeed nilpotent and maps the special NS states within themselves:
\begin{equation}
    \left|0,0\right> \xrightarrow{\quad G_{-1/2}^{\downarrow}\quad } \left|\tfrac{1}{10},\tfrac{2}{5} \right> \xrightarrow{\quad G_{-1/2}^{\downarrow}\quad } \left| \tfrac{6}{10},\tfrac{2}{5} \right> \xrightarrow{\quad G_{-1/2}^{\downarrow}\quad } \left| \tfrac{3}{2},0 \right>
\end{equation}
This complex has a target-space interpretation as the $\G_{2}$ Dolbeault complex \eqref{eq:G2_Dolbeault} which we will describe in the following section. The physical states should therefore be in the cohomology of this complex. In addition, a heuristic argument was given that the path integral localises on constant maps $x\colon\Sigma \rightarrow M$ and so one would not expect instanton corrections at finite volume.

One finds completely analogous results for the right-moving sector and so the total BRST operator should be
\begin{equation}
    Q = G^{\downarrow}_{-1/2} + \bar{G}^{\downarrow}_{-1/2},
\end{equation}
with the physical states given by tensor products of left- and right-moving states, each in \eqref{eq:G2_NS_states}, that are annihilated by $Q$. This poses the question: what is the target-space interpretation of the BRST double complex \eqref{eq:BRST_double_complex}? One of the results of this paper is to show that there exists a double complex on any $\G_{2}$ manifold which naturally represents this worldsheet complex. Moreover, we will examine its relation to the 1-loop partition function and compare our results to those found in \cite{deBoer08c}.

\subsection{The \texorpdfstring{$\Spin(7)$}{Spin(7)} string}\label{sec:spin7_string_review}

The topological $\Spin(7)$ string was also conjectured to exist in \cite{Shatashvili:1994zw} but there has been little further study since then.\footnote{Though see, for example, \cite{Benjamin:2014kna,Braun:2019lnn,Fiset:2020lmg,Fiset:2021ruv}} We will now outline some of the evidence for its existence.

As before, a target space with $\Spin(7)$ holonomy implies an extended worldsheet symmetry which is required for the twisting procedure. If $\Theta \in \Omega^{4}(M)$ is the self-dual 4-form defining the $\Spin(7)$ structure, we can define the operators
\begin{align}
    \tilde{X} &= \frac{1}{4!}\Theta_{\mu\nu\rho\sigma}\psi^{\mu}\psi^{\nu}\psi^{\rho}\psi^{\sigma} + \frac{1}{2}g_{\mu\nu}\psi^{\mu}\del\psi^{\nu}, \\
    \tilde{M} &= \frac{1}{3!}\Theta_{\mu\nu\rho\sigma}\psi^{\mu}\psi^{\nu}\psi^{\rho}\del x^{\sigma} - \frac{1}{2}g_{\mu\nu}\del x^{\mu}\del\psi^{\nu} + \frac{1}{2} g_{\mu\nu}\psi^{\mu}\del^{2}x^{\nu}.
\end{align}
These, together with the $N=1$ superconformal generators $(T,G)$, form a closed algebra. The rescaled operator $T_{I} = \tfrac{1}{8}\tilde{X}$ generates a Virasoro algebra with central charge $c=\tfrac{1}{2}$, known as the bosonic Ising model. This plays the same role as the tri-critical Ising model in the $\G_{2}$ string and will be important for the putative twisting procedure.

Once again, it is easier to understand the theory via the states. We can write the total energy-momentum tensor as
\begin{equation}
    T = T_{I} + T_{r}, \qquad T_{I}(z)T_{r}(w) = \text{regular},
\end{equation}
where $T_{r}$ defines a Virasoro algebra commuting with $T_{I}$ of central charge $c=\tfrac{23}{2}$. We can therefore label states as $\left|\Delta_{I},\Delta_{r}\right>$ with respect to their weights under $T_{I}$ and $T_{r}$. Since the bosonic Ising model is a minimal model, we know the possible weights are given by
\begin{equation}
    \Delta_{I}\colon \quad 0, \quad \tfrac{1}{16}, \quad \tfrac{1}{2}.
\end{equation}
Since the total weight of the R ground states must be equal to $\tfrac{d}{16} = \tfrac{8}{16}=\tfrac{1}{2}$, one finds that they must be
\begin{equation}
    \left|0,\tfrac{1}{2} \right>, \qquad \left|\tfrac{1}{16},\tfrac{7}{16}\right>, \qquad \left| \tfrac{1}{2},0 \right>.
\end{equation}
Once again, we find a state with only non-vanishing bosonic Ising weight which we can use as a spectral flow-like operator to define a map between the R ground states and certain NS highest-weight states. Indeed, using the fusion rules, the states in the NS sector are
\begin{equation}
    \left| 0 , 0 \right>,  \qquad \left|\tfrac{1}{16},\tfrac{7}{16}\right>, \qquad \left| \tfrac{1}{2},\tfrac{1}{2} \right>.
\end{equation}
By examining the total weight, we see that these states are generated by the operators\footnote{Once again, we do not include terms with $\del\phi$ or $\del\psi$ as they cannot be scalars with respect to a twisted Lorentz symmetry.}
\begin{equation}
    f(x), \qquad A_{\mu}(x)\psi^{\mu} ,\qquad B_{\mu\nu}(x)\psi^{\mu}\psi^{\nu}.
\end{equation}
Hence, the states are related to target-space 0-, 1-, and 2-forms. To ensure the states have the correct $T_{I}$ weights, we find that $B_{\mu\nu}$ must be restricted to lie in the $\rep{7}$ of $\Spin(7)$. Intriguingly, these representations are precisely those that appear in the $\Spin(7)$ instanton complex \cite{Carrion98a}.

We can also use the R ground state to form a twisted energy-momentum tensor with vanishing central charge. Indeed, bosonising the theory, one can write
\begin{align}
    \tilde{X} &= (\del\rho)^{2} + \frac{1}{4\sqrt{3}}\del^{2}\rho, \\
    \left| \tfrac{1}{2}, 0 \right> &= \exp\left(\tfrac{3\ii}{2\sqrt{3}}\rho\right) .\label{eq:Spin(7)_spectral_flow}
\end{align}
The insertion of $2g-2$ copies of \eqref{eq:Spin(7)_spectral_flow} into correlators is equivalent to twisting the energy-momentum tensor to
\begin{equation}
    \tilde{X}_{\text{twist}} = (\del\rho)^{2} + \frac{5}{4\sqrt{3}}\del^{2}\rho .
\end{equation}
Taking the twist of the full theory to be $T_{\text{twist}} = \tfrac{1}{8}\tilde{X}_{\text{twist}} + T_{r}$, one finds a Virasoro algebra with vanishing central charge. Furthermore, the weights of the NS states under this twisted algebra become
\begin{equation}
    \left| 0 , 0 \right>,  \qquad \left|-\tfrac{7}{16},\tfrac{7}{16}\right>, \qquad \left| -\tfrac{1}{2},\tfrac{1}{2} \right>.
\end{equation}
These have total weight zero under the twisted Lorentz symmetry, a necessary condition for the physical states of a topological theory.

It is still unknown whether there is an appropriate nilpotent operator $Q$ such that $T_{\text{twist}}$ is $Q$-exact. However, we find it highly suggestive that states of weight zero in the NS sector define the vector spaces of the $\Spin(7)$ instanton complex of \cite{Carrion98a}, much like we saw for the $\G_{2}$ string.\footnote{In fact, it is also possible to formulate the left- and right-moving sectors of the A- and B-models in terms of the instanton complex for $\SU(n)$ structures.} We therefore expect that the correct operator is some sub-operator of $G$, suitably projected so that one gets the correct target-space complex. We will provide some evidence for this in Section \ref{sec:top_spin7_string}. The full theory contains states that are tensor products of the left- and right-moving sectors, and the physical operators in the chiral ring again correspond to cohomology classes of $Q=Q_{L}+Q_{R}$.

Despite not knowing the precise worldsheet theory, we will show that there exists a natural double complex on any $\Spin(7)$ target space that seems to encode the left- and right-moving states and gives candidates for the left- and right-moving BRST operators. We will use this to make a conjecture for the partition function at 1-loop.


\section{\texorpdfstring{$G$}{G}-structure complexes for special holonomy manifolds}\label{sec:G_complexes}

It is very striking that the left- and right-moving states selected by the topological twist precisely form the vector spaces in the instanton complexes of \cite{Carrion98a}. These are particular complexes that arise on manifolds with $G$-structure $G\subset \Orth{d}$ as a subcomplex of de Rham.\footnote{The cohomology of these complexes is also related to the moduli space of $G$-instantons on these manifolds.}
Given the appearance of these complexes in topological strings, we will briefly review them for $\G_{2}$ and $\Spin(7)$ holonomy manifolds and analyse their Hodge theory. We will find a doubled version of these complexes in later sections by lifting to $\Orth{d,d}\times \bbR^{+}$ geometry, and match them to the BRST double complex. We will mirror the techniques used in this section when analysing the properties of these double complexes.

\subsection{A \texorpdfstring{$\G_{2}$}{G2} complex and Hodge theory}
\label{sec:g2-complex-adjoints}

Let $(M,\gphi)$ be a seven-dimensional manifold with a (possibly torsionful) $\G_{2}$ structure. The $\G_{2}$ structure defines a unique metric $g$ and hence a Hodge star operator $*$. The intrinsic torsion of the structure is encoded by $\dd\gphi$ and $\dd*\gphi$, which both vanish if and only if the intrinsic torsion vanishes \cite{Fernandez82a}. Any such manifold admits a decomposition of differential forms into irreducible $\G_{2}$ representations as \cite{Bryant87a}
\begin{align}
    \ext^{0}T^{*} &= \ext^{0}_{\rep{1}}T^{*}, \label{eq:G2_decomp_a} \\
    \ext^{1}T^{*} &= \ext^{1}_{\rep{7}}T^{*} ,\label{eq:G2_decomp_b} \\
    \ext^{2}T^{*} &= \ext^{2}_{\rep{7}}T^{*} \oplus \ext^{2}_{\rep{14}}T^{*}, \label{eq:G2_decomp_c} \\
    \ext^{3}T^{*} &= \ext^{3}_{\rep{1}}T^{*} \oplus \ext^{3}_{\rep{7}}T^{*} \oplus \ext^{3}_{\rep{27}}T^{*},\label{eq:G2_decomp_d}
\end{align}
where the subscript denotes the dimension of the $\G_{2}$ representation and we are using the shorthand $T^*\equiv T^*M$. Higher-degree differential forms have similar decompositions via Hodge duality. A precise definition of the subspaces in terms of $(\gphi,*\gphi)$ is given in Appendix \ref{app:G2_identities}. We will denote the space of sections of $p$-forms in the $\mathrm{r}$-dimensional representation as $\Omega^{p}_{\rep{r}}(M)$, and the projection onto those subspaces by $\Pr{p}{r}$.

Given such a decomposition, consider the following sequence of maps defined by composing the de Rham differential with certain projections \cite{Fernandez98a,salamon1989}
\begin{equation}\label{eq:G2_Dolbeault}
    \dc \colon \Omega^{0}_{\rep{1}}(M) \xrightarrow{\quad \dd \quad} \Omega^{1}_{\rep{7}}(M) \xrightarrow{\quad \Pr{2}{7} \dd \quad} \Omega^{2}_{\rep{7}}(M)  \xrightarrow{\quad \Pr{3}{1} \dd \quad} \Omega^{3}_{\rep{1}}(M) 
\end{equation}
Provided the intrinsic torsion of the $\G_{2}$ structure has no component in the $\rep{14}$, one finds $\dc^2=0$ and so the above sequence is actually a complex -- we will then refer to \eqref{eq:G2_Dolbeault} as the ``$\G_{2}$ complex'' \cite{Carrion98a}. For the remainder of this section we will restrict to torsion-free $\G_{2}$ structures, i.e.~those with $\dd\gphi = \dd*\gphi = 0$, and hence $G_2$ holonomy.

Given the $\G_{2}$ complex, we can introduce an inner product on each of the vector spaces and consider the Laplacian defined by $\Deltac = \dc\dc^{\dagger} + \dc^{\dagger}\dc$. We do this in a way that will allow for comparison to the usual de Rham Laplacian. First, we define isomorphisms between spaces with the same $\G_{2}$ representation as
\begin{equation}
\arraycolsep = 1.4pt
\begin{array}{rclcrcl}
    \theta_{\rep{1}} \colon \ext^{0}_{\rep{1}}T^{*} &\longrightarrow& \ext^{3}_{\rep{1}}T^{*}, & \qquad \qquad & \theta_{\rep{7}} \colon \ext^{1}_{\rep{7}}T^{*} & \longrightarrow & \ext^{2}_{\rep{7}}T^{*}, \\
    f & \longmapsto & k_{1}f\gphi, & & \lambda & \longmapsto & k_{2}\lambda^{a}\gphi_{abc},
\end{array}
\end{equation}
where $k_{1}$ and $k_{2}$ are constants we will determine later and indices are raised and lowered using the $\G_{2}$ metric defined by $\gphi$. Next we fix an inner product $(\cdot,\cdot )$ to be the standard inner product on 0- and 1-forms:
\begin{equation}
(f,g)_{0}=\int_{M}\vol fg,\qquad (\lambda,\nu )_{1}=\int_{M}\vol\lambda\lrcorner\nu.
\end{equation}
We extend this to an inner product on the whole complex by demanding that it depends only on the representation and not the degree of the $p$-form:
\begin{equation}\label{eq:G2_inner_product}
    (\theta_{\rep{1}}(f),\theta_{\rep{1}}(f'))_{3} = (f,f')_{0}, \qquad (\theta_{\rep{7}}(\lambda),\theta_{\rep{7}}(\lambda'))_{2} = (\lambda,\lambda')_{1},
\end{equation}
where $(\cdot,\cdot)_{p}$ denotes restriction to $p$-forms. Note that this forces $(\cdot,\cdot)$ to be the usual inner product on differential forms, up to possible multiplicative constants that are determined by $k_{1}$ and $k_{2}$.

We can fix the constants by demanding that the following diagram commutes
\begin{equation}
    \begin{tikzcd}[column sep= huge, row sep = huge]
    \Omega^{0}_{\rep{1}} \arrow[r, "\dc"] \arrow[d, "\theta_{\rep{1}}"] & \Omega^{1}_{\rep{7}} \arrow[d, "\theta_{\rep{7}}"] \arrow[r, "\dc"] & \Omega^{2}_{\rep{7}} \arrow[d, "\theta_{\rep{7}}^{-1}"] \arrow[r,"\dc"] & \Omega^{3}_{\rep{1}} \arrow[d, "\theta_{\rep{1}}^{-1}"]  \\
    \Omega^{3}_{\rep{1}} \arrow[r, "\dc^{\dagger}"] & \Omega^{2}_{\rep{7}}  \arrow[r, "\dc^{\dagger}"] & \Omega^{1}_{\rep{7}} \arrow[r, "\dc^{\dagger}"] & \Omega^{0}_{\rep{1}}
    \end{tikzcd}
\end{equation}
If this holds the Laplacians acting on isomorphic $\G_{2}$ representations are equivalent in the sense that
\begin{equation}
    \theta_{\rep{1}}\Deltac^{0} = \Deltac^{3}\theta_{\rep{1}}, \qquad \theta_{\rep{7}}\Deltac^{1} = \Deltac^{2}\theta_{\rep{7}},
\end{equation}
where $\Deltac^{p}$ is the restriction of $\Deltac$ to $p$-forms. We can therefore unambiguously write $\Deltac_{\rep{r}}$ for the Laplacian acting on differential forms in the $\G_{2}$ representation $\rep{r}$. This will be important later when we consider determinants of these Laplacians. A quick calculation shows that the diagram commutes and the Laplacians are isomorphic for
\begin{equation}\label{eq:k_ratio}
    \frac{k_{1}}{k_{2}} = -\frac{3}{7}.
\end{equation}
Finally, we can fix the coefficient $k_{2}$ (and hence $k_{1}$) up to an irrelevant overall sign by demanding that
\begin{equation}\label{eq:G2_Laplacian_constraint}
    \Deltac_{\rep{7}} = \Delta_{\rep{7}},
\end{equation}
where $\Delta = \dd\dd^{\dagger} + \dd^{\dagger}\dd$ is the de Rham Laplacian.\footnote{We emphasise that the de Rham adjoint $\dd^{\dagger}$ is defined by the usual inner product on forms and not the rescaled inner product we have defined for the $\dc$ complex.} It is possible to show that $\Delta$ also commutes with the projection operators and only depends on the $\G_{2}$ representation of the form, not the degree, and hence \eqref{eq:G2_Laplacian_constraint} is well defined. Note that this is a non-trivial constraint since $\Delta_{\rep{7}}$ contains terms coming from $\Pr{2}{14}\dd$, while $\Deltac_{\rep{7}}$ does not. Fortunately, for a torsion-free $\G_{2}$ structure and any $\lambda \in \Omega^{1}_{\rep{7}}$, we have~\cite{bryant2003a}
\begin{equation}
    \dd^{\dagger}\Pr{2}{14}\dd\lambda = 2 \dd^{\dagger}\Pr{2}{7}\dd\lambda .
\end{equation}
With this it is easy to check that \eqref{eq:k_ratio} and \eqref{eq:G2_Laplacian_constraint} impose
\begin{equation}
    k_{2} = -\frac{1}{3},\qquad \qquad k_{1} = \frac{1}{7}.
\end{equation}
The inner product is then given by
\begin{equation}
    (\alpha,\beta )_p = \kappa_{p} \int_{M} \alpha\wedge * \beta, \qquad \kappa_{p} = \begin{cases}
    1 & p = 0,1, \\
    3 & p = 2, \\
    7 & p = 3.
    \end{cases}
\end{equation}

Having fixed the coefficients $k_{1}$ and $k_{2}$ we can now define explicitly the operators $(\dc,\dc^{\dagger},\Deltac)$. Since we have assumed that the $\G_{2}$ structure is torsion-free, we can replace the de Rham differential in the definition of $\dc$ in \eqref{eq:G2_Dolbeault} with the Levi-Civita connection $\nabla$ compatible with the $G_2$ structure. This simplifies calculations as derivatives will then commute with the projection operators since the $\Pr{p}{r}$ are defined in terms of $\gphi$ and $*\gphi$ which are covariantly constant (see Appendix \ref{app:G2_identities} for more details).

In terms of the Levi-Civita connection, the $\dc$ operator acting on $p$-forms becomes
\begin{equation}
    (\dc \omega)_{a_{1}\dots a_{p+1}} = (p+1)(\Pr{p+1}{r})_{a_{1}\dots a_{p+1}}{}^{b_{1}\dots b_{p+1}} \nabla_{[b_{1}}\omega_{b_{2}\dots b_{p+1}]} .
\end{equation}
The adjoint operator $\dc^{\dagger}$ becomes
\begin{align}
    (\dc^{\dagger}\omega)_{a_{1}\dots a_{p-1}} & = - C_{p} \nabla^{b}\omega_{ba_{1}\dots a_{p-1}}, \qquad C_{p} = \begin{cases}
    1 & p = 1, \\
    3 & p = 2, \\
    \frac{7}{3} & p = 3.
    \end{cases}
\end{align}
Note that we do not need to include projectors in the definition of the adjoint operator as we are assuming that $\omega$ lives in one of the spaces in \eqref{eq:G2_Dolbeault}. Again, these definitions have the useful properties that the natural differential operators that one can construct depend only on the $\Gx 2$ representation and not the $p$-form degree of the object on which they act. For example, acting on 1-, 2- or 3-forms we have\footnote{Cf.~these with the relations given by Bryant~\cite{bryant2003a} using the de Rham differentials, which in the notation of that paper read $(\dd^{\rep{p}}_{\rep{q}})^{\dagger}=\dd^{\rep{q}}_{\rep{p}}$.}
\begin{equation}
\dc^{\dagger}|_{1}=\theta_{\rep 1}^{-1}\dc\theta_{\rep 7},\qquad\dc^{\dagger}|_{2}=\theta_{\rep 7}^{-1}\dc\theta_{\rep 7}^{-1},\qquad\dc^{\dagger}|_{3}=\theta_{\rep 7}\dc\theta_{\rep 1}^{-1}.\label{eq:d_dagger}
\end{equation}
Finally, the Laplacian $\Deltac$ can be written as
\begin{align}
    \Deltac^{0} f &= -\nabla^{a}\nabla_{a} f && =\Delta_{\rep{1}}f, \\
    \Deltac^{1} \lambda &= -6 (\Pr{2}{7})_{ba}{}^{cd}\nabla^{b}\nabla_{c}\lambda_{d} - \nabla_{a}\nabla^{b}\lambda_{b} && =\Delta_{\rep{7}}\lambda, \\
    \Deltac^{2} \mu &= -7(\Pr{3}{1})_{ba_{1}a_{2}}{}^{def}\nabla^{b}\nabla_{d}\mu_{ef} - 6 (\Pr{2}{7})_{a_{1}a_{2}}{}^{bc}\nabla_{b}\nabla^{d}\mu_{dc} &&= \theta_{\rep{7}}\Delta_{\rep{7}}\theta_{\rep{7}}^{-1} \mu, \\
    \Deltac^{3}\rho &= -7 (\Pr{3}{1})_{abc}{}^{def}\nabla_{d}\nabla^{g}\rho_{gef} &&= \theta_{\rep{1}} \Delta_{\rep{1}}\theta_{\rep{1}}^{-1} \rho.
\end{align}

Given these operators, it is natural to ask if there is some kind of Hodge decomposition for the spaces $\Omega^{p}_{\rep{r}}$. We defined the inner product $(\cdot,\cdot)$ in terms of the usual inner product on differential forms, differing on 2- and 3-forms by a positive multiplicative factor. The usual inner product is positive definite and is $\G_{2}$ invariant,\footnote{This is inherited from the $\SL(7,\bbR)$ invariance of the inner product.} hence it reduces to a positive-definite inner product on the irreducible $\G_{2}$ representations in \eqref{eq:G2_decomp_a}--\eqref{eq:G2_decomp_d}. In particular, it is positive definite on the spaces $(\Omega^{0}_{\rep{1}}, \Omega^{1}_{\rep{7}},\Omega^{2}_{\rep{7}},\Omega^{3}_{\rep{1}})$, and hence $(\cdot,\cdot)$ is positive definite. We therefore have a decomposition of the spaces in \eqref{eq:G2_Dolbeault} as
\begin{equation}
    \Omega^{p}_{\rep{r}} = \check{H}^{p} \oplus \dc \Omega^{p-1}_{\rep{r}'} \oplus \dc^{\dagger}\Omega^{p-1}_{\rep{r}''},
\end{equation}
where $\check{H}^{p}$ is the space of $\Deltac$-harmonic $p$-forms. Of course, with our choice of Laplacian, we have
\begin{equation}
    \check{H}^{0} \simeq \check{H}^{3} \simeq H_{\rep{1}} \simeq \bbR, \qquad \check{H}^{1} \simeq \check{H}^{2} \simeq H_{\rep{7}} = 0,
\end{equation}
where $H_{\rep{r}}$ is the de Rham cohomology group restricted to forms in the $\G_{2}$ representation $\rep{r}$. The fact that the cohomology group depends only on the representation follows from the equivalent statement for $\Delta$. The final equality holds on any manifold of $\G_{2}$ holonomy~\cite{joyce2000}.

\subsection{A \texorpdfstring{$\Spin(7)$}{Spin(7)} complex and Hodge theory}

An eight-dimensional $\Spin(7)$ manifold $M$ is defined by a 4-form $\Theta$ which, unlike the $\G_{2}$ case, is not stable but instead lives in a particular $\GL(8)$ orbit. Any such admissible 4-form defines a metric $g$ as in \cite{2007arXiv0709.4594K} with respect to which we have $\Theta = *\Theta$. We then say the $\Spin(7)$ structure is integrable if there exists a torsion-free compatible connection\footnote{Since $\Theta$ defines a metric $g$, this is equivalent to saying that the Levi-Civita connection is compatible and hence has $\Spin(7)$ holonomy.} which is the case if and only if $\dd \Theta = 0$. As in the $\G_{2}$ case, we can decompose the exterior algebra into $\Spin(7)$ representations:
\begin{align}
    \ext^{0}T^{*} &= \ext^{0}_{\rep{1}}T^{*}, \\
    \ext^{1}T^{*} &= \ext^{1}_{\rep{8}}T^{*}, \\
    \ext^{2}T^{*} &= \ext^{2}_{\rep{7}}T^{*} \oplus \ext^{2}_{\rep{21}}T^{*}, \\
    \ext^{3}T^{*} &= \ext^{3}_{\rep{8}}T^{*} \oplus \ext^{3}_{\rep{48}}T^{*}, \\
    \ext^{4}T^{*} &= \ext^{4}_{\rep{1}}T^{*} \oplus \ext^{4}_{\rep{7}}T^{*} \oplus \ext^{4}_{\rep{27}}T^{*} \oplus \ext^{4}_{\rep{35}}T^{*}.
\end{align}
The definition of these spaces along with the relevant projectors are given in Appendix \ref{app:G2_identities}. Following \cite{Carrion98a}, one can define a sequence of maps built from the de Rham differential and suitable projectors:
\begin{equation}\label{eq:Spin(7)_complex}
    \dc\colon\Omega^{0}_{\rep{1}}(M) \xrightarrow{\quad\, \dd \,\quad} \Omega^{1}_{\rep{8}}(M) \xrightarrow{\quad \Pr{2}{7}\dd \quad} \Omega^{2}_{\rep{7}}(M)
\end{equation}
which, for integrable $\Spin(7)$ structures, defines a complex.

Again, we would like to define an inner product on this complex such that the induced Laplacians $\Deltac$ match the conventional Laplacians $\Delta$ evaluated on $\Omega^{p}_{\rep{r}}$, possibly up to an overall scaling (which drops out when evaluating determinants of $\Deltac$). To do so, we adapt the arguments made in \cite{bryant2003a} for $\G_{2}$ manifolds and find that
\begin{align}
    \dd^{\dagger} \Pr{2}{21} \dd|_{\Omega^{1}_{\rep{8}}} = 3\dd^{\dagger}\Pr{2}{7}\dd|_{\Omega^{1}_{\rep{8}}}, \qquad \Pr{2}{7}\dd^{\dagger}\Pr{3}{48}\dd|_{\Omega^{2}_{\rep{7}}} = \tfrac{12}{7}\Pr{2}{7}\dd^{\dagger}\Pr{3}{8}\dd|_{\Omega^{2}_{\rep{7}}}.
\end{align}
One can then use these identities to show
\begin{equation}
    \Delta^{0}_{\rep{1}} = \dd^{\dagger}\dd , \qquad
    \Delta^{1}_{\rep{8}} = \dd\dd^{\dagger} + 4\dd^{\dagger}\Pr{2}{7}\dd , \qquad
    \Delta^{2}_{\rep{7}} = 4 \Pr{2}{7} \dd \dd^{\dagger}.
\end{equation}
Taking the following inner product on \eqref{eq:Spin(7)_complex}, it is easy to check that $\Deltac^{p}_{\rep{r}} = \Delta^{p}_{\rep{r}}$ as required:
\begin{equation}
    (\alpha,\beta )_p = \kappa_{p} \int_{M} \alpha\wedge * \beta, \qquad \kappa_{p} = \begin{cases}
    1 & p = 0,1, \\
    4 & p = 2.
    \end{cases}
\end{equation}
In terms of the Levi-Civita connection, the $\dc$ operator acting on $p$-forms becomes
\begin{equation}
    (\dc \omega)_{a_{1}\dots a_{p+1}} = (p+1)(\Pr{p+1}{r})_{a_{1}\dots a_{p+1}}{}^{b_{1}\dots b_{p+1}} \nabla_{[b_{1}}\omega_{b_{2}\dots b_{p+1}]} ,
\end{equation}
with the adjoint operator $\dc^{\dagger}$ given by
\begin{align}
    (\dc^{\dagger}\omega)_{a_{1}\dots a_{p-1}} & = - C_{p} \nabla^{b}\omega_{ba_{1}\dots a_{p-1}}, \qquad C_{p} = \begin{cases}
    1 & p = 1, \\
    4 & p = 2.
    \end{cases}
\end{align}


\section{The \texorpdfstring{$\gxg$}{G2 x G2} complex}\label{sec:G2G2_double_complex}

In the previous section we reviewed the $G_2$ complex and its Hodge theory in the case where the $G_2$ structure is torsion-free. In this section, we will give an extension of these ideas which is relevant for type II backgrounds with NSNS flux. In particular, we will see that the relevant geometric structure is that of a torsion-free $\gxg$ structure, which naturally gives rise to a double complex and Laplace-type operators that will turn out to capture information about the topological $G_2$ string. This will be described using the formalism of $O(7,7)\times\bbR^+$ generalised geometry

Generalised geometry has been of great use for understanding supergravity backgrounds that preserve some amount of supersymmetry and thus admit generalised $G$-structures. These structures are characterised by the presence of additional objects, usually in the form of globally defined non-vanishing tensors, that reduce the structure group of the generalised tangent bundle from $\Orth{d,d}\times \bbR^{+}$ to some subgroup $G$. For example, the generalised complex / Calabi--Yau structures of Hitchin and Gualtieri \cite{hitchin2003,gualtieri2004} are respectively $\Uni(\frac{d}{2},\frac{d}{2})$ and $\SU(\frac{d}{2},\frac{d}{2})$ structures. These have found many applications in string theory including formulating topological strings \cite{Kapustin:2004gv,Zucchini:2004ta,Pestun:2005rp,Pestun:2006rj}. We mostly follow the conventions of~\cite{Coimbra:2011nw}, and provide a brief review of the key concepts we will be using in Appendix~\ref{app:gg}.

\subsection{Generalised \texorpdfstring{$\gxg$}{G2 x G2} structures}\label{sec:gen-g2g2}

Let $M$ be a seven-dimensional Riemannian spin manifold and $E$ its $O(7,7)\times\bbR^+$ generalised tangent bundle. Introducing an $O(7)\times O(7)$ generalised metric $G$, or equivalently a Riemannian metric $g$, a two-form gauge field $B$ and a scalar $\dil$, corresponds to specifying an orthogonal decomposition $E=C_+\oplus C_-$, with each $C_{\pm}\cong T$. Let us now also assume that there exist two globally defined real spinors $\epsilon_+ \in S(C_+)$ and  $\epsilon_- \in S(C_-)$. Each define a $\G_{2}$ structure on $M$ given by $\gphi_{\pm}$.\footnote{These are sometimes labelled $\G_{2\pm}$. Moving forward we will mostly omit the signs on $\gphi_{\pm}$ since they can generally be deduced from the context.} When the spinors $\epsilon_{\pm}$ are linearly independent, the $\G_{2}$ structures are orthogonal and intersect on an $\SU(3)$ structure. There may, however, be points on the manifold where the $\epsilon_{\pm}$ align and hence the $\G_{2}$ structures coincide. In this case, the manifold does not admit a conventional global $G$-structure. However, within generalised geometry they define a single global generalised $\gxg$ structure on $E$.\footnote{Note that this is different from the $SU(7)$ structure defined in \cite{Ashmore:2019qii}, which generalises $\Gx2$ geometry to M-theory or string backgrounds with RR flux.}

It turns out that certain supersymmetric backgrounds of string theory compactified to three dimensions can be described by such $\gxg$ structures. For concreteness, consider a type IIB NSNS background of the form $\bbR^{2,1}\times M$ where $M$ is seven-dimensional -- this is the case first described in~\cite{Jeschek2005}. We have two supersymmetry parameters of opposite chirality $\varepsilon_{\pm}$ which decompose under the reduction $\Spin(9,1)\rightarrow \Spin(2,1)\times \Spin(7)$ as
\begin{equation}\label{eq:SUSY_param_decomp}
    \varepsilon_{\pm} = \zeta_{\pm}\otimes \epsilon_{\pm},
\end{equation}
where $\zeta_{\pm}$ and $\epsilon_{\pm}$ are irreducible $\Spin(2,1)$ and $\Spin(7)$ spinors respectively. 

For the background to preserve supersymmetry, the variations of the gravitinos and dilatinos under $\epsilon_{\pm}$ must vanish. These conditions give the Killing spinor equations for the supersymmetry parameters. Under the decomposition \eqref{eq:SUSY_param_decomp}, for vanishing RR fields these equations impose that $\zeta_{\pm}$ is a constant spinor on $\bbR^{2,1}$, and on $M$ we need
\begin{equation}
\begin{aligned}\label{eq:killing-spinor}
    \left( \gamma^{\mu}\del_{\mu}\dil \mp \tfrac{1}{12}\gamma^{\mu\nu\rho}H_{\mu\nu\rho} \right) \epsilon_{\pm} &=0, \\
    \left( \nabla_{\mu} \mp \tfrac{1}{8}\gamma^{\nu\rho}H_{\mu\nu\rho} \right)\epsilon_{\pm} &= 0,
\end{aligned}
\end{equation}
where the $\gamma_\mu$ are gamma matrices for the $\Orth{7}$ structure defined by $g$, and $\nabla$ is the associated Levi-Civita connection. As was shown in \cite{Jeschek2005}, these equations are satisfied if and only if $\epsilon_{\pm}$ define a generalised torsion-free $\gxg$ structure. For generic $(g,H,\dil)$, these equations describe a background preserving minimal supersymmetry in three dimensions. However, when $H$ vanishes and $\dil$ is constant, as must be the case for compact backgrounds~\cite{Gauntlett:2002sc}, these equations imply the preservation of four supercharges or $N=2$ supersymmetry in three dimensions.

\subsection{Torsion-free generalised \texorpdfstring{$\gxg$}{G2 x G2} structures}

Recall that one can always find a torsion-free generalised connection that is compatible with the $O(7)\times O(7)$ generalised metric structure on $M$, giving the analogue of the Levi-Civita connection in generalised geometry. As we review in Appendix~\ref{app:gg}, this connection is not uniquely defined, but there are certain combinations of it which give a unique generalised Ricci tensor and scalar. The generic form of a generalised Levi-Civita connection $D$ in terms of the background fields is given in~\eqref{eq:gen-LC}. 

We begin by finding the conditions that this generalised Levi-Civita connection must satisfy in order to be compatible with a $\gxg$ structure. Since the generalised Levi-Civita connection is torsion-free, the resulting $\gxg$-compatible connection will also be torsion-free.
However, unlike generalised metric structures, the existence of such a compatible connection is, in general, obstructed by the intrinsic torsion of the $\gxg$ structure.
That is, if $D$ is a generalised Levi-Civita connection, the conditions $D\epsilon_+ = D\epsilon_- = 0$ can be solved only if the generalised intrinsic torsion vanishes.

Using similar logic to \cite{Coimbra:2014uxa,Coimbra:2016ydd}, it can be shown that this constraint is equivalent to the background preserving minimal supersymmetry with vanishing RR fluxes, i.e.~that equations~\eqref{eq:killing-spinor} are satisfied. 
Using the expression for a generalised Levi-Civita connection given in~\eqref{eq:gen-LC}, one has that the compatibility conditions which must be imposed are
\begin{equation}
\begin{aligned}
   & D_{a}\epsilon_{+}=  \LC_a \epsilon_{+} - \tfrac{1}{24} H_{abc} \gamma^{bc }\epsilon_{+} -
		 \tfrac{1}{6}  \der_b\dil \gamma_a{}^b\epsilon_{+} 
           + \tfrac14 A^+_{abc}\gamma^{bc}\epsilon_{+} = 0, \\
   & D_{\ba} \epsilon_{+}=\nabla_{\ba}\epsilon_{+}-\tfrac{1}{8}H_{\ba bc}\gamma^{bc} \epsilon_{+} = 0,
\end{aligned}
\end{equation}
which ensure that the connection is compatible with the $G_{2}$ structure defined by $\epsilon_{+}$. There are then similar conditions for compatibility with the $G_2$ structure defined by $\epsilon_{-}$. 

The second equation should be familiar, as it says that $D_{\ba}$ must act on $C_+$ as the $\epsilon_+$-preserving Hull connection $\nabla^{-}$. This connection exists if and only if the ordinary intrinsic torsion of the $G_{2+}$ structure has no component in the $\rep{14}$~\cite{Friedrich:2001nh}. We can combine the two equations and derive a purely algebraic relation between the components of the generalised connection:
\begin{equation}
X_{abc}\gamma^{bc}\epsilon_+ \equiv
\left(\tfrac{1}{12}H_{abc}\gamma^{bc}-\tfrac{1}{6}\partial_{b}\dil\gamma_{a}{}^{b}+\tfrac{1}{4}A^+_{abc}\gamma^{bc}\right)\epsilon_{+}=0.\label{eq:D_a_xi_+}
\end{equation}
Note that this equation holds only when $X_{abc}$ acts on the structure-defining spinor, not for a generic spinor. To find the constraints this imposes, we can use $G_2$ representation theory (for the $G_2$ factor defined by $\epsilon_+$). In general, $X$ is a 1-form taking values in the 21-dimensional adjoint representation of $\Spin(7)$, so under $G_2$ it decomposes as $X\in \rep{7}\times\rep{7} +\rep{7}\times \rep{14} $. The second term gives a 1-form  valued in the adjoint of $G_2$, i.e.~it is the component of $X$ that is compatible with $\epsilon_+$, and so drops out of~\eqref{eq:D_a_xi_+}. Therefore, it is the components of $X$ in the $\rep{7}\times\rep{7}$ that must be set to vanish. Now consider the $G_2$ decompositions of the fields
\begin{equation}
\partial\dil\in\rep 7,\qquad H\in\rep 1+\rep 7+\rep{27},\qquad A^+\in\rep{14}+\rep{27}+\rep{64}.
\end{equation}
One can quickly check that the representations $\rep{1}$ and $\rep{14}$ occur only in the tensor product $\rep{7}\times\rep{7}$ while the $\rep{64}$ is only in $\rep{7}\times\rep{14}$, and the remainder may appear in both. As a result, we immediately conclude that \eqref{eq:D_a_xi_+} sets: 1) $A^+|_{\rep{14}} = 0$ -- recall that the $A^+$ tensor simply parametrises the freedom one has within the family of generalised Levi-Civita connections, and so this is not a constraint on the background; 2) $H|_{\rep{1}} = 0$ -- this is an actual constraint on the structure.\footnote{The vanishing of the singlet component of the $H$ flux matches the physical observation that this component of the torsion can be related to the cosmological constant in a supersymmetric background, and so it must be set to zero for the Minkowski solutions that we are considering.} On the other hand, since the component $A^+|_{\rep{64}}$ drops out entirely from~\eqref{eq:D_a_xi_+}, it is left unconstrained, implying that $\gxg$-compatible torsion-free connections, if they exist, are not unique.

For the $\rep 7$ components, we isolate the relevant terms by writing
\begin{equation}
\partial_{a}\dil|_{\rep 7}=\partial_{a}\dil,\qquad H_{abc}|_{\rep 7}=(*\gphi)_{abc}{}^{d}H_{d},
\end{equation}
which gives
\begin{equation}
\left(\tfrac{1}{12}(*\gphi)_{ad}{}^{bc}H^{d}\gamma_{bc}-\tfrac{1}{6}\partial_{b}\dil\gamma_{a}{}^{b}\right)\epsilon_{+} = 0.
\end{equation}
Next we note that $\zeta^{\text{T}}\gamma_{ab}\epsilon_{+}\in\rep{7}$ for any spinor $\zeta$, and so we can use the expression \eqref{eq:G2_projectors} for the projector onto the $\rep{7}$ representation to write
\begin{equation}
(*\gphi)_{ab}{}^{cd}\gamma_{cd}\epsilon_{+}=4\gamma_{ab}\epsilon_{+}.\label{eq:14_identity}
\end{equation}
We then have
\begin{equation}
\tfrac{1}{3}\left(H_{b}-\tfrac{1}{2}\partial_{b}\dil\right)\gamma_{a}{}^{b}\epsilon_{+} = 0.
\end{equation}
This will vanish for $H_{a}=\tfrac{1}{2}\partial_{a}\dil$, which must thus be the choice which is necessary for a $\gxg$-compatible connection. 

Next consider the $\rep{27}$ components, which we pick out by writing
\begin{equation}
H_{abc}|_{\rep{27}}=H_{e[a}\gphi^{e}{}_{bc]},\qquad A^+_{abc}|_{\rep{27}}=A^+_{ea}\gphi^{e}{}_{bc}-A^+_{e[a}\gphi^{e}{}_{bc]},
\end{equation}
where $H_{ab}$ and $A^+_{ab}$ are symmetric and traceless. Plugging this into the expression for $X$, we then have
\begin{equation}
\begin{split}& \tfrac{1}{4}\left(\tfrac{1}{3}H_{e[a}\gphi^{e}{}_{bc]}+(A^+_{ea}\gphi^{e}{}_{bc}-A^+_{e[a}\gphi^{e}{}_{bc]})\right)\gamma^{bc}\epsilon_{+}\\
 & =\tfrac{1}{4}\left(\alpha_{ae}\gphi^{e}{}_{bc}+\beta_{be}\gphi^{e}{}_{ca}\right)\gamma^{bc}\epsilon_{+} = 0,
\end{split}
\label{eq:27s}
\end{equation}
where
\begin{equation}
\alpha_{ae}=\tfrac{1}{9}H_{ae}+\tfrac{2}{3}A^+_{ae},\qquad\beta_{be}=\tfrac{2}{9}H_{be}-\tfrac{2}{3}A^+_{be}.
\end{equation}
To see how these two terms are related, one can contract \eqref{eq:14_identity} with $\alpha^{a}{}_{e}\gphi^{efb}$ and use \eqref{eq:B4_identity} to show that
\begin{equation}
(\alpha_{ae}\gphi^{e}{}_{bc}+6\alpha_{be}\gphi^{e}{}_{ca})\gamma^{bc}\epsilon^{+}=0.
\end{equation}
Thus the precise combination of the $\rep{27}$s which appears in (\ref{eq:27s}) is $\beta-6\alpha$, which vanishes for
\begin{equation}
A^+_{ab}=-\tfrac{2}{21}H_{ab}.
\end{equation}
This is the choice which is necessary for a connection compatible with $\epsilon_+$. Note that since we are simply using the freedom in choosing the $A^+$ tensor to obtain this cancellation, the background flux $H|_{\rep{27}}$ is entirely unconstrained, in agreement with the $G$-structure analysis of \cite{Friedrich:2001yp,Gauntlett:2002sc}.

The calculations for compatibility with $\epsilon_-$ are analogous, with the result
\begin{equation}
        D\epsilon_- = 0 \quad\Leftrightarrow\quad          
        \begin{aligned}
            \nabla^{+}\gphi_- = 0,\qquad        H_{\ba\bb\bc}\gphi^{\ba\bb\bc} = 0,\\
        H_{\ba}=-\tfrac{1}{2}\partial_{\ba}\dil,\qquad
        A^-_{\ba\bb}= \tfrac{2}{21}H_{\ba\bb}  .
        \end{aligned}
\end{equation}
The remaining unfixed components of the connection are the parts of $A^+$ and $A^-$ in the $(\rep{64},\rep 1)+(\rep 1,\rep{64})$. These simply parametrise the family of torsion-free connections which are compatible with the same $\gxg$ structure.

Putting this all together, a compatible, torsion-free $\gxg$-generalised connection takes the form
\begin{equation} \label{eq:torsion-free-G2-connection}
\begin{aligned}
    D_{a}v^{b} & =\nabla_{a}v^{b}-\tfrac{5}{42}\gphi_{bcd}H^d{}_a v^c -\tfrac{1}{42}\gphi_{cad}H^d{}_b v^c
    -\tfrac{1}{42}\gphi_{abd}H^d{}_c v^c \\
    & \eqspace -\tfrac{1}{12} (*\gphi)_{abcd}\partial^{d}\dil v^c-\tfrac13 \delta^b_{a}\partial_{c}\dil v^c +\tfrac13 \partial^{b}\dil v_a
    +(A^+|_{\rep{64}})_{a}{}^{b}{}_{c}v^{c},\\
    D_{\bar{a}}v^{b} & =\nabla_{\bar{a}}^{-}v^{b}\equiv\nabla_{\bar{a}}v^{b}-\tfrac{1}{2}H_{\bar{a}}{}^{b}{}_{c}v^{c},\\
    D_{a}v^{\bar{b}} & =\nabla_{a}^{+}v^{\bar{b}}\equiv\nabla_{a}v^{\bar{b}}+\tfrac{1}{2}H_{a}{}^{\bar{b}}{}_{\bar{c}}v^{\bar{c}},\\
    D_{\bar{a}}v^{\bar{b}} & =\nabla_{\bar{a}}v^{\bar{b}}+\tfrac{5}{42}\gphi_{\bb\bc\bd}H^{\bd}{}_{\ba} v^{\bc} +\tfrac{1}{42}\gphi_{\bc\ba\bd}H^{\bd}{}_{\bb} v^{\bc}
    +\tfrac{1}{42}\gphi_{\ba\bb\bd}H^{\bd}{}_{\bc} v^{\bc} \\
    & \eqspace -\tfrac{1}{12}(*\gphi)_{\ba\bb\bc\bd}\partial^{\bd}\dil v^{\bc}-\tfrac13 \delta_{\ba}^{\bb}\partial_{\bc}\dil v^{\bc} +\tfrac13 \partial^{\bb}\dil v_{\ba} +(A^-|_{\rep{64}})_{\bar{a}}{}^{\bar{b}}{}_{\bar{c}}v^{\bar{c}}.
\end{aligned}
\end{equation}

\subsection{The double complex}\label{sec:G2_double_complex}

We now introduce the analogue of the $G_2$ complex within $O(7,7)\times\bbR^+$ generalised geometry. Given a $\gxg$ structure, we can consider a decomposition of $\ext^{n}E$ into irreducible representations of $\gxg$. In particular, we will be interested in the spaces
\begin{equation}\label{eq:Apq_def}
    \A^{p,q}_{\rep{m},\rep{n}} \coloneqq \Gamma(\ext^{p}_{\rep{m}}C_{+}\wedge \ext^{q}_{\rep{n}}C_{-}),
\end{equation}
where $\rep{m}$ and $\rep{n}$ correspond to irreducible $\G_{2\pm}$ representations defined by $\gphi_{\pm}$. 
We write $(p,q)$-forms $\omega\in\A^{p,q}$ as
\begin{equation}
\omega=\tfrac{1}{p!q!}\omega_{a_{1}\ldots a_{p}\bar{b}_{1}\ldots\bar{b}_{q}}E^{+a_{1}\ldots a_{p}}\otimes E^{-\bar{b}_{1}\ldots\bar{b}_{q}},
\end{equation}
where $\{E^{+a}\}$ and $\{E^{-\bar{b}}\}$ are a basis for $C_{+}$ and $C_{-}$ respectively.

Moreover, using a generalised connection we can build maps between the spaces to give the following diagram
\begin{equation}\label{eq:double_complex}
    \begin{tikzcd}[column sep = tiny, row sep = tiny]
    & & \arrow{dl}[swap]{\dd_{+}} & \A^{0,0}_{\rep{1},\rep{1}} \arrow[dl] \arrow[dr] & \arrow{dr}{\dd_{-}} & & \\
    &\left.\right. & \A^{1,0}_{\rep{7},\rep{1}} \arrow[dl] \arrow[dr] & &  
    \A^{0,1}_{\rep{1},\rep{7}} \arrow[dl] \arrow[dr] & \left.\right. & \\
    & \A^{2,0}_{\rep{7},\rep{1}} \arrow[dl] \arrow[dr] & & 
    \A^{1,1}_{\rep{7},\rep{7}} \arrow[dl] \arrow[dr] & &
    \A^{0,2}_{\rep{1},\rep{7}} \arrow[dl] \arrow[dr] & \\
    \A^{3,0}_{\rep{1},\rep{1}} \arrow[dr] & &
    \A^{2,1}_{\rep{7},\rep{7}} \arrow[dl] \arrow[dr] & & 
    \A^{1,2}_{\rep{7},\rep{7}} \arrow[dl] \arrow[dr] & &
    \A^{0,3}_{\rep{1},\rep{1}} \arrow[dl] \\
    & \A^{3,1}_{\rep{1},\rep{7}} \arrow[dr] & &
    \A^{2,2}_{\rep{7},\rep{7}} \arrow[dl] \arrow[dr] & &
    \A^{1,3}_{\rep{7},\rep{1}} \arrow[dl] & \\
    & & \A^{3,2}_{\rep{1},\rep{7}} \arrow[dr] & & 
    \A^{2,3}_{\rep{7},\rep{1}} \arrow[dl]  & & \\
    & & & \A^{3,3}_{\rep{1},\rep{1}} & & &
    \end{tikzcd}
\end{equation}
where we have defined
\begin{align}\label{eq:pm-operators}
    (\dd_{+}\omega)_{a_{1}\dots a_{p+1}\bar{a}_{1}\dots \bar{a}_{q}} &=(p+1) (\Pr{+}{m})_{a_{1}\dots a_{p+1}}{}^{b_{1}\dots b_{p+1}}D_{b_{1}}\omega_{b_{2}\dots b_{p+1}\bar{a}_{1}\dots \bar{a}_{q}}, \\
    (\dd_{-}\omega)_{a_{1}\dots a_{p}\bar{a}_{1}\dots \bar{a}_{q+1}} &= (-1)^{p}(q+1)(\Pr{-}{m})_{\bar{a}_{1}\dots \bar{a}_{q+1}}{}^{\bar{b}_{1}\dots \bar{b}_{p+1}}D_{\bar{b}_{1}}\omega_{a_{1}\dots a_{p}\bar{b}_{2}\dots \bar{b}_{q+1}},
\end{align}
where $\omega \in \A^{p,q}_{\rep{m},\rep{n}}$, and $\Pr{\pm}{m}$ are the projectors onto the relevant $\G_{2\pm}$ representation as given in Appendix \ref{app:G2_identities}. Here we assume that $D$ is a $\gxg$-compatible connection so that it commutes with the projectors -- such a connection always exists (though it may not be torsion-free).

We now ask when \eqref{eq:double_complex} is actually a double complex. That is, when do we have
\begin{equation}\label{eq:nilpotent}
    \dd_{\pm}^{2} = 0, \qquad \dd_{+}\dd_{-} + \dd_{-}\dd_{+} = 0.
\end{equation}
We will show that a sufficient condition is that the $\gxg$ structure is torsion-free, which corresponds physically to a supersymmetric NSNS Minkowski background. Then we can take the generalised connections in~\eqref{eq:pm-operators} to be of the form~\eqref{eq:torsion-free-G2-connection}. 

At first sight this statement might worry the reader -- since these connections are not uniquely determined, it would seem that we need some extra information (beyond that of the supergravity background) to further constrain the connection, as otherwise the operators might not be uniquely defined. As we mentioned earlier, the generalised Levi-Civita connection is also not unique however one can construct unique operators from it, such as the generalised versions of the Ricci tensor and scalar. Something similar happens here, namely the $\dd_{\pm}$ operators do not depend on the undetermined  $\repp{64}{1}+\repp{1}{64}$ components of the connection and so they are actually unique, i.e.~they depend only on the data of the torsion-free $G_{2}\times G_{2}$ structure itself. To see this, note that the double complex consists solely of maps between the $\gxg$ representations $\repp{1}{1}$, $\repp{7}{1}$, $\repp{1}{7}$ and $\repp{7}{7}$. Now, simple representation theory tells us that the $\repp{64}{1}$ or $\repp{1}{64}$ cannot give such maps. In other words, any tensor transforming in those representations must be projected out. Therefore, we can compute the double complex with any choice of $A^{\pm}|_\rep{64}$ tensor in~\eqref{eq:torsion-free-G2-connection} and obtain a unique answer.

Another useful result for the torsion-free case is that one may  use the following ``simplified'' connection to define the $\dd_{\pm}$ operators:
\begin{equation}
\label{eq:basic-connection}
\begin{aligned}
	\hat{D}_a v^b  & = \LC_a v^b ,\\
	\hat{D}_{\ba} v^b  & = \LC^{-}_{\ba} v^{b} = \LC_{\ba} v^{b} - \tfrac12 H_{\ba}{}^{b}{}_{c} v^{c} ,\\
	\hat{D}_{a} v^{\bb}  & = \LC^{+}_{a} v^{\bb} 
		= \LC_{a} v^{\bb} + \tfrac12 H_{a}{}^{\bb}{}_{\bc} v^{\bc} ,\\
	\hat{D}_{\ba} v^{\bb}  & = \LC_{\ba} v^{\bb} ,
\end{aligned}
\end{equation}
where the Hull connections $\nabla^{\mp}$ are assumed to preserve the $G_2$ structures $\gphi_{\pm}$.  As a generalised connection $\hat{D}$ is neither torsion-free nor is it compatible with the $\gxg$ structure, and yet the operators  $\dd_{\pm}$ defined from it coincide with the ones defined using $\Dgen$. Remarkably, this means that in the torsion-free case, the double complex can be described using just the ordinary Levi-Civita and Hull connections.

To verify  this, take for example $\alpha \in \A^{1,1}_{\rep{7},\rep{7}}$, and let $D$ be a generalised Levi-Civita connection of the form~\eqref{eq:gen-LC}. Then we have
\begin{equation}
    \begin{aligned}
        \tfrac12(\dd^D_+\alpha - \dd^{\hat{D}}_+\alpha)_{ab\ba} & = \mathcal{P}_{ab}{}^{cd}(D_c - \hat{D}_c)\alpha_{d\ba}\\
        & = \mathcal{P}_{ab}{}^{cd}(\tfrac16 H_c{}^{e}{}_d\alpha_{e\ba} - \tfrac13 \partial_c\dil \alpha_{d\ba} -(A^+)_c{}^e{}_d \alpha_{e\ba} ) \\
        & =  \mathcal{P}_{ab}{}^{cd}(\tfrac16 H \gphi_c{}^{e}{}_d\alpha_{e\ba} 
        -\tfrac16 H^f (*\gphi)_{cdef}\alpha^e{}_{\ba} 
        -\tfrac16 H^f{}_{[c} \gphi_{de]f}\alpha^e{}_{\ba} \\
        &\eqspace- \tfrac13 \partial_c\dil \alpha_{d\ba}  +(A^+)^f{}_{c}\gphi_{def} \alpha^e{}_{\ba}
        -(A^+)^f{}_{[c}\gphi_{de]f} \alpha^e{}_{\ba} ) \\
        & = \tfrac16 H \gphi_a{}^{e}{}_b\alpha_{e\ba}
         +  \tfrac13 \mathcal{P}_{ab}{}^{cd} (2 H_c  - \partial_c\dil ) \alpha_{d\ba}\\
        &\eqspace-\tfrac{1}{18}(H^{cd}-\tfrac{21}{2} (A^+)^{cd})\gphi_{abc}v_d .
    \end{aligned}
\end{equation}
The difference between the operators vanishes precisely when $D$ is a $\gxg$-compatible, torsion-free connection (these are the same conditions we found in the previous section). It should also be clear from this calculation that one could consider the action on any other element of the complex \eqref{eq:double_complex} and obtain analogous constraints. Thus, the two operators coincide if and only if the generalised structure has vanishing torsion. Assuming this is the case, we see that the operators agree and so we are free to use $\hat{D}$ to define $\dd_\pm$.

This simplified connection makes checking the nilpotency conditions~\eqref{eq:nilpotent} substantially easier. First consider $\dd^2_+$. We have that the simplified connection satisfies
\begin{equation} \label{eq:com-simpD}
\begin{aligned}[]
	 &[\hat{D}_{[a_1}, \hat{D}_{a_2}] \omega_{a_3 \dots a_{p+2}] \bb_1 \dots \bb_q}
		\\  &\qquad= - p \Riem_{[a_1 a_2}{}^{e}{}_{a_3} \omega_{|e|a_4 \dots a_k] \bb_1 \dots \bb_q} 
		- q \Riem^{+}_{[a_1 a_2|}{}^{\bc}{}_{[\bb_1} \omega_{|a_3 \dots a_k] |\bc| \bb_1 \dots \bb_q]} \\ 
	 &\qquad= - q \Riem^{+}_{[a_1 a_2|}{}^{\bc}{}_{[\bb_1} \omega_{|a_3 \dots a_k] |\bc| \bb_1 \dots \bb_q]} ,
\end{aligned}
\end{equation}
where we have used \eqref{eq:Riem} to write the commutator of connections in terms of curvatures. Now notice that because $\nabla^+$ is compatible with $\gphi_-$ it follows that $\Riem^{+} \in \Lambda^2T\otimes \mathfrak{g}_2^-$, and similarly $\Riem^{-} \in \Lambda^2T\otimes \mathfrak{g}_2^+$. But since $\dd H = 0$, we have that $\Riem^{+}_{a_1a_2 b_1b_2} = \Riem^{-}_{b_1b_2 a_1a_2}$ and so actually  $\Riem^{+} \in \mathfrak{g}_2^{+} \otimes \mathfrak{g}_2^{-}$ (and $\Riem^{-} \in \mathfrak{g}_2^{-} \otimes \mathfrak{g}_2^{+}$). Therefore,~\eqref{eq:com-simpD} vanishes when the projectors in the definition of $\dd_+$ are applied to it. So for $\omega \in \A^{0,q}$ one has
\begin{equation}
    \begin{aligned}
        \tfrac12(\dd_+^2\omega)_{a_1a_2 \bb_1\dots \bb_q} &= \mathcal{P}_{a_1a_2}{}^{c_1c_2}D_{c_1}D_{c_2}\omega_{\bb_1\dots \bb_q} \\
        &=\mathcal{P}_{a_1a_2}{}^{c_1c_2}\hat{D}_{[c_1}\hat{D}_{c_2]}\omega_{\bb_1\dots \bb_q} = 0 ,
    \end{aligned}
\end{equation}
and if $\omega\in \A^{1,q}$
\begin{equation}
    \begin{aligned}
        \tfrac16(\dd_+^2\omega)_{a_1a_2a_3 \bb_1\dots \bb_q} &= \mathcal{P}_{a_1a_2a_3}{}^{d_1d_2d_3}D_{d_1}\mathcal{P}_{d_2d_3}{}^{c_1c_2}\hat{D}_{c_1}\omega_{c_2\bb_1\dots \bb_q} \\
        &=\mathcal{P}_{a_1a_2a_3}{}^{d_1d_2d_3}\mathcal{P}_{d_2d_3}{}^{c_1c_2}D_{d_1}\hat{D}_{c_1}\omega_{c_2\bb_1\dots \bb_q}\\
        &=\mathcal{P}_{a_1a_2a_3}{}^{c_1c_2c_3}D_{c_1}\hat{D}_{c_2}\omega_{c_3\bb_1\dots \bb_q}\\
        &=\mathcal{P}_{a_1a_2a_3}{}^{c_1c_2c_3}\hat{D}_{[c_1}\hat{D}_{c_2}\omega_{c_3]\bb_1\dots \bb_q} = 0 ,
    \end{aligned}
\end{equation}
where we have used that the projectors commute with the compatible connection $D$ and that $(\mathcal{P}_{\rep{1}})_{a_1a_2a_3}{}^{b_1b_2b_3}(\mathcal{P}_{\rep{7}})_{b_1b_2}{}^{c_1c_2} =(\mathcal{P}_{\rep{1}})_{a_1a_2a_3}{}^{c_1c_2b_3}$. Obviously, if $\omega \in \A^{p>1,q}$  then $\dd_+^2\omega = 0$ trivially, and one can repeat this reasoning to also conclude that $\dd_-^2 = 0$.

To see that $\left\{ \dd_+, \dd_- \right\} = 0$, consider first $\alpha \in \A_{\rep{7},\rep{7}}^{1,1}$. Then
\begin{equation}
\begin{aligned}[]
\label{eq:commutator}
	[\hat{D}_{a}, \hat{D}_{\bb}]  \alpha_{c \bd}
	&= [\LC_{a}, \LC_{\bb}]  \alpha_{c \bd} + \tfrac12 (\LC_a H_{\bb}{}^e{}_c) \alpha_{e\bd}
		+ \tfrac12 (\LC_{\bb} H_a{}^{\be}_{\bd} ) \alpha_{c\be} \\
	&\eqspace (1-1) \tfrac12 H_{\bb}{}^e{}_c \alpha_{e\bd} 
		+ (1-1)\tfrac12 H_a{}^{\be}{}_{\bd} \LC_{\bb} \alpha_{c\be} \\
	&\eqspace -\tfrac12 \left( H_a{}^{\be}{}_{\bb} \LC_{\be}\alpha_{c\bd} 
		+ H_{\bb}{}^{e}{}_{a} \LC_e \alpha_{c\bd} \right) \\
	&= - \left( \Riem_{a\bb}{}^e{}_c - \tfrac12 \LC_a H_{\bb}{}^e{}_c 
		+ \tfrac14 H_{a}{}^{\bar{f}}{}_{\bb} H_{\bar{f}}{}^e{}_c \right) \alpha_{e\bd} \\
	&\eqspace - \left( \Riem_{a\bb}{}^{\be}{}_{\bd} - \tfrac12 \LC_{\bb} H_{a}{}^{\be}{}_{\bd} 
		+ \tfrac14 H_{\bb}{}^{{f}}{}_{a} H_{{f}}{}^{\be}{}_{\bd} \right) \alpha_{c\be} .\\
\end{aligned}
\end{equation}
Antisymmetrising on $[ac]$ and $[\bb\bd]$, this becomes
\begin{equation}
\begin{aligned}
	 -\tfrac12 \Riem^{+}_{ac}{}^e{}_{\bb} \alpha_{e\bd} 
		+ \tfrac12 \Riem^{-}_{\bb\bd}{}^{\be}{}_{a} \alpha_{c\be} .
\end{aligned}
\end{equation}
Again, due to the representations that $\Riem^+$ and $\Riem^-$ live in, this is projected out in $\left\{ \dd_+, \dd_- \right\}\alpha$.

To show that on an element $\beta\in\A^{2,0}_{\rep{7},\rep{1}}$ we also have $\left\{ \dd_+, \dd_- \right\} \beta = 0$, it is actually simpler to use a torsion-free compatible connection. One then has
\begin{equation}\label{eq:nilpotent-pm-A2}
    \begin{aligned}
        \gphi^{abc}(D_a D_{\ba} - D_{\ba} D_a)\beta_{bc} &= [D_a, D_{\ba}]\gphi^{abc}\beta_{bc}  = \GenRic_{a\ba}\gphi^{abc}\beta_{bc} = 0,
    \end{aligned}
\end{equation}
where in the first equality we used compatibility to commute the $G_2$ 3-form through the derivatives, then we used the definition of the generalised Ricci tensor of a torsion-free connection~\eqref{eq:genRicTensor}, and finally we used the fact that generalised torsion-free $\gxg$ manifolds are generalised Ricci-flat.

The action of $\left\{ \dd_+, \dd_- \right\}$ on the remaining spaces of the double complex can be computed similarly to these two examples and leads to the same result. One can therefore conclude that if a generalised $\gxg$ structure has vanishing intrinsic torsion, then there exists a double complex of the form~\eqref{eq:double_complex}.


\subsection{Hodge theory}\label{sec:G2G2_Hodge}

Let us now see how several concepts familiar from the conventional $\G_2$ complex naturally generalise to the $\gxg$ double complex.

We start by defining the adjoint operators in direct analogy with Section~\ref{sec:g2-complex-adjoints}. Consider two tensors in $\A^{1,0}$
\begin{equation}
\alpha=\alpha_{a}E^{+a},\qquad\beta=\beta_{a}E^{+a}.
\end{equation}
We can define an inner product between them as
\begin{equation}
\begin{split}(\alpha,\beta) & =\int_{M}\Phi\,G(\alpha,\beta)=\int_{M}\Phi\,\alpha_{a}\beta_{b}\,\eta(E^{+a},E^{+b})\\
 & =\int_{M}\Phi\,\alpha_{a}\beta_{b}\,\eta^{ab}=\int_{M}\ee^{-2\dil}\vol\,\alpha\lrcorner\beta,
\end{split}
\end{equation}
where both $\Phi$ and $G$ are $\Orth d\times\Orth d$ invariant, so that the inner product is also invariant. More generally for $\alpha,\beta\in \A^{p,q}$ one has
\begin{align}
\int_{M}\Phi\,G(\alpha,\beta)
 & =\kappa_{p}\kappa_{q}\int_{M}\ee^{-2\dil}\vol\,\alpha\lrcorner\beta ,\label{eq:G2_double_complex_metric}
\end{align}
where the constants $\kappa_{p}$ and $\kappa_{q}$ are defined by
\begin{equation}
\kappa_{p}=\begin{cases}
1 & p=0,1,\\
3 & p=2,\\
7 & p=3.
\end{cases}
\end{equation}
This ensures that this inner product agrees with the usual inner product for each $\Gx 2$ factor. Given this definition, one can check that the adjoint operators, defined by $(\alpha,\dd_{\pm}\beta)=(\dd_{\pm}^{\dagger}\alpha,\beta)$, act on $(p,q)$-forms as
\begin{align}
(\dd_{+}^{\dagger}\alpha)_{c_{2}\ldots c_{p}\bar{d}_{1}\ldots\bar{d}_{q}} & =-\gamma_{p}D^{c_{1}}\alpha_{c_{1}\ldots c_{p}\bar{d}_{1}\ldots\bar{d}_{q}},\\
(\dd_{-}^{\dagger}\alpha)_{c_{1}\ldots c_{p}\bar{d}_{2}\ldots\bar{d}_{q}} & =\gamma_{q}(-1)^{p+1}D^{\bar{d}_{1}}\alpha_{c_{1}\ldots c_{p}\bar{d}_{1}\ldots\bar{d}_{q}},
\end{align}
where
\begin{equation}
\gamma_{p}=\begin{cases}
1 & p=1,\\
3 & p=2,\\
\tfrac{7}{3} & p=3.
\end{cases}
\end{equation}
In this way, the adjoint operators inherit many of the properties of the those for the usual $G_2$ complex, as described in Section~\ref{sec:g2-complex-adjoints}. For example, using~\eqref{eq:d_dagger}, acting on a tensor $\alpha \in \A^{3,1}$, we have that $\dd^{\dagger}_+\alpha = (\theta_+)_{\rep{7}}\dd_+(\theta_+)^{-1}_{\rep{1}}\alpha$.


\subsubsection{K\"ahler identities}

There exist useful anticommutation relations between the differentials and their adjoints, which are the $\gxg$ analogues of the K\"ahler identities of the Dolbeault complex. Taking $\lambda\in \A^{1,0}$, in components we have that
\begin{equation}
        (\dd^{\dagger}_+ \dd_-  \lambda)_{\bar{a}} = D^a D_{\bar{a}} \lambda_a ,\qquad
        (\dd_- \dd^{\dagger}_+  \lambda)_{\bar{a}} = - D_{\bar{a}} D^a   \lambda_a ,
\end{equation} 
and so
\begin{equation} \label{eq:khaler-id-1form}
       ( \dd^{\dagger}_+ \dd_-\lambda+\dd_- \dd^{\dagger}_+ \lambda)_{\bar{b}}= [D_a,D_{\bb}]\lambda^a= \GenRic_{a\bar{b}}  \lambda^a \equiv 0,
\end{equation}
since the generalised Ricci tensor vanishes for a torsion-free $\gxg$ structure. Note that this is essentially the same calculation as~\eqref{eq:nilpotent-pm-A2}. Indeed, because of the isomorphisms~\eqref{eq:d_dagger}, the  K\"ahler identities are automatically implied by the anticommutation relations of the $\dd_\pm$ operators, and vice-versa. For example, acting on $\mu \in \A^{2,0}$ satisfying $\mu = (\theta_+)_{\rep 7} \lambda$ for some $\lambda \in \A^{1,0}$, we have
\begin{equation}
(\dd^{\dagger}_+ \dd_- + \dd_-\dd^{\dagger}_+)\mu = (\dd^{\dagger}_+ \dd_- + \dd_-\dd^{\dagger}_+)(\theta_+)_{\rep{7}}\lambda = (\theta_+)^{-1}_{\rep{7}}(\dd_+ \dd_- + \dd_-\dd_+)\lambda = 0 ,
\end{equation}
since $\left\{ \dd_+, \dd_- \right\}\lambda = 0$. 

We conclude that the differentials and their adjoints over the $\gxg$ complex satisfy  
\begin{equation}
\dd^2_{\pm}=(\dd_{\pm}^{\dagger})^2=\left\{ \dd_\pm, \dd_\mp \right\}=\{\dd^{\dagger}_{\pm},\dd_{\mp}\}=0.
\end{equation} 


\subsubsection{Laplacians}\label{sec:g2g2_laplacians}

We define Laplacians for both the ``plus'' and ``minus'' differentials as usual:
\begin{equation}
\Delta_{\pm} = \dd^{\dagger}_{\pm}\dd_{\pm} +\dd_{\pm} \dd^{\dagger}_{\pm} .
\end{equation}
Much as for the $G_2$ complex in Section~\ref{sec:g2-complex-adjoints}, it follows from the properties of the adjoints that the Laplacians depend only on the $\gxg$ representation of the object on which they act and not on the $(p,q)$ degree of the form. For instance, taking $\alpha \in \A^{2,3}_{\rep{7},\rep{1}}$, there exists some $\beta \in  \A^{1,0}_{\rep{7},\rep{1}}$ such that $\alpha = (\theta_+)_{\rep{7}}(\theta_-)_{\rep{1}}\beta$. Then 
\begin{equation}
\begin{aligned}
\Delta_+ \alpha &= \left(\dd_+^{\dagger} \dd_+ + \dd_+ \dd^{\dagger}_+ \right)(\theta_+)_{\rep{7}}(\theta_-)_{\rep{1}}\beta = (\theta_-)_{\rep{1}} \left(\dd_+^{\dagger} \dd_+ + \dd_+ \dd^{\dagger}_+ \right)(\theta_+)_{\rep{7}}\beta \\
&= (\theta_-)_{\rep{1}} \left(\dd_+^{\dagger} (\theta_+)^{-1}_{\rep{7}}\dd^{\dagger} _+ + \dd_+ (\theta_+)^{-1}_{\rep{7}}\dd_+ \right)\beta \\
&=  (\theta_-)_{\rep{1}} (\theta_+)_{\rep{7}}\left(\dd_+ \dd^{\dagger} _+ + \dd^{\dagger}_+  \dd_+ \right)\beta  = (\theta_-)_{\rep{1}} (\theta_+)_{\rep{7}} \Delta_+\beta .
\end{aligned}
\end{equation}
Note as well that if we consider the combined differential $\hat\dd =  \dd_+ + \dd_-$, the K\"ahler identities imply that its Laplacian coincides with the sum of the Laplacians for $\dd_+$ and $\dd_-$:
\begin{equation}\label{eq:combined-Lap}
    \hat\Delta = \hat\dd^{\dagger} \hat\dd + \hat\dd \hat\dd^{\dagger} = \Delta_+ + \Delta_-.
\end{equation}

We will now show that the $\Delta_+$ and $\Delta_-$ Laplacians are in fact equal, as is the case for $\Delta_\partial$ and $\Delta_{\bar\partial}$ of the Dolbeault complex. Considering first $f\in \A^{0,0}$, we have that
\begin{equation}
    \begin{aligned}
    \Delta_+ f = \dd^{\dagger}_+\dd_+ f = -D^{a}D_{a}f = -\nabla^2 f +2 \partial^{a}\dil\nabla_{a}f ,\\
    \Delta_- f = \dd^{\dagger}_-\dd_- f = -D^{\Bar{a}}D_{\Bar{a}}f = -\nabla^2 f +2 \partial^{\Bar{a}}\dil\nabla_{\Bar{a}}f ,
    \end{aligned}
\end{equation}
and so $\Delta_+ f = \Delta_- f$. Now take $\lambda\in \A^{1,0}$. For $\Delta_-$ we have
\begin{equation}
\begin{split}
(\Delta_-\lambda)_b &=  (\dd^{\dagger}_-\dd_-\lambda)_b  = -D^{\bar{a}}D_{\bar{a}}\lambda_b   \\
&=-\nabla^2 \lambda_b -H_{\bar{a}cb}\nabla^{\bar{a}}\lambda^c+2\partial^{\bar{a}}\dil\nabla_{\bar{a}}\lambda_b + \tfrac14 H^{\bar{a}d}{}_bH_{\bar{a}dc}\lambda^c \\
&\eqspace -\tfrac12(\nabla^{\bar{a}}H_{\bar{a}bc} -2\partial^{\bar{a}}\dil H_{\bar{a}bc})\lambda^c .
\end{split}
\end{equation}
For the two terms in $\Delta_+$, we find
\begin{equation}
(\dd_+\dd^{\dagger}_+\lambda)_b = -D_b D^a \lambda_a =-\nabla_b\nabla^a \lambda_a 
+2\partial^a\dil\nabla_b \lambda_a +2 (\nabla_b\nabla_a\dil)\lambda^a ,
\end{equation}
and
\begin{equation}
\begin{aligned}
(\dd^{\dagger}_+\dd_+\lambda)_b &=  -2 D^aD_{[a}\lambda_{b]} - \varphi_{abcd}D^aD^c\lambda^d 
\\ 
  &=  -\nabla^2\lambda_b + \nabla^a\nabla_b \lambda_a  + 4\partial^a\dil \nabla_{[a}\lambda_{b]}  \\
  &\eqspace-\tfrac12 (*\gphi)_{bcde} \partial^c\dil  \nabla^d\lambda^e - \gphi_{c[bd} H^c{}_{e]}  \nabla^{d}\lambda^{e}  \\
   &=  -\nabla^2\lambda_b + \nabla^a\nabla_b \lambda_a  + 4\partial^a\dil \nabla_{[a}\lambda_{b]}  -H_{bcd}\nabla^c\lambda^d ,
\end{aligned}
\end{equation}
where we used the compatibility condition $\tfrac12 \partial_a \dil = H_a$ and the $G_2$ decomposition of $H$ in reverse $H_{abc} =  (*\gphi)_{abcd}H^d + \gphi_{e[ab}H^e{}_{c]}$ for the final step. Putting this together, we can compare the two Laplacians to find
\begin{equation}
\begin{aligned}
(\Delta_-\lambda - \Delta_+\lambda)_b &= 2(D^aD_{[a}\lambda_{b]} +\tfrac12 (*\gphi)_{abcd}D^aD^c \lambda^d) + D_b D^a \lambda_a -D^{\bar{a}}D_{\bar{a}}\lambda_b  \\
&=  [\nabla_b,\nabla_a]\lambda^a -2 (\nabla_a\nabla_b\dil) \lambda^a + \tfrac14 H^{\bar{a}d}{}_bH_{\bar{a}dc}\lambda^c   \\
&\eqspace -\tfrac12(\nabla^{\bar{a}}H_{\bar{a}bc} -2\partial^{\bar{a}}\dil H_{\bar{a}bc})\lambda^c +H_{\bar{a}bc}\nabla^{\bar{a}}\lambda^c +H_{bcd}\nabla^c\lambda^d  \\
&\eqspace+2\partial^{\bar{a}}\dil\nabla_{\bar{a}}\lambda_b -2\partial^a\dil\nabla_b\lambda_a- 2\partial^a\dil \nabla_{[a}\lambda_{b]}  \\
&= 0 ,
\end{aligned}
\end{equation}
which can be checked in a gauge where the $C_{\pm}$ frames are aligned, and using the fact that the equations of motion are automatically satisfied for a generalised  $\gxg$ background. It should also be clear that the computation of the actions of $\Delta_{\pm}$ on an element $\tilde{\lambda}\in \A^{0,1}$ would be entirely symmetrical, and so $\Delta_+ \tilde{\lambda} = \Delta_-\tilde{\lambda}$ as well.

Now let us consider the action of the Laplacians on an element $\zeta\in \A^{1,1}$.
Using $(*\gphi)^{cd}{}_{ef}\Riem^+_{cdab}= -2 \Riem^+_{efab}$, which follows from $\Riem^{+} \in \mathfrak{g}_2^{+} \otimes \mathfrak{g}_2^{-}$, we perform a similar calculation to find the ``plus'' Laplacian
\begin{equation}
\begin{aligned}
(\Delta_+ \zeta)_{a\bar{a}} &= -D_a D^b \zeta_{b\bar{a}} - 2 D^aD_{[a}\zeta_{b]\bar{a}} - \varphi_{abcd}D^aD^c\zeta^d{}_{\bar{a}} \\
&=     - \nabla^2 \zeta_{a\bar{a}}  +  H_{acb}H^c{}_{\bar{a}\bar{b}}w^{b\bar{b}}+ 2\der^b\dil \nabla_{ b} \zeta_{a \bar{a}} -  H_{abc}\nabla^b w^c{}_{\bar{a}} +  H_{b\bar{b}\bar{a}}\nabla^bw_a{}^{\bar{b}} \\&
 \eqspace-[\nabla_a,\nabla_b] \zeta^b{}_{\bar{a}} +2 (\nabla_a\nabla_b\dil)\zeta^b{}_{\bar{a}} - \tfrac14 H^b{}_{\bar{c}\bar{a}}H_{b}{}^{\bar{b}\bar{c}}\zeta_{a\bar{b}}  \\
 &\eqspace  +  \Riem^+_{ab\bar{b}\bar{a}}\zeta^{b\bar{b}}  +  \nabla_{[a} H_{b]\bar{b}\bar{a}} \zeta^{b\bar{b}}  +\tfrac12 H_{[b}{}^{\bar{c}}{}_{|\bar{a}|}H_{a ]\bar{b}\bar{c}}w^{b\bar{b}}  \\
 &=   -\nabla^2 \zeta_{a\bar{a}} +  H_{acb}H^c{}_{\bar{a}\bar{b}}\zeta^{b\bar{b}} +  2\Riem^+_{ab\bar{b}\bar{a}}\zeta^{b\bar{b}}  \\
 &\eqspace + 2\der^b\dil \nabla_{ b} \zeta_{a \bar{a}} -  H_{abc}\nabla^b \zeta^c{}_{\bar{a}} +  H_{b\bar{b}\bar{a}}\nabla^b \zeta_a{}^{\bar{b}} .
\end{aligned}
\end{equation}
We can immediately deduce the ``minus'' Laplacian by exchanging barred and unbarred indices and taking $H \rightarrow -H$:
\begin{equation}
\begin{aligned}
(\Delta_- \zeta)_{a\bar{a}} &=      - \nabla^2 \zeta_{a\bar{a}}  +  H_{\bar{a}\bar{c}\bar{b}}H^{\bar{c}}{}_{ab}\zeta^{b\bar{b}}  +  2\Riem^-_{\ba\bb b a}\zeta^{b\bar{b}} \\
&\eqspace+ 2\der^{\bar{b}}\dil \nabla_{\bar{ b}} \zeta_{a \bar{a}} + H_{\bar{a}\bar{b}\bar{c}}\nabla^{\bar{b}} \zeta_a{}^{\bar{c}} -  H_{\bar{b}ba}\nabla^{\bar{b}}\zeta^{b}{}_{\bar{a}},
\end{aligned}
\end{equation}
and so we directly observe that $\Delta_+\zeta = \Delta_-\zeta$ too.

Finally, since the Laplacians depend only on the $\gxg$ representation, the cases we have covered are actually sufficient to conclude that over the entire double complex $\Delta_+ = \Delta_- = \tfrac12 \hat\Delta$.


\section{Relation to the topological \texorpdfstring{$\G_{2}$}{G2} string}\label{sec:relation_to_G2_string}

In this section we will show that the double complex \eqref{eq:double_complex} is the target-space realisation of the worldsheet BRST complex of the topological $\G_{2}$ string. Indeed, if one studies the left- and right-moving sectors separately, one finds that the states in the topological theory are the following \cite{Shatashvili:1994zw,deBoer:2005pt,deBoer08c,deBoer08b}
\begin{equation}\label{eq:G2_string_LR_states}
\begin{array}{c}
    \left| 0,0 \right>  \\
    \Omega^{0}_{\rep{1}} 
    \end{array}
\,,\qquad 
\begin{array}{c}
    \left|\tfrac{1}{10},\tfrac{2}{5} \right> \\
     \Omega^{1}_{\rep{7}} 
    \end{array}
\,,\qquad 
\begin{array}{c}
     \left|\tfrac{6}{10}, \tfrac{2}{5} \right> \\
    \Omega^{2}_{\rep{7}} 
    \end{array}
\,,\qquad 
\begin{array}{c}
    \left|\tfrac{3}{2},0\right> \\
    \Omega^{3}_{\rep{1}}
    \end{array} \, .
\end{equation}
Here, the states are labelled as in Section \ref{sec:review_top_strings} and the second row shows the interpretation of these states as differential forms on the target space. By studying the OPE of the supercurrent $G^+$ with states of the form $A_{\mu_{1}\dots \mu_{k}}(X)\psi^{\mu_{1}}\dots \psi^{\mu_{k}}$, one can show that the left-moving BRST operator $Q_{L} = G_{-1/2}^{\downarrow}\sim\dc$ acts as
\begin{equation}\label{eq:G2_BRSTL}
\begin{array}{ccccccc}
    \left| 0,0 \right> & \xrightarrow{\quad Q_{L} \quad} & \left|\tfrac{1}{10},\tfrac{2}{5} \right> & \xrightarrow{\quad Q_{L} \quad} & \left|\tfrac{6}{10}, \tfrac{2}{5} \right> & \xrightarrow{\quad Q_{L} \quad} & \left|\tfrac{3}{2},0\right> \\
    \Omega^{0}_{\rep{1}} &\xrightarrow{\quad \dc \quad}& \Omega^{1}_{\rep{7}} &\xrightarrow{\quad \dc \quad}& \Omega^{2}_{\rep{7}} &\xrightarrow{\quad \dc \quad}& \Omega^{3}_{\rep{1}}
\end{array}
\end{equation}
Similar results hold for the right-moving sector as well, with $Q_{R} = \bar{G}_{-1/2}^{\downarrow}$.

The full string states are tensor products of left- and right-moving states from \eqref{eq:G2_string_LR_states}. As target-space tensors, we find that the string states correspond to
\begin{equation}
    \Omega^{p}_{\rep{m}}\otimes\Omega^{q}_{\rep{n}} = \A^{p,q}_{\rep{m},\rep{n}},
\end{equation}
where the spaces $\A^{p,q}_{\rep{m},\rep{n}}$ are as in \eqref{eq:Apq_def}. Moreover, given \eqref{eq:G2_BRSTL} and the fact that $Q_{L}^{2} = Q_{R}^{2} = \{Q_{L},Q_{R}\}=0$, we see that the natural target-space identification of the BRST operators is
\begin{equation}
    Q_{L} \sim \dd_{+}, \qquad Q_{R} \sim \dd_{-}.
\end{equation}
Physical states then correspond to cohomology classes of $Q = Q_{L}+Q_{R}$ in the Hilbert space, or equivalently, harmonic forms under the Laplacian $\hat{\Delta}$ given in~\eqref{eq:combined-Lap}. By the analysis in Section \ref{sec:G2G2_Hodge}, we see that these are precisely the harmonic forms of $\Delta_{\pm}$.

\subsection{1-loop partition function}\label{eq:g2_one_loop}

We can use these observations to understand the 1-loop partition function of the topological $\G_{2}$ string from the target space. As was shown in \cite{deBoer08c}, one can find the 1-loop partition function using the standard formula for the 1-loop free energy of the topological string~\cite{Bershadsky:1993cx}:
\begin{equation}
    \mathcal{F}_{1} = \frac{1}{2}\int\frac{\dd\tau \dd\bar{\tau}}{\tau_{2}} \tr \left((-1)^{F}F_{L}F_{R}\,\ee^{2\pi \ii \tau H_{L} - 2\pi \ii \bar{\tau} H_{R}} \right),
\end{equation}
where $F_{L}$ and $F_{R}$ are the left- and right-moving fermion number operators respectively, $F = F_{L}+F_{R}$ is the total fermion number operator, $H_{L}=\{Q_{L},Q_{L}^{\dagger}\}$ is the left-moving Hamiltonian, and similarly for $H_{R}$. Taking the domain of $\tau$ to be the upper half plane, evaluating the integral gives
\begin{equation}
    \mathcal{F}_{1} = \frac{1}{2}\delta(H_{L}-H_{R}) \log\left[ \prod_{F_{L},F_{R}} \det\bigl(2\pi(H_{L}+H_{R})\bigr)^{(-1)^{F}F_{L}F_{R}} \right].
\end{equation}
Our target space picture provides a clear interpretation of this object. It is precisely the product\footnote{We are using the $\zeta$-regularised determinant of the Laplacians with zero modes removed, denoted by $\det{}'$.}
\begin{equation}\label{eq:G2_free_energy}
    \mathcal{F}_{1} = \frac{1}{2} \log \left[ \prod_{p,q} (\det{}'\hat{\Delta}^{p,q})^{(-1)^{p+q}pq} \right].
\end{equation}

It is instructive to compare this with the analogous result for the topological B-model~\cite{Pestun:2005rp,Bershadsky:1993cx}. Indeed, the free energy in \eqref{eq:G2_free_energy} is of precisely the same form as that for the B-model on a Calabi--Yau threefold, but with the Dolbeault complex replaced with the $\gxg$ complex found in the previous section. This striking fact will become important when we consider the topological $\Spin(7)$ string in Section \ref{sec:top_spin7_string}, about which far less is known.

Using the usual normalisation of the partition function in terms of the free energy, $Z = \ee^{-\mathcal{F}}$, the corresponding 1-loop partition function is
\begin{align}
    Z_{1} &= \left[ \prod_{p,q}(\det{}'\hat{\Delta}^{p,q})^{(-1)^{p+q}pq} \right]^{-1/2} \label{eq:Z_1_G2} \\
    &= (\det{}'\hat{\Delta}_{\rep{1},\rep{1}})^{-9/2} (\det{}' \hat{\Delta}_{\rep{7},\rep{1}})^{3/2}(\det{}'\hat{\Delta}_{\rep{1},\rep{7}})^{3/2} (\det{}'\hat{\Delta}_{\rep{7},\rep{7}})^{-1/2},\label{eq:Z1_g2_gen}
\end{align}
where in the second line the subscript denotes the $\gxg$ representation that $\hat{\Delta}$ acts on, and we have used the fact that the determinant depends only on the representation on which $\hat{\Delta}$ acts. Comparing \eqref{eq:Z_1_G2} to \eqref{eq:A-model_1-loop}, we see that, much like in the A/B-model, the 1-loop partition function calculates the analytic torsion of the $\G_{2}$ double complex.

\begin{figure}
\centering
\begin{tikzpicture}[scale=1.75]
\draw[black] (0,0) grid [rotate=45] (3,3);
\node[circle,fill=white,draw=white,align=center,text=black,inner sep=0pt,minimum size=0pt] at (0*1.414,3*1.414) {$(0,0)$};
\node[circle,fill=white,draw=white,align=center,text=black,inner sep=0pt,minimum size=0pt] at (-.5*1.414,2.5*1.414) {$(1,0)$};
\node[circle,fill=white,draw=white,align=center,text=black,inner sep=0pt,minimum size=0pt] at (.5*1.414,2.5*1.414) {$(0,1)$};
\node[circle,fill=white,draw=white,align=center,text=black,inner sep=0pt,minimum size=0pt] at (1*1.414,2*1.414) {$(0,2)$};
\node[circle,fill=white,draw=white,align=center,text=black,inner sep=0pt,minimum size=0pt] at (0*1.414,2*1.414) {$(1,1)$};
\node[circle,fill=white,draw=white,align=center,text=black,inner sep=0pt,minimum size=0pt] at (-1*1.414,2*1.414) {$(2,0)$};
\node[circle,fill=white,draw=white,align=center,text=black,inner sep=0pt,minimum size=0pt] at (-1.5*1.414,1.5*1.414) {$(3,0)$};
\node[circle,fill=white,draw=white,align=center,text=black,inner sep=0pt,minimum size=0pt] at (-0.5*1.414,1.5*1.414) {$(2,1)$};
\node[circle,fill=white,draw=white,align=center,text=black,inner sep=0pt,minimum size=0pt] at (0.5*1.414,1.5*1.414) {$(1,2)$};
\node[circle,fill=white,draw=white,align=center,text=black,inner sep=0pt,minimum size=0pt] at (1.5*1.414,1.5*1.414) {$(0,3)$};
\node[circle,fill=white,draw=white,align=center,text=black,inner sep=0pt,minimum size=0pt] at (1*1.414,1*1.414) {$(1,3)$};
\node[circle,fill=white,draw=white,align=center,text=black,inner sep=0pt,minimum size=0pt] at (0*1.414,1*1.414) {$(2,2)$};
\node[circle,fill=white,draw=white,align=center,text=black,inner sep=0pt,minimum size=0pt] at (-1*1.414,1*1.414) {$(3,1)$};
\node[circle,fill=white,draw=white,align=center,text=black,inner sep=0pt,minimum size=0pt] at (-.5*1.414,.5*1.414) {$(3,2)$};
\node[circle,fill=white,draw=white,align=center,text=black,inner sep=0pt,minimum size=0pt] at (.5*1.414,.5*1.414) {$(2,3)$};
\node[circle,fill=white,draw=white,align=center,text=black,inner sep=0pt,minimum size=0pt] at (0*1.414,0*1.414) {$(3,3)$};
\node[align=center] at (0*1.414,2.5*1.414) {\mbox{\Large$-$}\\$A$};
\node[align=center] at (0.5*1.414,2*1.414) {\mbox{\Large$+$}\\$B$};
\node[align=center] at (-0.5*1.414,2*1.414) {\mbox{\Large$+$}\\$B$};
\node[align=center] at (0*1.414,1.5*1.414) {\mbox{\Large$-$}\\$C$};
\node[align=center] at (1*1.414,1.5*1.414) {\mbox{\Large$-$}\\$A$};
\node[align=center] at (-1*1.414,1.5*1.414) {\mbox{\Large$-$}\\$A$};
\node[align=center] at (0.5*1.414,1*1.414) {\mbox{\Large$+$}\\$B$};
\node[align=center] at (-0.5*1.414,1*1.414) {\mbox{\Large$+$}\\$B$};
\node[align=center] at (0*1.414,0.5*1.414) {\mbox{\Large$-$}\\$A$};
%
\end{tikzpicture}
\caption{For a global $G_2$ structure with vanishing flux, equality of $\Delta_\pm$ and the isomorphisms provided by the 3-form $\gphi$ mean that $\det{}' \Delta^{p,q}$ can be expressed in terms of three independent determinants. For example, $\det{}' \hat\Delta^{0,0}=\det{}' \hat\Delta_{\rep1,\rep1}=\det{}' \Delta_{\rep1}\equiv A$, $\det{}' \hat\Delta^{1,0}=\det{}' \hat\Delta_{\rep7,\rep1}=\det{}' \Delta_{\rep7}\equiv AB$, and $\det{}' \hat\Delta^{1,1}=\det{}' \hat\Delta_{\rep7,\rep7}=\det{}' \Delta_{\rep1}\det{}' \Delta_{\rep{21}}\det{}' \Delta_{\rep{27}}\equiv AB^2C$. The analytic torsion (the 1-loop partition function) is then given by $(A^{-4}B^4 C^{-1})^{1/2}$, in agreement with \eqref{eq:g2_partition}.}
\label{fig:G2_diagram}
\end{figure}
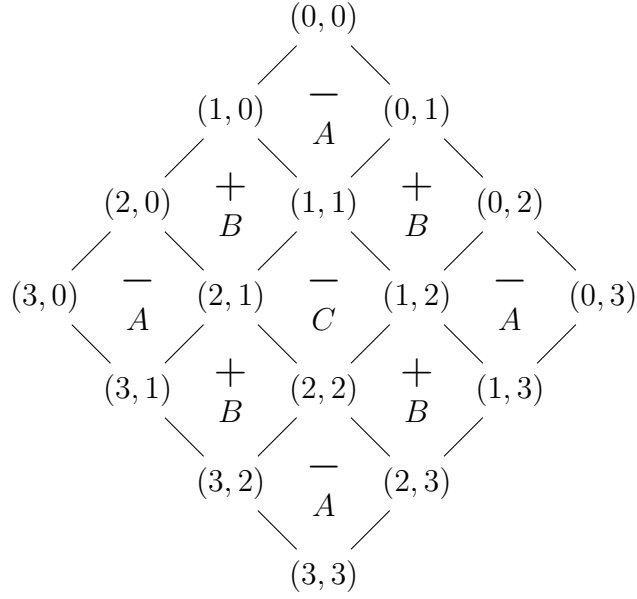

In the case of the topological $G_2$ string, the target manifold has $\G_{2}$ holonomy with vanishing $H$-flux. This means we can further simplify the partition function by considering the diagonal subgroup $G_2\subset \gxg$, and then using the decomposition $\rep{7}\times \rep{7} = \rep{1}+\rep{7}+\rep{14}+\rep{27}$. With this, and the fact that $\hat{\Delta} \simeq \Delta$ on these subspaces, the 1-loop partition function of the topological $G_2$ string is given by
\begin{equation}\label{eq:g2_partition}
    Z_{1} = (\det{}'\Delta_{\rep{1}})^{-5}(\det{}'\Delta_{\rep{7}})^{5/2}(\det{}'\Delta_{\rep{14}})^{-1/2}(\det{}'\Delta_{\rep{27}})^{-1/2},
\end{equation}
which exactly matches the expression given in \cite{deBoer08c}. Much like in the A/B-models, we can read this result off immediately from the double complex, as shown in Figure \ref{fig:G2_diagram}. For the pure $\G_{2}$ case, we find three independent Laplacians assigned to the faces of the squares in the diamond. These once again correspond to determinants of Laplacians restricted to the subspaces in the Hodge decomposition of $\A^{p,q}$. The partition function is then given by the product of these values with alternating powers of $\pm\tfrac{1}{2}$ in a checkerboard pattern, as shown in the figure.

We can use the work of Pestun and Witten~\cite{Pestun:2005rp} on the B-model to identify the target-space theory that reproduces this 1-loop expression. An attempt was made in \cite{deBoer08c} to describe the topological $G_2$ string in terms of a target-space theory defined by a Hitchin-like functional (see Equation~\eqref{eq:gen-Hitchin-g2}), but this did not reproduce the partition function that one calculates from the worldsheet. Rather than starting from an invariant functional, we will simply write down a target-space action whose BV quantisation matches $Z_{1}$.

\subsection{A quadratic target-space action}\label{sec:G2-quad-action}
Using our double complex \eqref{eq:double_complex}, and by direct analogy with the Dolbeault complex of complex geometry and Pestun and Witten~\cite{Pestun:2005rp}, we propose the following quadratic target-space action:
\begin{align}
    S_0 &= \int_{M}\Phi\, (\theta_+)_{\rep{1}}^{-1} (\theta_-)_{\rep{1}}^{-1}\left(\tfrac12\,b_{11}\wedge \dd_{+}\dd_{-} b_{11} + a_{00}\wedge \dd_{+}\dd_{-} c_{22}\right) \label{eq:G2_action} \\
    &\overset{\text{s.h}}{\sim} \int_{M} \ee^{-2\dil}\vol\, \varphi^{mnp}\varphi^{qrs}\left( -\tfrac{1}{2}(b_{11})_{mq}\nabla_{n}\nabla_{r}(b_{11})_{ps} + \tfrac{1}{4} a_{00}\nabla_{m}\nabla_{q}(c_{22})_{nprs} \right),
\end{align}
where the fields $a_{00}$, $b_{11}$ and $c_{22}$ are real elements of $\A^{p,q}$. The integrand now sits in $\A^{3,3}_{\rep{1},\rep{1}}$ multiplied by the $\gxg$ volume form $\Phi=\ee^{-2\dil}\sqrt{g}$ (which ensures one can integrate by parts). In the second line, we have written the action for the case of $\G_{2}$ special holonomy, with $\nabla$ denoting the Levi-Civita connection. The volume form and projections will be omitted from hereon and taken as part of the integration measure. Since in what follows we will be tackling the two terms in the action separately, we will denote the first term involving $c_{11}$ by $S_0^a$, and the second term by $S_0^b$. 

The idea then is that the partition function of this theory should match the 1-loop partition function calculated in \eqref{eq:Z1_g2_gen}. We can compute this partition function in two ways. The first is to use the standard BRST-BV quantisation procedure \cite{Batalin:1981jr}, in analogy with \cite{Pestun:2005rp}. We will follow this path here, assuming all relevant cohomologies of $M$ vanish to simplify the presentation. The second approach is by direct computation, as also demonstrated in \cite{Pestun:2005rp}. Given a quadratic action, this method is usually robust and perhaps more illustrative if the reader is unfamiliar with the BV approach. We illustrate this computation in Appendix~\ref{app:DirectOneLoop}. 

\subsubsection*{\texorpdfstring{$b_{11}\wedge \dd_+\dd_- b_{11}$}{c11 d+ d- c11}}

We begin by considering the first term $S_0^a$ involving the field $b_{11}$ and constructing the BV action. The gauge symmetries of this term lead to the following BRST transformations:
\begin{equation}
\begin{split}
  Qb_{11}&=\dd_+b_{01}+\dd_-b_{10},\\
  Qb_{10}&=\dd_+b_{00},\\
  Qb_{01}&=\dd_-b_{00}.
\end{split}
\end{equation}
These are similar to the transformations of \cite{Pestun:2005rp} but without any reality constraints as the fields involved are real. We have introduced ghosts $b_{10}$ and $b_{01}$, and a ghost for ghosts $b_{00}$. The fields $b_{pq}$ have statistics $(-1)^{(p+q)}$.

As in \cite{Pestun:2005rp} we introduce antifields. The antifield of $b_{pq}$ is a field $b^*_{(3-p)(3-q)}$ of ghost number $p+q-3$ and statistics $(-1)^{p+q+1}$. The master action then takes the form
\begin{equation}
S=S^a_0+\sum_{p,q}\int_M b^*_{(3-p)(3-q)}\wedge Qb_{pq}.
\end{equation}
This action reduces to $S^a_0$ when the antifields are zero, and satisfies the usual requirements that
\begin{equation}
    \{S,S\}=0,\qquad
    \{S,\Psi_i\}=Q\Psi_i,
\end{equation}
for any field $\Psi_i$, where the antibracket between two functionals $F$ and $G$ is given by
\begin{equation}
    \{F,G\}=\sum_i\,\left(\frac{\delta F}{\delta\Psi_i}\cdot\frac{\delta G}{\delta\Psi_i^*}-\frac{\delta F}{\delta\Psi_i^*}\cdot\frac{\delta G}{\delta\Psi_i}\right),
\end{equation}
where $\Psi_i^*$ is the antifield for $\Psi_i$. Using the master action, one can also derive the BRST transformations of the antifields via $Q\Psi_i^*=\{S,\Psi_i^*\}$:
\begin{equation}
    \begin{split}
        Qb^*_{22}&=\dd_+\dd_-b_{11},\\
        Qb^*_{23}&=\dd_-b^*_{22},\\
        Qb^*_{32}&=\dd_+b^*_{32}.
    \end{split}
\end{equation}

Next one chooses a Lagrangian submanifold. We choose this so as to remove the kernels of kinetic terms in the master action. This is done by projecting each field onto a subspace orthogonal to its variation under gauge transformations. From the classical part $S_0^a$, the Hodge decomposition implies that we should set
\begin{equation}
    b_{11}=\dd_+^\dagger\dd_-^\dagger d_{22}.
\end{equation}
The term involving fermionic ghosts and antifields reads
\begin{equation}\label{eq:Action1_Ghost1}
    S^a_1=\int_M b^*_{22}\wedge\left(\dd_+b_{01}+\dd_-b_{10}\right).
\end{equation}
Note that $\dd_+$ acting on $b_{01}$ has no kernel (assuming vanishing cohomologies), and likewise for $\dd_-$ on $b_{10}$. This implies that we can decompose $b_{01}$ and $b_{10}$ as
\begin{equation}
\begin{split}
     b_{01}&=\dd_-^\dagger d_{02}+\dd_-d_{00},\\
     b_{10}&=\dd_+^\dagger d_{20}+\dd_+\tilde d_{00}.
\end{split}
\end{equation}
Plugging this into \eqref{eq:Action1_Ghost1}, we see that terms involving the adjoint operators are orthogonal and so cannot cancel. However, terms involving the differentials cancel when $\tilde d_{00}=d_{00}$. To remove the kernel, we should set $\tilde d_{00}=-d_{00}$.
Finally, the bosonic action involving ghosts of ghosts reads
\begin{equation}
    S_2^a=\int_M\left(b^*_{23}\wedge\dd_+b_{00}+b^*_{32}\wedge\dd_-b_{00}\right).
\end{equation}
Again assuming vanishing cohomologies, this term puts no constraints on $b_{00}$.

The antifields are constrained by demanding
\begin{equation}
    \sum_i\int_M\Psi^*_i\wedge\Psi_i=0.
\end{equation}
As $b_{00}$ is unconstrained, we are forced to set $b^*_{33}=0$. From the constraint on $b_{11}$, we can derive a constraint on $b^*_{22}$:
\begin{equation}
    b^*_{22}=\dd_+^\dagger d^*_{32}+\dd_-^\dagger d^*_{23}.
\end{equation}
The constraints on $b^*_{23}$ and $b^*_{32}$ come from requiring that
\begin{equation}
\int_M\left(b^*_{32}\wedge b_{01}+b^*_{23}\wedge b_{10}\right)=0.    
\end{equation}
This holds provided we set 
\begin{equation}
        b^*_{32}=\dd_-^\dagger d_{33}, \qquad
        b^*_{23}=\dd_+^\dagger d_{33}.
\end{equation}
To show this requires an integration by parts, and the fact that the Laplacians $\Delta_+$ and $\Delta_-$ are equal.

We now compute contribution to the partition function of each term in the master action of $S_0^a$. For the classical term, the result is $(\det{}'\dd_+\dd_-)^{1/2}$ as the term is quadratic in $b_{11}$, which is bosonic. Note however that with the projection onto the Lagrangian submanifold, the operator $\dd_+\dd_-$ should be thought of as acting on $\dd_+^\dagger\dd_-^\dagger$-exact $(1,1)$-forms. The determinant is then
\begin{equation}
    \left(\det{}'(\dd_+\dd_-)\right)^{-1/2}=\left(\det{}'(\dd_+^\dagger\dd_-^\dagger\dd_+\dd_-)\right)^{-1/4}=\left(\det{}'\underset{\bullet}{\hat\Delta}^{1,1}\right)^{-1/2},
\end{equation}
as again the Laplacians are equal. The dot below denotes the fact that we are acting on $\dd_+^\dagger\dd_-^\dagger$-exact forms, as explained in Appendix \ref{app:zeta-dets}. This is also the determinant we have referred to as $C$ in Figure \ref{fig:G2_diagram}.

Next, let us compute the partition function of the bosonic action $S_2^a$ involving ghosts and ghosts of ghosts. We can write the ghost of ghost action as
\begin{equation}
\int_M\left(b^*_{23}+b^*_{32}\right)\wedge(\dd_++\dd_-)b_{00}.
\end{equation}
With the projection onto the Lagrangian submanifold, the result is \begin{equation}
   \left(\det{}'(\dd_++\dd_-)\right)^{-1}=\left(\det{}'\underset{\bullet}{\hat\Delta}^{0,0}\right)^{-1/2}=A^{-1/2}.
\end{equation}
Finally, the fermionic ghost term reads
\begin{equation}
    S^a_1=\int_M\left(b_{01}+b_{10}\right)\wedge(\dd_++\dd_-)b^*_{22}.
\end{equation}
The Lagrangian submanifold projects $b^*_{22}$ onto $\AomAC^{2,2}\oplus\AomCD^{2,2}\oplus\AomBD^{2,2}$, as defined in Appendix \ref{app:zeta-dets}. The contribution to the partition function of this term is therefore
\begin{equation}
    \det{}'(\dd_++\dd_-)=\left(\det{}'\underset{\bullet}{\hat\Delta}^{2,2}\det{}'\raisebox{1pt}{$\scriptstyle \bullet$}\hat\Delta^{2,2}\det{}' \hat\Delta\raisebox{1pt}{$\scriptstyle \bullet$}^{2,2}\right)^{1/2}=\left(AB'B\right)^{1/2},
\end{equation}
where in the general $\gxg$ case with flux there is a potential asymmetry between $B$ and $B'$ when mirroring the diagram of Figure \ref{fig:G2_diagram} about the vertical axis. In this case we denote the Laplacians of the middle upper-left square and the middle lower-right square $B$, and the middle upper-right and lower-left squares are denoted $B'$. For a global $G_2$ structure, one has $B=B'$, which is the case shown in Figure \ref{fig:G2_diagram}. Putting this together, the partition function of the term $S_0^a$ in the classical action is
\begin{equation}\label{eq:Za}
    Z^a=\left(\frac{BB'}{C}\right)^{1/2}.
\end{equation}


\subsubsection*{\texorpdfstring{$a_{00}\wedge \dd_+ \dd_- b_{22}$}{a00 d+ d- c22}}

Next we consider the term $S_0^b$ in the classical action. The BRST transformations read
\begin{equation}
    \begin{aligned}
        Qc_{22}&=\dd_+c_{12}+\dd_-c_{21}, &\qquad Qc_{20}&=\dd_+c_{10},\\
        Qc_{12}&=\dd_+c_{02}+\dd_-c_{11}, & Qc_{02}&=\dd_-c_{01},\\
        Qc_{21}&=\dd_-c_{20}+\dd_+c_{11}, & Qc_{10}&=\dd_+c_{00},\\
        Qc_{11}&=\dd_+c_{01}+\dd_-c_{10}, & Qc_{01}&=\dd_+c_{00},
    \end{aligned}
\end{equation}
while the field $a_{00}$ is gauge invariant. For each field and ghost $c_{pq}$ we again introduce an antifield $c^*_{(3-p)(3-q)}$ of statistics $(-1)^{p+q+1}$ and ghost number $p+q-5$. We also introduce a fermionic antifield $a^*_{33}$ for the field $a_{00}$.

The master action is now given as
\begin{equation}
    S=S_0^b+\sum_{p,q}\int_M c^*_{(3-p)(3-q)}\wedge Qc_{pq},
\end{equation}
which is easily checked to satisfy $\{S,S\}=0$, and generates the BRST transformations as $Qc_{pq}=\{S,c_{pq}\}$. Similarly the BRST transformations of the antifields are given as $Qc^*_{pq}=\{S,c^*_{pq}\}$ and $Qa^*_{33}=\{S,a^*_{33}\}$.
We proceed by introducing a Lagrangian submanifold to project out kernels of kinetic terms in the master action. This is again done by projecting each field onto a subspace orthogonal to its variation under gauge transformations. Assuming vanishing cohomologies, for the classical fields $c_{22}$ and $a_{00}$ we get
\begin{equation}
c_{22}=\dd^\dagger_+\dd^\dagger_- d_{33},    
\end{equation}
or $c_{22}\in\AomCD^{2,2}$ and with no conditions on $a_{00}$. We can immediately compute the partition function of $S^b_0$:
\begin{equation}
    Z^b_0=A^{-1}.
\end{equation}

Next consider the first-level fermionic ghost action
\begin{equation}
    S^b_1=\int_Mc^*_{11}\wedge\left(\dd_+c_{12}+\dd_-c_{21}\right).
\end{equation}
The contribution from this is most easily computed by considering the BRST transformation of the antifield $c^*_{11}$:
\begin{equation}
    Qc^*_{11}=\{S,c^*_{11}\}=\dd_+\dd_- c_{00}.
\end{equation}
The Lagrangian submanifold should hence project $c^*_{11}$ to $\AomAC^{1,1}\oplus\AomCD^{1,1}\oplus\AomBD^{1,1}$, and the contribution from $S^b_1$ is straightforwardly computed as
\begin{equation}
    Z^b_1=(CBB')^{1/2}.
\end{equation}

The second-level bosonic ghost action reads
\begin{equation}
    S^b_2=\int_M c^*_{21}\wedge\left(\dd_+c_{02}+\dd_-c_{11}\right)+\int_M c^*_{12}\wedge\left(\dd_-c_{20}+\dd_+c_{11}\right).
\end{equation}
Consider first the terms involving the field $c_{11}$. The gauge transformation of $c_{11}$ requires us to project $c_{11}$ to $\AomCD^{1,1}$ giving a contribution $C^{-1/2}$ from these terms. The gauge transformation of $c_{02}$ suggests projecting to $\AomCD^{0,2}$, while $c_{20}$ is projected to $\AomCD^{2,0}$. Both of these terms hence contribute a factor $A^{-1/2}$, giving
\begin{equation}
    Z_2^b=A^{-1}C^{-1/2}.
\end{equation}

The third-level fermionic action reads
\begin{equation}
    S_3^b=\int_M\left(c^*_{22}\wedge(\dd_+c_{01}+\dd_-c_{10})+c^*_{31}\wedge\dd_-c_{01}+c^*_{13}\wedge\dd_+c_{10}\right).
\end{equation}
The contribution from this action is again most easily computed by considering the BRST transformation of the antifields. We have
\begin{equation}
    Qc^*_{22}=\dd_+c_{12}+\dd_-c_{21},
\end{equation}
which tells us that we should project $c^*_{22}$ to $\AomCD^{2,2}$. The term involving $c^*_{22}$ then contributes a factor $A^{1/2}$ to the partition function. Similarly, the BRST transformation of $c^*_{31}$ is
\begin{equation}
    Qc^*_{31}=\dd_+c^*_{21}.
\end{equation}
Neglecting cohomologies, we find that we should set $c^*_{31}=0$ as part of the Lagrangian projection. Similarly, we also set $c^*_{13}=0$. The third-level action thus contributes
\begin{equation}
    Z_3^b=A^{1/2}
\end{equation}
to the partition function. 

The final term to consider is the bosonic action
\begin{equation}
    S^b_4=\int_M\left(c^*_{23}+c^*_{32}\right)\wedge(\dd_++\dd_-)c_{00}.
\end{equation}
As the ghost field $c_{00}$ is gauge invariant we have no constraints on this field. This term hence contributes
\begin{equation}
    Z_4^b=A^{-1/2}.
\end{equation}
Collecting all contributions, we thus find 
\begin{equation}\label{eq:Zb}
    Z^b=\frac{(BB')^{1/2}}{A^2}.
\end{equation}

\subsubsection*{Final result}

Putting together \eqref{eq:Za} and \eqref{eq:Zb}, the full partition function for the target-space action \eqref{eq:G2_action} is
\begin{equation}
    Z=Z^aZ^b=\frac{BB'}{C^{1/2}A^2}.
\end{equation}
It is straightforward to check that this expression agrees with the general $\gxg$ expression \eqref{eq:Z_1_G2}, and in particular the expression \eqref{eq:g2_partition} for the special case of a global $G_2$ structure where $B'=B$.


\section{The \texorpdfstring{$\sxs$}{Spin(7) x Spin(7)} complex}\label{spin7xspin7}

So far we have seen that we can use $\Orth{d,d}\times\bbR^+$ generalised geometry to build a double complex that gives the target-space analogue of the worldsheet BRST complex. We used this double complex to compute the 1-loop partition function of the topological $G_2$ string and then identify a target-space action which reproduces this result. These arguments can, in fact, be extended to any subgroup $G \subset \Orth{d}$ identified in \cite{Carrion98a} where one ``doubles'' the complexes discussed in that work. These should provide the 1-loop partition function to a suitably twisted $\sigma$-model on a target space with the corresponding $G$-structure. Here, we will focus on the $\Spin(7)$ case and use our results to provide a prediction for the conjectured topological $\Spin(7)$ string~\cite{Shatashvili:1994zw}. The calculations here are entirely analogous to those in the $\gxg$ case, so we will be light on details and simply sketch out some of the proofs while stating the key results.

\subsection{Generalised \texorpdfstring{$\sxs$}{Spin(7) x Spin(7)} structures}

The set up is much like Section \ref{sec:gen-g2g2}, except now we take $M$ to be eight-dimensional, so that we are working in $\Orth{8,8}\times \bbR^{+}$ geometry. In this case, two globally non-vanishing chiral\footnote{Note that the subscript $\pm$ identifies which spinor bundle the spinors are sections of (as in Section \ref{sec:gen-g2g2}), and \emph{not} the chirality of the spinors.} spinors $\epsilon_{\pm}\in S(C_{\pm})$ each define a $\Spin(7)$ structure given by $\Theta_{\pm}$. When the spinors are linearly independent, the $\Spin(7)$ structures are orthogonal and intersect on a $\G_{2}$ or $\SU(4)$ structure depending on the relative chirality of $\epsilon_{\pm}$ \cite{Witt:2004vr,VARADARAJAN2001163}. Much like in the $\G_{2}$ case, however, there may be places where the spinors align and the structure degenerates to $\Spin(7)$. If this is the case, the manifold admits only a local conventional $G$-structure, however the spinors still define a global $\Spin(7)\times \Spin(7)$ structure within generalised geometry.

One can describe certain backgrounds of type II strings compactified down to two dimensions in terms of $\Spin(7)\times \Spin(7)$ structures. In this case, we assume that we have a decomposition of the chiral ten-dimensional spinors as $\varepsilon_{\pm} = \zeta_{\pm}\otimes \epsilon_{\pm}$ into irreducible $\Spin(1,1)$ and $\Spin(8)$ spinors respectively. For the background to preserve supersymmetry, we need the supersymmetry variations of the gravitinos and dilatinos to vanish under $\varepsilon_{\pm}$. Under the decomposition above, and the assumption of vanishing RR flux, we find that we need $\zeta_{\pm}$ to be constant spinors on $\bbR^{1,1}$, and that $\epsilon_{\pm}$ must satisfy the Killing spinor equations \eqref{eq:killing-spinor} on $M$. It was shown in \cite{Witt:2004vr} that these equations hold if and only if $\epsilon_{\pm}$ define a torsion-free $\Spin(7)\times \Spin(7)$ structure.

\subsection{The double complex}

A generalised $\Spin(7)\times \Spin(7)$ structure defines a generalised metric and hence a decomposition $E = C_{+}\oplus C_{-}$ as discussed in Appendix \ref{sec:gen_metrics}. As we did for $\gxg$ structures, we can use the $G$-structure to define a refinement of the exterior algebra of this space and take
\begin{equation}
    \A^{p,q}_{\rep{m},\rep{n}} = \Gamma(\ext^{p}_{\rep{m}}C_{+}\oplus \ext^{q}_{\rep{n}}C_{-}),
\end{equation}
where $\rep{m}$ and $\rep{n}$ now denote $\Spin(7)$ representations. Then, given some compatible connection $D$, we can define a doubling of the complex \eqref{eq:Spin(7)_complex} through the following diagram:
\begin{equation}\label{eq:Spin(7)_double_complex}
    \begin{tikzcd}[column sep = small, row sep = small]
     & \arrow{dl}[swap]{\dd_{+}} & \A^{0,0}_{\rep{1},\rep{1}} \arrow[dl] \arrow[dr] & \arrow{dr}{\dd_{-}} &  \\
    \left. \right. & \A^{1,0}_{\rep{8},\rep{1}} \arrow[dl] \arrow[dr] & &  
    \A^{0,1}_{\rep{1},\rep{8}} \arrow[dl] \arrow[dr] & \left. \right.  \\
    \A^{2,0}_{\rep{7},\rep{1}} \arrow[dr] & & 
    \A^{1,1}_{\rep{8},\rep{8}} \arrow[dl] \arrow[dr] & &
    \A^{0,2}_{\rep{1},\rep{7}} \arrow[dl]  \\
    & \A^{2,1}_{\rep{7},\rep{8}}  \arrow[dr] & &
    \A^{1,2}_{\rep{8},\rep{7}} \arrow[dl]  &  \\
 & & \A^{2,2}_{\rep{7},\rep{7}} & & 
\end{tikzcd}
\end{equation}
where for $\omega \in \A^{p,q}_{\rep{m},\rep{n}}$ we have defined
\begin{align}
    (\dd_{+}\omega)_{a_{1}\dots a_{p+1}\bar{a}_{1}\dots \bar{a}_{q}} &=(p+1) (\Pr{+}{m'})_{a_{1}\dots a_{p+1}}{}^{b_{1}\dots b_{p+1}}D_{b_{1}}\omega_{b_{2}\dots b_{p+1}\bar{a}_{1}\dots \bar{a}_{q}}, \\
    (\dd_{-}\omega)_{a_{1}\dots a_{p}\bar{a}_{1}\dots \bar{a}_{q+1}} &= (-1)^{p}(q+1)(\Pr{-}{n'})_{\bar{a}_{1}\dots \bar{a}_{q+1}}{}^{\bar{b}_{1}\dots \bar{b}_{p+1}}D_{\bar{b}_{1}}\omega_{a_{1}\dots a_{p}\bar{b}_{2}\dots \bar{b}_{q+1}}.
\end{align}
We will see that, when $D$ is torsion-free, \eqref{eq:Spin(7)_double_complex} defines a double complex and the restriction to $(\A^{\bullet,0},\dd_{+})$ is isomorphic to \eqref{eq:Spin(7)_complex}.

First, we find the condition on the components of a generalised Levi-Civita connection for it to be $\Spin(7)\times \Spin(7)$ compatible. As before, this comes from taking a type IIB NSNS background of the form $\bbR^{1,1}\times M$ where $M$ is eight-dimensional, and then considering the Killing spinor equations on $M$. The conditions on $(H,\dil, A^{\pm})$ that we need to impose are
\begin{align}
    & D_{a}\epsilon_{+}=  \LC_a \epsilon_{+} - \tfrac{1}{24} H_{abc} \gamma^{bc }\epsilon_{+} -
		 \tfrac{1}{7}  \der^b\dil \gamma_a{}^b\epsilon_{+} 
           + \tfrac14 A^+_{abc}\gamma^{bc}\epsilon_{+} = 0, \\
   & D_{\ba} \epsilon_{+}=\nabla_{\ba}\epsilon_{+}-\tfrac{1}{8}H_{\ba bc}\gamma^{bc} \epsilon_{+} = 0,
\end{align}
where $\epsilon_{\pm}$ are the internal spinors that appear in the Killing spinor equations and hence define the $\Spin(7)\times \Spin(7)$ structure. The conditions above imply compatibility with the first $\Spin(7)$ factor, while the analogous conditions for $\epsilon_{-}$ imply compatibility with the second $\Spin(7)$ factor.

Decomposing the fields under the first $\Spin(7)$, one finds
\begin{align}
    \del\dil \in \rep{8}, \qquad H \in \rep{8} + \rep{48}, \qquad A^{+} \in \rep{48} + \rep{112}.
\end{align}
We can therefore write
\begin{align}
    H_{abc} &= H^{d}\Theta_{dabc} + \tilde{H}_{abc}, \\
    A^{+}_{abc} &= \left(\tilde{A}_{ade}\Theta_{bc}{}^{de} - \tilde{A}_{[a|de}\Theta_{|bc]}{}^{de}\right) + \hat{A}_{abc},
\end{align}
where $H_{d}\in \Omega^{1}_{\rep{8}}$, $\tilde{H},\tilde{A}\in \Omega^{3}_{\rep{48}}$, and $\hat{A}$ transforms in the $\rep{112}$ representation. Using the fact that
\begin{equation}
    (\Pr{2}{21})_{ab}{}^{cd}\gamma_{cd}\epsilon_{+} = 0, \qquad \tilde{A}_{[a|de}\Theta_{|bc]}{}^{de} = \tfrac{2}{3}\tilde{A}_{abc},
\end{equation}
we find that the conditions for $D$ to be a compatible connection are
\begin{equation}
    H_{d} = \tfrac{2}{7}\del_{d}\dil, \qquad \tilde{H} = 20\tilde{A},
\end{equation}
while $\hat{A}$ is unfixed. One finds analogous relations for the second $\Spin(7)$ factor.

As in the $\gxg$ case, we find that the operators in the double complex can be defined in terms of a ``simplified'' connection $\hat{D}$ which is neither torsion-free, nor compatible. Nonetheless, it can be used to check the nilpotency and anticommutivity of $\dd_{\pm}$. The simplified connection is
\begin{equation}
\label{eq:basic-connection-2}
\begin{aligned}
	\hat{D}_a v^b  & = \LC_a v^b ,\\
	\hat{D}_{\ba} v^b  & = \LC^{-}_{\ba} v^{b} = \LC_{\ba} v^{b} - \tfrac12 H_{\ba}{}^{b}{}_{c} v^{c} ,\\
	\hat{D}_{a} v^{\bb}  & = \LC^{+}_{a} v^{\bb} 
		= \LC_{a} v^{\bb} + \tfrac12 H_{a}{}^{\bb}{}_{\bc} v^{\bc} ,\\
	\hat{D}_{\ba} v^{\bb}  & = \LC_{\ba} v^{\bb} .
\end{aligned}
\end{equation}
Note that the first line of \eqref{eq:basic-connection-2} immediately implies that $(\A^{*,0},\dd_{+})$ is isomorphic to \eqref{eq:Spin(7)_complex} as required.

We now check the conditions for \eqref{eq:Spin(7)_complex} to be a double complex. Firstly note that, as in \eqref{eq:com-simpD}, we find that $\dd_{+}^{2}$ acting on $\omega\in \A^{0,p}$ is given by $(\Pr{2}{7})_{+}$ acting on the following:
\begin{equation}
   [\hat{D}_{a_{1}},\hat{D}_{a_{2}}]\omega_{\bar{b}_{1}\bar{b}_{2}} = -q\Riem^{+}_{a_{1}a_{2}}{}^{\bar{c}}{}_{[\bar{b}_{1}|}\omega_{\bar{c}|\bar{b}_{2}]}.
\end{equation}
However, after taking the projection, this vanishes because $\Riem^{+} \in \fspin^{+}\otimes \fspin^{-}$. It is clear therefore, that $\dd_{+}^{2} = 0$ acting on any vertex of \eqref{eq:Spin(7)_double_complex}. The $\dd_{-}^{2}=0$ condition follows analogously, and so we just need to check $\{\dd_{+},\dd_{-}\} = 0$ on $\A^{0,0}$, $\A^{1,0}$, $\A^{0,1}$ and $\A^{1,1}$. The calculation is very similar to those done in Section \ref{sec:G2_double_complex} and so we will demonstrate it only for $\omega \in \A^{1,0}_{\rep{8},\rep{1}}$. First consider
\begin{equation}
        [\hat{D}_{a},\hat{D} _{\bar{b}}]\omega_{c\bar{d}} = (\Riem_{a\bar{b}}{}^{e}{}_{c} - \tfrac{1}{2}\nabla_{a}H_{\bar{b}c}{}^{e} - \tfrac{1}{4}H_{a\bar{b}}{}^{\bar{d}}H_{\bar{d}c}{}^{e}) \omega_{e}.
\end{equation}
Antisymmetrising on $[ac]$, this becomes
\begin{equation}
    -\tfrac{1}{2}\Riem^{+}_{ac}{}^{e}{}_{\bar{b}}\omega_{e},
\end{equation}
which is annihilated by $(\Pr{2}{7})_{+}$. The others follow similarly and hence we see that if $D$ is a torsion-free $\Spin(7)\times \Spin(7)$ connection, \eqref{eq:Spin(7)_complex} defines a double complex.

\subsection{Hodge theory}

Next, we define Laplacians and analyse the Hodge theory. To do so, we introduce a metric on the complex as we did in \eqref{eq:G2_double_complex_metric}. That is, for $\alpha,\beta \in\A^{p,q}$, we have
\begin{equation}
    (\alpha,\beta) = \int_{M}\Phi\, G(\alpha,\beta) = \kappa_{p}\kappa_{q}\int_{M}\ee^{-2\dil}\vol\, \alpha\lrcorner \beta,
\end{equation}
where now
\begin{equation}
    \kappa_{p} = \begin{cases}
        1 & p=0,1, \\
        4 & p = 2.
    \end{cases}
\end{equation}
Defining adjoint operators $\dd_{\pm}^{\dagger}$ with this metric, we find their action on $(p,q)$-forms to be
\begin{align}
    \dd_{+}^{\dagger}\alpha &= -\gamma_{p}D^{a_{1}}\alpha_{a_{1}\dots a_{p}\bar{b}_{1}\dots \bar{b}_{q}}, \\
    \dd_{-}^{\dagger}\alpha &= (-1)^{p+1}\gamma_{q}D^{\bar{b}_{1}}\alpha_{a_{1}\dots a_{p}\bar{b}_{1}\dots \bar{b}_{q}},
\end{align}
where
\begin{equation}
    \gamma_{p} = \begin{cases}
    1 & p=1, \\
    4 & p=2.
    \end{cases}
\end{equation}

\subsubsection{K\"ahler identities}

As in the $\gxg$ case, the operators $\dd_{\pm}$ satisfy the analogue of the K\"ahler identities:
\begin{equation}\label{eq:Spin(7)_kahler_identities}
    \dd_{\pm}^{2} = (\dd_{\pm}^{\dagger})^{2} = \{\dd_{\pm}^{\dagger},\dd_{\mp}\} = 0.
\end{equation}
Unlike in the $\gxg$ case, however, we do not have isomorphisms connecting different vertices in the complex. We must therefore check the relations on all vertices. For the condition $\{\dd_{+}^{\dagger},\dd_{-}\} = 0$, the only non-trivial checks are on $\A^{1,0}, \A^{1,1}, \A^{2,0}$ and $\A^{2,1}$. For $\lambda_{1,0} \in \A^{1,0}$, the condition follows simply from \eqref{eq:genRicTensor} and the fact that $\GenRic_{ab} = 0$ for torsion-free $\Spin(7)\times \Spin(7)$ manifolds:
\begin{equation}
        \bigl( \{\dd_{+}^{\dagger},\dd_{-} \} \lambda_{1,0}\bigr)_{\bar{a}} = [D_{a},D_{\bar{a}}] \lambda^{a}= \GenRic_{a\bar{a}}\lambda^{a}=0.
\end{equation}
For $\lambda_{1,1} \in \A^{1,1}$ we find
\begin{equation}
    \begin{split}
        \bigl( \{\dd_{+}^{\dagger},\dd_{-}\} \lambda_{1,1})\bigr)_{\bar{a}\bar{b}} &= 2(\Pr{2}{7})_{\bar{a}\bar{b}}{}^{\bar{c}\bar{d}} \Bigl( \nabla_{\bar{c}}\nabla_{a} \lambda^{a}{}_{\bar{d}} - \nabla_{a}\nabla_{\bar{c}}\lambda^{a}{}_{\bar{d}} \\
        & \eqspace + (-2\nabla_{\bar{c}}\nabla_{a}\dil + \tfrac{1}{2}\nabla_{b}H_{\bar{c}}{}^{b}{}_{a} + \tfrac{1}{4} H_{b\bar{c}}{}^{\bar{e}}H_{\bar{e}}{}^{b}{}_{a} + \del_{b}\dil \, H_{\bar{c}}{}^{b}{}_{a})\lambda^{a}{}_{\bar{d}} \\
        & \eqspace + ( \tfrac{1}{2} \nabla_{\bar{c}}H_{a\bar{d}}{}^{\bar{e}} + \tfrac{1}{4}H_{b\bar{d}}{}^{\bar{e}}H_{\bar{c}}{}^{b}{}_{a} ) \lambda^{a}{}_{\bar{e}} \Bigr) \\
        &= 2(\Pr{2}{7})_{\bar{a}\bar{b}}{}^{\bar{c}\bar{d}} \bigl( -\GenRic_{\bar{c}a}\lambda^{a}{}_{\bar{d}} + \tfrac{1}{2} \Riem^{+}_{\bar{c}\bar{d}}{}^{\bar{e}}{}_{a}\lambda^{a}{}_{\bar{e}} \bigr) = 0.
    \end{split}
\end{equation}
The first term vanishes as before, and the second term vanishes because $\Riem^{+} \in \fspin^{+}\otimes \fspin^{-}$ and hence is annihilated by $\Pr{2}{7}$. For the remaining vertices, $\A^{2,0}$ and  $\A^{2,1}$, the calculation is similar:
\begin{align}
    \begin{split}
        \bigl( \{\dd_{+}^{\dagger},\dd_{-}\} \lambda_{2,0} \bigr)_{a\bar{a}} &= -4\GenRic _{\bar{a}b}\lambda^{b}{}_{a} - 2\Riem^{-}_{bc\bar{a}a}\lambda^{bc} \\
        &= 0,
    \end{split} \\
    \begin{split}
        \bigl( \{\dd_{+}^{\dagger},\dd_{-}\} \lambda_{2,1} \bigr)_{a\bar{a}\bar{b}} &= 8(\Pr{2}{7})_{\bar{a}\bar{b}}{}^{\bar{c}\bar{d}} \bigl( -\GenRic_{\bar{c}b}\lambda^{b}{}_{a\bar{d}} - \tfrac{1}{2} \Riem^{-}_{bca\bar{c}}\lambda^{bc}{}_{\bar{d}} - \tfrac{1}{2}\Riem^{+}_{\bar{c}\bar{d}}{}^{\bar{e}}{}_{b}\lambda^{b}{}_{a\bar{e}} \bigr) \\
        &= 0.
    \end{split}
\end{align}
The terms involving $\Riem^{-}$ vanish since contraction with $\lambda$ gives a contraction between the $\rep{21}$ and $\rep{7}$ which is necessarily zero. The analysis for $\{\dd_{-}^{\dagger},\dd_{+} \}$ then mirrors what we have done above. Putting this all together, one sees
\begin{equation}
    \{ \dd_{\pm}^{\dagger},\dd_{\mp}\} = 0.
\end{equation}

One can also check that $(\dd_{\pm}^{\dagger})^{2} = 0$. Unlike in the $\gxg$ case, this does not immediately follow from $\dd_{\pm}^{2} = 0$ and certain automorphisms of the complex. Instead, one needs to do the calculation explicitly. For example, for $\lambda\in \A^{2,q}$, one finds
\begin{equation}
    \bigl((\dd^{\dagger}_{+})^{2}\lambda\bigr)_{\bar{a}_{1}\dots \bar{a}_{q}} = -4\GenRic_{ab}\lambda^{ab}{}_{\bar{a}_{1}..\bar{a}_{q}} - 2q\Riem^{-}_{ab}{}^{\bar{b}}{}_{[\bar{a}_{1}|} \lambda^{ab}{}_{\bar{b}|\dots \bar{a}_{q}]} = 0,
\end{equation}
and similarly for $(\dd_{-}^{\dagger})^{2}$. This proves the K\"ahler identities in \eqref{eq:Spin(7)_kahler_identities}.

\subsubsection{Laplacians}

Finally, we can define the Laplacians of the ``plus'' and ``minus'' differentials via
\begin{equation}
    \Delta_{\pm} = \dd_{\pm}^{\dagger}\dd_{\pm} + \dd_{\pm} \dd_{\pm}^{\dagger}.
\end{equation}
Taking the combined differential $\hat{\dd} = \dd_{+} + \dd_{-}$, we find that the K\"ahler identities imply $\hat{\dd}^{2} = 0$, and that
\begin{equation}
    \hat{\Delta} = \hat{\dd}^{\dagger}\hat{\dd} + \hat{\dd}\hat{\dd}^{\dagger} = \Delta_{+} + \Delta_{-}.
\end{equation}
While we omit the rather lengthy calculations, one can follow the same arguments as in Section \ref{sec:g2g2_laplacians} to show that the Laplacians are again equal: $\Delta_{+} = \Delta_{-} = \tfrac{1}{2}\hat{\Delta}$.


\section{Relation to the topological \texorpdfstring{$\Spin(7)$}{Spin(7)} string}\label{sec:top_spin7_string}

The topological $\Spin(7)$ string was postulated in \cite{Shatashvili:1994zw} but is far less understood than its $\G_{2}$ counterpart. Despite this, we can use the double complex we have derived, along with intuition gained from the A/B-model and the $\G_{2}$-string, to provide a conjecture for its 1-loop partition function.

The key idea is to take the double complex above as the target space interpretation of the BRST complex and assume \eqref{eq:Z_1_G2} holds for any topological string. That is, we assume that the 1-loop partition function calculates the analytic torsion of the $\Spin(7)$ double complex:
\begin{equation}
    Z_{1} = \left[ \prod_{p,q} (\det{}'\hat{\Delta}^{p,q})^{(-1)^{p+q}pq} \right]^{-1/2},
\end{equation}
where now the product is taken over the vertices in the $\Spin(7)$ double complex. Again, using the fact that the determinants depend only on the $\Spin(7)\times \Spin(7)$ representation, we find
\begin{equation}
    Z_{1} = (\det{}'\hat{\Delta}_{\rep{7},\rep{7}})^{-2} (\det{}'\hat{\Delta}_{\rep{8},\rep{7}})(\det{}'\hat{\Delta}_{\rep{7},\rep{8}}) (\det{}' \hat{\Delta}_{\rep{8},\rep{8}})^{-1/2}.
\end{equation}
As before, the subscripts denote the $\Spin(7)\times \Spin(7)$ representation that $\hat{\Delta}$ is acting on. 

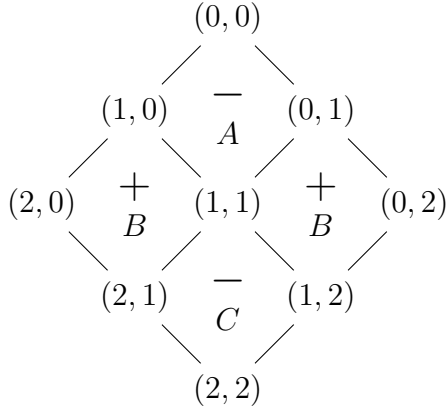
\begin{figure}
\centering
\begin{tikzpicture}[scale=1.75]
\draw[black] (0,0) grid [rotate=45] (2,2);
\node[circle,fill=white,draw=white,align=center,text=black,inner sep=0pt,minimum size=0pt] at (0*1.414,2*1.414) {$(0,0)$};
\node[circle,fill=white,draw=white,align=center,text=black,inner sep=0pt,minimum size=0pt] at (-.5*1.414,1.5*1.414) {$(1,0)$};
\node[circle,fill=white,draw=white,align=center,text=black,inner sep=0pt,minimum size=0pt] at (.5*1.414,1.5*1.414) {$(0,1)$};
\node[circle,fill=white,draw=white,align=center,text=black,inner sep=0pt,minimum size=0pt] at (1*1.414,1*1.414) {$(0,2)$};
\node[circle,fill=white,draw=white,align=center,text=black,inner sep=0pt,minimum size=0pt] at (0*1.414,1*1.414) {$(1,1)$};
\node[circle,fill=white,draw=white,align=center,text=black,inner sep=0pt,minimum size=0pt] at (-1*1.414,1*1.414) {$(2,0)$};
\node[circle,fill=white,draw=white,align=center,text=black,inner sep=0pt,minimum size=0pt] at (-0.5*1.414,.5*1.414) {$(2,1)$};
\node[circle,fill=white,draw=white,align=center,text=black,inner sep=0pt,minimum size=0pt] at (0.5*1.414,.5*1.414) {$(1,2)$};
\node[circle,fill=white,draw=white,align=center,text=black,inner sep=0pt,minimum size=0pt] at (0*1.414,0*1.414) {$(2,2)$};
\node[align=center] at (0*1.414,1.5*1.414) {\mbox{\Large$-$}\\$A$};
\node[align=center] at (0.5*1.414,1*1.414) {\mbox{\Large$+$}\\$B$};
\node[align=center] at (-0.5*1.414,1*1.414) {\mbox{\Large$+$}\\$B$};
\node[align=center] at (0*1.414,.5*1.414) {\mbox{\Large$-$}\\$C$};
%
\end{tikzpicture}
\caption{For a global $Spin(7)$ structure with vanishing flux, equality of $\Delta_\pm$ means that all $\det{}' \Delta^{p,q}$ can be expressed in terms of three independent determinants. For example, $\det{}' \hat\Delta^{0,0}=\det{}' \hat\Delta_{\rep1,\rep1}=\det{}' \Delta_{\rep1}\equiv A$, $\det{}' \hat\Delta^{2,0}=\det{}' \hat\Delta_{\rep7,\rep1}=\det{}' \Delta_{\rep7}\equiv B$, and $\det{}' \hat\Delta^{2,2}=\det{}' \hat\Delta_{\rep7,\rep7}=\det{}' \Delta_{\rep1}\det{}' \Delta_{\rep{21}}\det{}' \Delta_{\rep{27}}\equiv C$. The analytic torsion (the 1-loop partition function) is then given by $(A^{-1}B^2 C^{-1})^{1/2}$, in agreement with \eqref{eq:Spin(7)_partition_function}.}
\label{fig:Spin7_diagram}
\end{figure}

The topological $\Spin(7)$ string should correspond to the case where we have a global $\Spin(7)$ structure with vanishing flux, so that the metric on $M$ has special holonomy. The above expression can then be rewritten using representations of the diagonal $\Spin(7)$ and the fact that, by construction, $\hat{\Delta}\simeq \Delta$ on these subspaces:
\begin{equation}
\begin{split}
    Z_{1} &= (\det{}' \Delta_{\rep{1}})^{-5/2}(\det{}' \Delta_{\rep{7}})^{-1/2} (\det{}' \Delta_{\rep{8}})^{2} (\det{}' \Delta_{\rep{21}})^{-5/2}(\det{}'\Delta_{\rep{27}})^{-2}\\
    &\eqspace (\det{}'\Delta_{\rep{35}})^{-1/2} (\det{}'\Delta_{\rep{48}})^{2}.
    \end{split}
\end{equation}
It is possible to further simplify this as, for a $\Spin(7)$ manifold, the Laplacians are not independent. Using the relations outlined in Appendix \ref{app:Spin(7)_identities} we find that
\begin{equation}\label{eq:Spin(7)_partition_function}
    Z_{1} = (\det{}'\Delta_{\rep{1}})^{-1}(\det{}'\Delta_{\rep{7}})(\det{}'\Delta_{\rep{21}})^{-1/2}(\det{}'\Delta_{\rep{27}})^{-1/2}.
\end{equation}
Once more, this result can be read off directly from the double complex by assigning values to the squares in the diamond, using the isomorphisms to see which are equal, and multiplying them together with alternating powers of $\pm\tfrac{1}{2}$. This is illustrated in Figure \ref{fig:Spin7_diagram}.

Before providing more motivation for why this is might be the correct 1-loop partition function, we make some brief comments about the result. Firstly, this combination of determinants does not define a topological invariant of the $\Spin(7)$ manifold, much like the $\G_{2}$ case.\footnote{One can calculate the combinations of determinants that are topological, much like in \cite{deBoer08c}. For $Spin(7)$ there are four independent combinations. We find that $Z_1$ in \eqref{eq:Spin(7)_partition_function} cannot be written as a combination of these invariants. } This is not a surprise given that the $G$-structure defines a unique metric and the partition functions clearly depend on the $G$-structure. Secondly, there does not appear to be a quadratic target-space action whose partition function reproduces \eqref{eq:Spin(7)_partition_function} since the natural ``top form'' one would write down transforms in the $(\rep{7},\rep{7})$ and hence is not a $\Spin(7)\times \Spin(7)$ invariant. Despite this, our analysis provides a natural geometric interpretation of the 1-loop partition function as the product of $\zeta$-regularised determinants of Laplacians acting on the double complex \eqref{eq:Spin(7)_double_complex}.

Let us now motivate the above result by finding worldsheet operators that act as $\dd_{\pm}$. As we saw in Section \ref{sec:spin7_string_review}, the special NS highest-weight states selected out by the topological string are the following:
\begin{equation}\label{eq:Spin7_NS_states}
    \begin{array}{c}
    \left|0,0\right>  \\
    \Omega^{0}_{\rep{1}} 
    \end{array} \, , \qquad    
    \begin{array}{c}
    \left|\tfrac{1}{16},\tfrac{7}{16} \right> \\
    \Omega ^{1}_{\rep{8}} 
    \end{array} \, , \qquad   
    \begin{array}{c}
     \left|\tfrac{1}{2},\tfrac{1}{2} \right> \\
    \Omega^{2}_{\rep{7}}
    \end{array} \, ,  
\end{equation}
where we have written their interpretation as target-space differential forms underneath. These states should be generated by the chiral ring of the theory, which in turn should be annihilated by some nilpotent operator $Q$ we are yet to find. Given the discussion in this and the previous section, we expect that this $Q$ should have a left-moving piece that corresponds to the differential $\dc$ described in \eqref{eq:Spin(7)_complex}.

Typically, the operator $Q$ is built from the supersymmetry generators of the theory.\footnote{At least this is true in the A/B-models, and there is evidence that a version of this is true for the $\G_{2}$ string.} We therefore examine the OPE of the supercurrent $G$ with the operators defining the NS states above, and find
\begin{align}
    G(z)f(w) &= \ldots  - \frac{\del_{\mu}f \psi^{\mu}(w)}{2(z-w)} + \dots  \\
    G(z)A_{\mu}\psi^{\mu}(w) &= \ldots - \frac{\del_{\mu}A_{\nu}\psi^{\mu}\psi^{\nu}(w)}{2(z-w)}+\frac{\del x^{\mu}A_{\mu}(w)}{(z-w)}+\dots 
\end{align}
where we have  expressed only the order-one poles which give the action of $G_{-1/2}$. We see that, up to some term that involves $\del x^{\mu}$, we have $G_{-1/2}\sim -\tfrac{1}{2}\dd$. In fact, if we decompose the expressions above according to their weight under $T_{I}=\tfrac{1}{8}\tilde{X}$, we find
\begin{align}
    \del_{\mu}f\psi & \sim \left[\tfrac{1}{16}\right], \label{eq:supersym_transf_1} \\
    (\Pr{2}{7})_{\mu\nu}{}^{\rho\sigma}\del_{\rho}A_{\sigma}\psi^{\mu}\psi^{\nu} &\sim \left[\tfrac{1}{2}\right], \label{eq:supersym_transf_2} \\
    -\tfrac{1}{2}(\Pr{2}{21})_{\mu\nu}{}^{\rho\sigma}\del_{\rho}A_{\sigma}\psi^{\mu}\psi^{\nu} + \del x^{\mu}A_{\mu} &\sim \left[0\right]. \label{eq:supersym_transf_3}
\end{align}
We would like to pick some sub-operator of $G_{-1/2}$ that selects \eqref{eq:supersym_transf_1} and \eqref{eq:supersym_transf_2}, but leaves out \eqref{eq:supersym_transf_3}.

To see what this could be, we need to understand better the fusion rules of the Ising model. The Ising model is a minimal model at $p=3$. It therefore has three states $\alpha_{n}$ for $n=1,2,3$, where the weight of $\alpha_{n}$ is
\begin{equation}
    \alpha_{n} \sim \frac{(3n-4)^{2}-1}{48} = \begin{cases}
    0 & n=1, \\
    \tfrac{1}{16} & n=2, \\
    \tfrac{1}{2} & n=3.
    \end{cases}
\end{equation}
Plugging in the different values for $n$ reproduces the weights we have written above as expected. By examining the fusion rules, one finds~\cite{Shatashvili:1994zw}
\begin{equation}
    \alpha_{1}\alpha_{2} \sim \alpha_{2}, \qquad \alpha_{2}\alpha_{2} \sim \alpha_{1} + \alpha_{3}, \qquad \alpha_{3}\alpha_{2} \sim \alpha_{2}.
\end{equation}
Using the conformal block decomposition of a state -- in which we view a state $\alpha_{n}$ as a collection of maps between states dictated by the fusion rules -- we find that we can write
\begin{equation}
    \alpha_{2} = \alpha_{2}^{+} + \alpha_{2}^{-}, \qquad \alpha_{2}^{\pm}\colon\alpha_{n} \to \alpha_{n\pm 1},
\end{equation}
where we take $\alpha_{n} = 0$ for $n \notin \{1,2,3\}$.

Looking at \eqref{eq:supersym_transf_1}--\eqref{eq:supersym_transf_3}, we see that, at least in the Ising sector, $G_{-1/2} \sim \alpha_{2}$. Decomposing the action of $G_{-1/2}$ into conformal blocks as we did for $\alpha_{2}$, the sub-operator that selects only the special NS states in \eqref{eq:Spin7_NS_states}, and hence the natural candidate for the left-moving BRST operator, is
\begin{equation}
    Q_{L} = G_{-1/2}^{+} \sim \dd_{-}.
\end{equation}
One finds similar results for the right-moving sector with $Q_{R} = \bar{G}_{-1/2}^{+}$, and hence we reproduce the $\Spin(7)\times\Spin(7)$ double complex of Section \ref{spin7xspin7}. Note that this construction is analogous to that of the BRST operator in \cite{deBoer:2005pt} for the topological $\G_{2}$ string. We view this as strong evidence that this, or a small variation of this, is the correct worldsheet and target-space interpretation of the BRST operator for the topological $\Spin(7)$ string.

\section{Some other examples}
\label{sec:other-examples}

Though we have focused on the cases of $\gxg$ and $\sxs$, and the diagonal subgroups relevant for topological strings, our construction is in fact far more general. Once one has identified a group $G\subset \Orth{d}$ and the relevant instanton complex from \cite{Carrion98a}, one can find a doubled version of the complex by lifting to a $G\times G \subset \Orth{d,d}$ structure and then using a torsion-free generalised connection that is compatible with the structure. Moreover, this construction will work for non-vanishing $H$-flux even if this breaks integrability of the conventional $G$-structure.

Since the proofs are essentially the same as for the $\G_{2}$ and $\Spin(7)$ case with the relevant groups and representations exchanged, we will not show explicitly that the diagrams we give are double complexes, that the K\"ahler identities hold, nor that the left and right Laplacians are equal. Instead, we simply write down the complexes in a few cases of interest and comment on connections to 1-loop partition functions.

\subsection{A- and B-models with background \texorpdfstring{$H$}{H}-flux}

As we have seen, the double complex exists only for certain choices of $H$-flux. While these choices may break integrability of the conventional $G$-structure, if the flux preserves integrability of the generalised $G\times G$ structure, and hence the background is still supersymmetric, the double complex is well defined.\footnote{Recall that integrability of the generalised structure forces some components of $H$ to vanish, while others are related to $\del\dil$ or left unconstrained.} This allows us to describe topological strings on backgrounds with non-vanishing flux where one necessarily needs to use generalised geometry and the doubled complex. 
In particular, it is interesting to see how our double complex describes the A- or B-models with flux.

First, let us build the double complex that applies to six-dimensional supersymmetric backgrounds. These backgrounds have an integrable $\SU(3)\times \SU(3)$ structure\footnote{In fact, one only needs the target space to be generalised K\"ahler, which is a slightly weaker structure. However, restricting to $\SU(3)\times \SU(3)$ ensures there are no anomalies in the twist \cite{Kapustin:2004gv}.} which means we can decompose the bundles $C_{\pm}$ under this group into complex conjugate representations
\begin{equation}\label{eq:SU3xSU3_decomp}
    C_{\pm} = C_{\pm}^{1,0} \oplus C_{\pm}^{0,1}.
\end{equation}
In this case, we can build two inequivalent double complexes out of this decomposition that we suggestively call the A- and B-complexes. These are respectively defined by
\begin{align}
    \A^{p,q}_{A} &= \Gamma(\ext^{p}C^{0,1}_{+}\otimes \ext^{q}C^{1,0}_{-} ),\label{eq:A_complex} \\
    \A^{p,q}_{B} &= \Gamma(\ext^{p}C^{0,1}_{+} \otimes \ext^{q}C^{0,1}_{-} ).\label{eq:B_complex}
\end{align}
We can build the corresponding differentials from a torsion-free compatible connection as before, this time with the projections mapping onto the vector spaces above. The 1-loop partition function then follows immediately from the analytic torsion formula we used previously, namely
\begin{equation}\label{eq:A/B_1-loop_with_flux}
    Z_{1} = \left[ \prod_{p,q}(\det{}'\hat\Delta^{p,q})^{(-1)^{p+q}pq} \right]^{-1/2}.
\end{equation}
Let us now see how this relates to the A- and B-models with flux, as studied in \cite{Kapustin:2004gv}.

The relation of the A/B-models to generalised geometry has been studied in great detail \cite{Lindstrom:2004eh,Pestun:2006rj,Kapustin:2003sg}. In general, a two-dimensional $N=(2,2)$ $\sigma$-model with $H$-flux has a target space with a twisted generalised K\"ahler structure~\cite{Gates:1984nk,gualtieri2004}. This is defined by two commuting, integrable generalised complex structures $\mathcal{J}_{1}$ and $\mathcal{J}_{2}$. Individually, they give a decomposition of the generalised tangent bundle into eigenbundles:
\begin{equation}\label{eq:gen_kahler_e-bundles}
    E_{\bbC} = L_{1}\oplus \overline{L_{1}} = L_{2} \oplus \overline{L_{2}},
\end{equation}
where $L_{i}$ is the $+\ii$ eigenbundle of $\mathcal{J}_{i}$. Since the two complex structures commute, we can find mutual eigenbundles and write
\begin{equation}
    E_{\bbC} = \underbrace{L_{1}^{+}}_{(1,1)} \oplus \underbrace{L_{1}^{-}}_{(1,-1)}\oplus \underbrace{\overline{L_{1}^{+}}}_{(-1,-1)} \oplus \underbrace{\overline{L_{1}^{-}}}_{(-1,1)},
\end{equation}
where we have indicated the charges under $(\mathcal{J}_{1},\mathcal{J}_{2})$. We see that $L_1$ and $L_2$ can be identified with
\begin{equation}
    L_{1} = L_{1}^{+} \oplus L_{1}^{-}, \qquad L_{2} = L_{1}^{+} \oplus \overline{L_{1}^{-}}.
\end{equation}

As usual, the two generalised complex structures define a generalised metric via $G=-\mathcal{J}_{1}\mathcal{J}_{2}$, which in turn defines the subspaces $C_{\pm}$. It turns out that the decompositions \eqref{eq:SU3xSU3_decomp} and \eqref{eq:gen_kahler_e-bundles} are then related via
\begin{equation}
    C^{1,0}_{+} = L_{1}^{+}, \qquad C^{1,0}_{-} = L_{1}^{-}.
\end{equation}
Using this, the vector spaces \eqref{eq:A_complex} and \eqref{eq:B_complex} that we have dubbed the A- and B-complexes are given by 
\begin{align}
    \A^{p,q}_{A} &= \Gamma(\ext^{p}\overline{L_{1}^{+}} \otimes \ext^{q} L_{1}^{-}) \subseteq \Gamma(\ext^{p+q}\overline{L_{2}}), \\
    \A^{p,q}_{B} &= \Gamma(\ext^{p} \overline{L_{1}^{+}} \otimes \ext^{q} \overline{L_{1}^{-}}) \subseteq \Gamma(\ext^{p+q}\overline{L_{1}}),
\end{align}
so that the total space of the A- and B-complexes are $\ext^{\bullet}L_{2}^{*}$ and $\ext^{\bullet}L_{1}^{*}$ respectively.

Using the results of \cite{gualtieri2004}, we know that for any generalised complex structure $\mathcal{J}$, its $+\ii$ eigenbundle $L$ defines a Lie algebroid under the Courant bracket. This means that there is an associated differential $\dd_{L}$ that makes $(\ext^{\bullet}L^{*},\dd_{L})$ a complex. Furthermore, if $\mathcal{J}$ is part of a generalised K\"ahler structure, then the differential splits as
\begin{equation}
    \dd_{L} = \del_{L}^{+} + \del_{L}^{-},
\end{equation}
with
\begin{equation}
    \del_{L}^{+}\colon\ext^{p,q}L^{*} \to \ext^{p+1,q}L^{*}, \qquad \del_{L}^{-}\colon\ext^{p,q}L^{*} \to \ext^{p,q+1}L^{*},
\end{equation}
where we have defined $\ext^{p,q}L^{*}_{1} = \ext^{p}\overline{L_{1}^{+}} \otimes \ext^{q} \overline{L_{1}^{-}}$, and similarly for $L_{2}$ (see also \cite{Cavalcanti:2012fr}). One can show that these differentials coincide with the those in the double complex defined via generalised connections, i.e.~$\del_{L}^{\pm} = \dd_{\pm}$. Hence, the total BRST operator is
\begin{align}
    Q &= Q_{L} + Q_{R}= \dd_{+} + \dd_{-}= \del_{L}^{+} + \del_{L}^{-} \\
    & = \dd_{L},
\end{align}
where $L$ is $L_{1}$ for the B-model and $L_{2}$ for the A-model. The chiral ring then corresponds to the cohomology associated to one of the generalised complex structures, with the choice of structure fixed by whether one uses the A or B twist. This exactly reproduces the results of \cite{Kapustin:2004gv}. We can now interpret the 1-loop partition function \eqref{eq:A/B_1-loop_with_flux} more concretely with the knowledge that the relevant Laplacian $\hat\Delta$ is that associated to the differential $\dd_{L}$.\footnote{As was shown in \cite{Kapustin:2004gv}, depending on whether one uses $\mathcal{J}_{1}$ or $\mathcal{J}_{2}$, $Z_1$ may receive instanton corrections at finite volume, and so we should view \eqref{eq:A/B_1-loop_with_flux} as the 1-loop partition function at infinite volume.}

For completeness, note that one can recover the chiral rings of the A- and B-models by defining a generalised K\"ahler structure from a conventional K\"ahler structure $(I, \omega)$ via
\begin{equation}
    \mathcal{J}_{1} = \begin{pmatrix}
        I & 0 \\
        0 & -I
    \end{pmatrix}, \qquad \mathcal{J}_{2} = \begin{pmatrix}
        0 & -\omega^{-1} \\
        \omega & 0
    \end{pmatrix}.
\end{equation}
With this choice, the A- and B-complexes in \eqref{eq:A_complex} and \eqref{eq:B_complex} reproduce the chiral rings of the A- and B-models respectively, with the 1-loop partition function in \eqref{eq:A/B_1-loop_with_flux} reducing to \eqref{eq:A-model_1-loop}.

\subsection{Topological strings on K3}

A K3 manifold has an $\SU(2)\subset \Spin(4)$ structure, for which the relevant instanton complex is isomorphic to the Dolbeault complex
\begin{equation}
    \Omega^{0,0} \xrightarrow{\quad \delb \quad} \Omega^{0,1} \xrightarrow{\quad \delb \quad} \Omega^{0,2},
\end{equation}
where we can choose any combination of the three commuting complex structures to define the $\delb$ operator. Lifting to a generalised $\SU(2)\times \SU(2) \subset \Spin(4,4)$ structure, we find two possible ways to define the doubling of the complex, corresponding to the A- and B-models on K3. In both cases, in the infinite-volume limit where instanton corrections can be neglected, the complex is naturally isomorphic to the Dolbeault complex:\footnote{By changing the choice of complex structure within the hyperk\"ahler structure, one can continuously interpolate between the A- and B-model \cite{Kapustin:2003sg}.}
\begin{equation}\label{eq:K3_double_complex}
    \begin{tikzcd}[column sep = small, row sep = small]
     & \arrow{dl}[swap]{\del} & \Omega^{0,0} \arrow[dl] \arrow[dr] & \arrow{dr}{\delb} &  \\
    \left. \right. & \Omega^{1,0} \arrow[dl] \arrow[dr] & &  
    \Omega^{0,1} \arrow[dl] \arrow[dr] & \left. \right.  \\
    \Omega^{2,0} \arrow[dr] & & 
    \Omega^{1,1}\arrow[dl] \arrow[dr] & &
    \Omega^{0,2} \arrow[dl]  \\
    & \Omega^{2,1}  \arrow[dr] & &
    \Omega^{1,2} \arrow[dl]  &  \\
 & & \Omega^{2,2} & & 
\end{tikzcd}
\end{equation}
One can then find the 1-loop partition function of the topological string on K3 using the analytic torsion \eqref{eq:A/B_1-loop_with_flux}.
Thanks to the Calabi--Yau structure of a K3, one finds that the $\zeta$-regularised determinants of Laplacians have many identifications:
\begin{align}
    \det{}'\Delta^{p,q} &= \det{}'\Delta^{q,p} = \det{}'\Delta^{2-p,q},\\
    \det{}'\Delta^{1,1} &= (\det{}'\Delta^{1,0})^{2} = (\det{}'\Delta^{0,0})^{4}.
\end{align}
Applying these to \eqref{eq:A/B_1-loop_with_flux}, we find that the 1-loop partition function is trivial, $Z^{\text{K3}}_{1} = 1$. This matches the result that the partition functions for the A- and B-models are trivial on a K3 in the large-volume limit~\cite{Bershadsky:1993cx}.\footnote{One needs to use the worldsheet theory to see that it is not possible to absorb fermion zero-modes in order to show that the partition function for the A-model remains trivial at finite volume.} Note that we were able to show this directly from the target-space geometry without a detailed description of the worldsheet theory.


\section{Conclusions and future directions}

In this paper, we have given a prescription for calculating the 1-loop partition function of certain topological string models whose target spaces admit torsion-free $G \times G$ structures within $O(d,d)\times\mathbb{R}^+$ generalised geometry. We reviewed how there are natural complexes for both $G_2$ and $Spin(7)$ structures. We then extended these to double complexes for $G_2\times G_2$ and $Spin(7)\times Spin(7)$, with the relevant differentials constructed from torsion-free compatible generalised connections. We showed that such connections exist provided the target space satisfies certain differential conditions that correspond to it being a supersymmetric NSNS background for a Minkowski spacetime. In each case, there existed an analogue of K\"ahler identities and Hodge theory which allowed us to define Laplacians acting on representations of $G \times G$. Starting from the conjecture that the 1-loop partition function is given by a certain alternating product of determinants of these Laplacians, we showed that our formalism reproduced the known worldsheet results for the A- and B-models and the $G_2$ string. Our result for the $Spin(7)$ string is new. Finally, as further examples, we discussed how our formalism captures topological strings on K3 surfaces and the A- and B-models with flux.

An overarching theme of our work is that $G\times G \subset O(d,d)\times\mathbb{R}^+$ structures within generalised geometry should be thought of as the correct target-space language for describing worldsheet models, with the left- and right-moving sectors captured by the spaces $C_+$ and $C_-$ respectively (as mentioned previously in \cite{Coimbra:2011nw,Strickland-Constable:2021afa}). By moving to the twisted theory, one is restricted to special subspaces of $C_{\pm}$ selected by the $G$-structure.

Unfortunately, we were not able to give a target-space action for the $Spin(7)$ string, nor for backgrounds with general $G\times G$ structure without special holonomy. Following the logic of \cite{Pestun:2005rp}, one might imagine that there are functionals whose quantisation leads to the 1-loop partition functions we have calculated in this paper. It may be that one needs to consider RR degrees of freedom and extend to exceptional generalised geometry~\cite{Hull:2007zu,PiresPacheco:2008qik,Grana:2009im,Coimbra:2011ky,Coimbra:2012af} in order understand these, or if one wants to understand the story in M-theory. We hope to tackle this in the future, for example by building on the work defining invariant functionals in~\cite{Ashmore:2015joa,Ashmore:2019qii,Ashmore:2019rkx,Tennyson:2021qwl}. As a first step, one could imagine quantising variations of the ``hypermultiplet structure'' of \cite{Ashmore:2015joa}, which in type IIA would give the analogue of \cite{Pestun:2005rp} but for the A-model (or K\"ahler gravity). However, in order to extend the $\gxg$ and $\sxs$ constructions introduced in this paper one would need to identify the corresponding structures in $E_{8(8)}$ and $E_{9(9)}$ generalised geometries respectively, which have not yet been formulated (though certain subsectors of $E_{8(8)}$ generalised geometry have been introduced in \cite{Strickland-Constable:2013xta} which might provide clues on how to build the invariant functionals).

There are a number of research directions opened up by our work. In \cite{deBoer08c}, the quantised $G_2$ target-space theory was compared with the results of Pestun and Witten~\cite{Pestun:2005rp} by reducing the theory on a circle. One could perform a similar check by reducing the $Spin(7)$ double complex to $G_2$ in the cases with and without $H$ flux. Staying in eight dimensions, as another direction one might consider embedding a global $SU(4)$ structure in $\sxs$ where it should give a non-critical version of the B-model on a fourfold.

Another relatively straightforward extension of this work would be to consider the cases where the generalised intrinsic torsion of the $\gxg$ or $\sxs$ structures does not (entirely) vanish. Recall that we used the fact that we were examining supersymmetric Minkowski backgrounds to immediately conclude that the $G\times G$ generalised structures must be torsion-free, and that was sufficient to prove the existence of the corresponding double complexes. However, it is possible that one may be able to weaken this constraint for other backgrounds -- in particular, supersymmetric AdS backgrounds are described in generalised geometry by constant singlet torsion~\cite{Coimbra:2015nha,Ashmore:2016qvs,Coimbra:2017fqv}. For the $\gxg$ case, one could then hope to use the concepts developed in this work to make contact with worldsheet computations for NSNS AdS$_3$ backgrounds \cite{Fiset:2021azq}.

It would also be worthwhile to understand whether there is a physical interpretation of the double complexes when the groups for right- and left-movers on the worldsheet are not matched. For example, one could imagine taking $SU(3)\times G_2 \subset O(7)\times O(7)$. Such a generalised structure would be defined by three global spinors, a pair of orthogonal $\epsilon^{1,2}_+$ and an $\epsilon_-$, so that the seven-dimensional manifold would generically have an $\SU(2)$ structure that becomes $\SU(3)$ wherever $\epsilon_-$ is parallel to either $\epsilon^i_+$. Again, one can write down the conditions for the structure to be torsion-free and construct differentials using the corresponding torsion-free compatible connection. Closely related to this would be considering the generalised geometric description of heterotic supergravity, where the relevant group is $O(d)\times O(d+n)$, with the gauge group $G$ embedding in the second factor~\cite{Coimbra:2014qaa,Ashmore:2018ybe,Ashmore:2019rkx,Garcia-Fernandez:2013gja,Garcia-Fernandez:2015hja, delaOssa:2017pqy, Clarke:2020erl, delaOssa:2021cgd}. If a double complex can be constructed in this case, one imagines it could be related to $G$-instantons.

Given the importance of the A- and B-models for understanding mirror symmetry on Calabi--Yau threefolds, one might wonder if these double complexes could be used to probe mirror symmetry on $G_2$ or $Spin(7)$ manifolds~\cite{Shatashvili:1994zw,Papadopoulos:1995da,Acharya:1996fx,Acharya:1997rh,Braun:2017ryx,Braun:2019lnn}, or if the 1-loop partition functions can be expressed in terms of the ``$\rho$-characteristic'' of \cite{Borsten:2021pte} which has special properties for self-mirror manifolds. As a consistency check, $G_2$ mirror symmetry appears on the worldsheet as a certain automorphism of the right-moving extended algebra~\cite{Becker:1996ay,Roiban:2002iv,Gaberdiel:2004vx}, suggesting that Figure \ref{fig:G2_diagram} should be symmetric when reflected along the diagonal from the top left to the bottom right, which indeed it is.

Our work may also have applications in K-theory and index formulae, which can be seen by reinterpreting the construction of the double complexes in terms of generalised spinors. Indeed, for $\Spin(7)\times \Spin(7)$, one can check that the total space of the double complex is isomorphic to the space of generalised spinors. The operators $\mathcal{D}_{\pm}=\dd_{\pm}+\dd_{\pm}^{\dagger}$ then define new elliptic operators on this space which are related to, but not exactly, (twisted) Dirac operators on $\Omega^{\bullet}(M)$. If these operators, or some construction related to them, has a parallel $\Cliff(8,8)$ action, it would descend to the finite-dimensional space $\ker\mathcal{D}_{\pm}$, giving it the structure of a $\Cliff(8,8)$ module. The residue of this, as defined in e.g.~\cite{Lawson:1998yr}, would give a $\mathbb{Z}$-valued index for the manifold which, under general arguments, should be invariant under continuous deformations of the operator~\cite{Atiyah:1966aaa,Karoubi:1978aaa}. We would like to see if and how this index is related to other indices on eight-dimensional manifolds. Something even more curious happens in the $\G_{2}\times \G_{2}$ case. Here, the total space of the double complex is isomorphic to two copies of the generalised spinors, possibly indicating that the correct description should be in terms of pinors. In any case, if one can find a parallel $\Cliff(7,7)$ action with respect to $\mathcal{D}_{\pm}$, then the finite-dimensional space $\ker\mathcal{D}_{\pm}$ becomes a $\Cliff(7,7)$ module. Due to the split signature of the Clifford algebra, the residue of this representation does not trivially vanish as one would expect for seven-dimensional manifolds. This may give a new index that could be used to distinguish $\G_{2}$ structures.

More speculatively, one might hope that higher-loop contributions to the partition function can also be captured by the generalised geometry of the target space. Similarly, one might wonder whether one can use the double complexes to compute twisted worldsheet indices in the spirit of Cecotti et al.~\cite{Cecotti:1992qh}. In another direction, there has also been recent progress in both the physics and mathematics literature in understanding instantons, invariants and enumerative geometry in the exceptional setting, see e.g.~\cite{donaldson2009gauge, Joyce:2016fij, doan2017counting, Braun:2018fdp, Acharya:2018nbo}. As mentioned, for instantons and their counting, the single complexes of \cite{Carrion98a} play a natural role. Then, in analogy with how the open-closed duality of the A-model and gauge theory can be used to compute Gromov--Witten invariants of Calabi--Yau manifolds~\cite{Aganagic:2003db}, one could ask about the relations between our double complexes and the counting of, for example, associative submanifolds and $G_2$ instantons.


\acknowledgments

We thank Bobby Acharya, Magdalena Larfors, Matthew Magill, Paul de Mederios, Xenia de la Ossa and Daniel Waldram for useful discussions. AA is supported by the European Union's Horizon 2020 research and innovation programme under the Marie Sk\l{}odowska-Curie grant agreement No.\,838776. AC is supported by the European Research Council under the European Union’s Horizon 2020 research and innovation programme (``Exceptional Quantum Gravity'', grant agreement No.\,740209). DT is supported by the EPSRC New Horizons Grant ``New geometry from string dualities'' EP/V049089/1.


\appendix

\section{Conventions and useful identities}\label{app:conventions}

In this appendix, we collect our conventions together with a number of useful identities and projectors for $G_2$ and $Spin(7)$.

\subsection{Conventional geometry}

Given a conventional connection $\nabla$ on $M$, we can express its torsion $T\in \Gamma(TM\otimes\ext^{2}T^{*}M)$ as
\begin{align}
\nabla_{m}v^{n} & =\partial_{m}v^{n}+\Gamma_{m}{}^{n}{}_{p}v^{p},\nonumber \\
T(v,w) & =\nabla_{v}w-\nabla_{w}v-[v,w],\\
T^{m}{}_{np} & =\Gamma_{n}{}^{m}{}_{p}-\Gamma_{p}{}^{m}{}_{n},\nonumber 
\end{align}
where $[\,,\,]$ is the Lie bracket. The curvature of $\nabla$ is then given by the Riemann tensor $\Riem\in\Gamma(\ext^{2}T^{*}M\otimes\op{End}TM)$, defined by
\begin{equation}
\begin{aligned}\label{eq:Riem}
\Riem(u,v)w & =[\nabla_{u},\nabla_{v}]w-\nabla_{[u,v]}w,\\
\Riem_{mn}{}^{p}{}_{q}w^{q} & =[\nabla_{m},\nabla_{n}]w^{p}-T^{q}{}_{mn}\nabla_{q}w^{p},
\end{aligned}
\end{equation}
with the Ricci tensor and Ricci scalar defined by
\begin{equation}
\Ric_{mn}=\Riem_{p m}{}^{p}{}_{n},\qquad \Scalar=g^{mn}\Ric_{mn}.
\end{equation}

We define the generalised Kronecker delta as
\begin{equation}
    \delta^{m_1\dots m_p}_{n_1\dots n_p}=\delta^{[m_1}_{n_1}\dots\delta^{m_p]}_{n_p},
\end{equation}
so that its components are zero or $\pm\tfrac{1}{p!}$. In particular, this convention implies
\begin{equation}
    \delta^{m_1\dots m_p}_{n_1\dots n_p}\alpha^{n_1\dots n_p}=\alpha^{[m_1\dots m_p]}.
\end{equation}

\subsection{\texorpdfstring{$G_2$}{G2}}\label{app:G2_identities}

We use a (conventional) orthonormal frame $g_{mn} = \delta_{mn}$ and take the seven-dimensional gamma matrices to furnish a representation of $\Cliff(7;\bbR)$ with $\gamma^{(8)} = \gamma^1 \dots \gamma^7 = - \ii \id$.\footnote{One can take the gamma matrices to be imaginary and antisymmetric, so that a Majorana spinor is real and obeys $\bar\epsilon = \epsilon^{\text{T}}$~\cite{Gauntlett:2003cy}.} We take the $G_2$ structure to be defined by a Majorana spinor $\epsilon$ normalised such that $\bar\epsilon \epsilon = 1$. The $G_2$-invariant 3-form $\gphi$ and its Hodge dual $*\gphi$ are defined as
\begin{equation}
	\gphi_{mnp} =  -\ii\bar\epsilon \gamma_{mnp} \epsilon,\qquad 	(*\gphi){}_{m_1 \dots m_4} = - \bar\epsilon \gamma_{m_1 \dots m_4} \epsilon.
\end{equation}
In an orthonormal frame, these can be written as
\begin{equation}
\begin{split}
	\gphi &=  e^{246} - e^{235} - e^{145} - e^{136} + e^{127} + e^{347} + e^{567},\\
	*\gphi&=  e^{1234} + e^{1256} + e^{3456} + e^{1357} - e^{1467} - e^{2367} - e^{2457}.
	\end{split}
\end{equation}

\subsubsection*{Identities}

Using Fierz identities on products of four $\epsilon$'s, one can show the following identities hold:
\begin{align}
\begin{split}
	\gphi^{m_1 m_2 p} \gphi_{n_1 n_2 p} &= 2 \delta^{m_1 m_2}_{n_1 n_2} 
		+  (*\gphi){}^{m_1 m_2}{}_{n_1 n_2}, \\
	\gphi^{mp_1 p_2} \gphi_{np_1 p_2} &= 6 \delta^{m}{}_{n}, \\
	\gphi^{m_1 m_2 m_3} \gphi_{m_1 m_2 m_3} &= 42,
\end{split}\\
\begin{split}\label{eq:B4_identity}
	(*\gphi){}^{m_1 m_2 m_3 p} \gphi_{n_1 n_2 p} 
		&= -6  \delta^{[m_1}{}_{[n_1} \gphi_{n_2]}{}^{m_2m_3]}, \\
	(*\gphi){}^{m_1 m_2 p_1 p_2} \gphi_{n p_1 p_2} 
		&= 4  \gphi^{m_1m_2}{}_{n}.
\end{split}
\end{align}
One also has
\begin{equation}
\begin{aligned}
	(*\gphi){}^{m_1 \dots m_4} (*\gphi){}_{n_1 \dots n_4} 
		&= 24 \delta^{m_1 \dots m_4}_{n_1 \dots n_4} 
			+ 72  \delta^{[m_1 m_2}_{[n_1 n_2} (*\gphi){}^{m_3 m_4]}{}_{n_3 n_4]}
			\\&\eqspace- 16 \delta^{[m_1}{}_{[n_1} \gphi^{m_2 m_3 m_4]} \gphi_{n_2 n_3 n_4]},\\
	(*\gphi){}^{m_1 m_2 m_3 p} (*\gphi){}_{n_1 n_2 n_3 p} 
		&= 6 \delta^{m_1 m_2 m_3}_{n_1 n_2 n_3} 
			+ 9  \delta^{[m_1}{}_{[n_1} (*\gphi){}^{m_2 m_3]}{}_{n_2 n_3]}
			-\gphi^{m_1 m_2 m_3} \gphi_{n_1 n_2 n_3},\\
	(*\gphi){}^{m_1 m_2 p_1 p_2} (*\gphi){}_{n_1 n_2 p_1 p_2} 
		&= 8 \delta^{m_1 m_2}_{n_1 n_2} 
			+ 2  (*\gphi){}^{m_1 m_2}{}_{n_1 n_2}, \\
	(*\gphi){}^{m p_1 p_2 p_3} (*\gphi){}_{n p_1 p_2 p_3} 
		&= 24 \delta^m{}_n, \\
	(*\gphi){}^{p_1 \dots p_4} (*\gphi){}_{p_1 \dots p_4} 
		&= 168.
\end{aligned}
\end{equation}
Other useful identities include
\begin{equation}
\begin{aligned}
	\gphi^{m_1 m_2 m_3} \gphi_{n_1 n_2 n_3} 
		&= 3\gphi^{[m_1 m_2}{}_{[n_1} \gphi^{m_3]}{}_{n_2 n_3]} 
		+ 6 \delta^{[m_1}{}_{[n_1} (*\gphi){}^{m_2 m_3]}{}_{n_2 n_3]}, \\
	\gphi_{[m_1 m_2 m_3} \gphi_{m_4 m_5]}{}^p 
		&=  (*\gphi){}_{[m_1 \dots m_4} \delta_{m_5]}{}^p.
\end{aligned}
\end{equation}

\subsubsection*{Projectors on forms}

It is useful to have explicit expressions for the various projectors onto representations of $G_2$. We define the projectors $\Pr{p}{r}$, which project a $p$-form onto the $\rep{r}$ representation, so that
\begin{equation}
\begin{split}
	\lambda_{mm'} &= \bigl( \Pr{2}{7} 
		+ \Pr{2}{14} \bigr)_{mm'}^{\phantom{mm'}nn'} \lambda_{nn'}, \\
	\sigma_{m_1 m_2 m_3} &= \bigl( \Pr{3}{1}
		+\Pr{3}{7}
		+\Pr{3}{27} \bigr)_{m_1 m_2 m_3}^{\phantom{m_1 m_2 m_3}n_1 n_2 n_3} \sigma_{n_1 n_2 n_3}.
\end{split}
\end{equation}
In indices, these projectors are given by
\begin{equation}
\begin{split}\label{eq:G2_projectors}
	(\Pr{2}{7})_{mm'}{}^{nn'} &= \tfrac13 \Bigl( \delta_{mm'}^{nn'} 
		+ \tfrac{1}{2} (*\gphi)_{mm'}{}^{nn'} \Bigr), \\
	(\Pr{2}{14})_{mm'}{} ^{nn'} &= \tfrac13 \Bigl( 2\,\delta_{mm'}^{nn'} 
		- \tfrac{1}{2} (*\gphi)_{mm'}{}^{nn'} \Bigr), \\
	(\Pr{3}{1})_{m_1 m_2 m_3}{}^{n_1 n_2 n_3} 
		&= \tfrac{1}{42} \gphi_{m_1 m_2 m_3} \gphi^{n_1 n_2 n_3}, \\
	(\Pr{3}{7})_{m_1 m_2 m_3}{}^{n_1 n_2 n_3} 
		&= \tfrac{1}{4} \tfrac{1}{3!} (*\gphi)_{m_1 m_2 m_3 q} (*\gphi)^{n_1 n_2 n_3q}, \\
	(\Pr{3}{27})_{m_1 m_2 m_3}{}^{n_1 n_2 n_3} 
		&= \tfrac34 \delta_{m_1 m_2 m_3}^{n_1 n_2 n_3}
			-\tfrac38  \delta_{[m_1}{}^{[n_1} (*\gphi)_{m_2 m_3]}{}^{n_2n_3]}
			+\tfrac{1}{56} \gphi_{m_1 m_2 m_3} \gphi^{n_1 n_2 n_3}.
\end{split}
\end{equation}

From these we can obtain useful relations like
\begin{equation}
    *\gphi_{mn}{}^{pq}(\lambda_{\rep{7}})_{pq} = 4(\lambda_{\rep{7}})_{mn}, \quad *\gphi_{mn}{}^{pq}(\lambda_{\rep{14}})_{pq} = -2(\lambda_{\rep{14}})_{mn} .
\end{equation}

\subsection{\texorpdfstring{$\Spin(7)$}{Spin(7)}}\label{app:Spin(7)_identities}

We use an orthonormal frame $g_{mn} = \delta_{mn}$ and take the eight-dimensional gamma matrices to furnish a representation of $\Cliff(8;\bbR)$ with $\gamma^{(9)} = \gamma^1 \dots \gamma^8 = \id$.\footnote{One can take the gamma matrices to be real and symmetric, so that a Majorana spinor obeys $\bar\epsilon = \epsilon^{\text{T}}$~\cite{Gauntlett:2003cy}.} We take the $Spin(7)$ structure to be defined by a chiral Majorana spinor $\epsilon$, with chirality $\gamma^{(9)}\epsilon=\epsilon$ normalised such that $\bar\epsilon \epsilon = 1$. The self-dual $Spin(7)$-invariant 4-form is defined as
\begin{equation}
	\Theta_{mnpq} =  \bar\epsilon \gamma_{mnpq} \epsilon.
\end{equation}
In an orthonormal frame, this can be written as
\begin{equation}
\begin{split}
	\Theta &=  -e^{1234} - e^{1256} - e^{1278} - e^{3456} - e^{3478} - e^{5678} - e^{1357}\\
	&\eqspace + e^{1368} + e^{1458} + e^{1467} + e^{2358} + e^{2367} + e^{2457} - e^{2468}.
	\end{split}
\end{equation}

\subsubsection*{Identities}

Again, using Fierz rearrangement one can show the following identities hold:
\begin{align}
    \Theta^{mnpq}\Theta_{mnpq} &= 336, \\
    \Theta^{qmnp}\Theta_{rmnp} &= 42 \delta^{q}_{r}, \\
    \Theta^{pqmn}\Theta_{rsmn} &= 12\delta^{pq}_{rs} - 4 \Theta^{pq}{}_{rs}, \\
    \Theta^{ijkm}\Theta_{pqrm} &= 6\delta^{ijk}_{pqr} -9 \Theta^{[ij}{}_{[pq}\delta^{k]}_{r]}.
\end{align}

\subsubsection*{Projectors on forms}

We define the projectors $\Pr{p}{r}$, which project a $p$-form onto the $\rep{r}$ representation of $Spin(7)$, so that
\begin{align}
    \lambda_{mm'} &= \bigl(\Pr{2}{7} + \Pr{2}{21}\bigr)_{mm'}^{\phantom{mm'}nn'}\lambda_{nn'}, \\
    \sigma_{m_{1}m_2 m_{3}} &= \bigl(\Pr{3}{8} + \Pr{3}{48}\bigr)_{m_{1}m_2 m_{3}}^{\phantom{m_1 m_2 m_3}n_{1}n_2 n_{3}} \sigma_{n_{1}n_2 n_{3}}, \\
    \tau_{m_{1}\dots m_{4}} &= \bigl(\Pr{4}{1} + \Pr{4}{7} + \Pr{4}{27} + \Pr{4}{35}\bigr)_{m_{1}\dots m_{4}}^{\phantom{m_{1}\dots m_{4}}n_{1}\dots n_{4}} \tau_{n_{1}\dots n_{4}}.
\end{align}
In indices, these projectors are given by
\begin{align}
    (\Pr{2}{7})_{mm'}{}^{nn'} &= \tfrac{1}{4}\left( \delta_{mm'}^{nn'} - \tfrac{1}{2}\Theta_{mm'}{}^{nn'} \right), \\
    (\Pr{2}{21})_{mm'}{}^{nn'} &= \tfrac{3}{4} \left( \delta_{mm'}^{nn'} + \tfrac{1}{6}\Theta_{mm'}{}^{nn'} \right), \\
    (\Pr{3}{8})_{m_{1}m_2 m_{3}}{}^{n_{1} m_2 n_{3}} &= \tfrac{1}{7}\left( \delta_{m_{1} m_2 m_{3}}^{n_{1} n_2 n_{3}}-\tfrac{3}{2}\Theta_{[m_{1}m_{2}}{}^{[n_{1}n_{2}}\delta^{n_{3}]}_{m_{3}]} \right), \\
    (\Pr{3}{48})_{m_{1} m_2 m_{3}}{}^{n_{1} n_2 n_{3}} &= \tfrac{1}{7}\left( 6\,\delta_{m_{1} m_2 m_{3}}^{n_{1} n_2 n_{3}} +\tfrac{3}{2}\Theta_{[m_{1}m_{2}}{}^{[n_{1}n_{2}}\delta^{n_{3}]}_{m_{3}]}\right), \\
    (\Pr{4}{1})_{m_{1}\dots m_{4}}{}^{n_{1}\dots n_{4}} &= \tfrac{1}{336}\Theta_{m_{1}\dots m_{4}}\Theta^{n_{1}\dots n_{4}}, \\
    \begin{split}
        (\Pr{4}{7})_{m_{1}\dots m_{4}}{}^{n_{1}\dots n_{4}} &= \tfrac{1}{8}\Bigl( \delta_{m_{1}\dots m_{4}}^{n_{1}\dots n_{4}} - \tfrac{3}{2}\Theta_{[m_{1}m_{2}}{}^{[n_{1}n_{2}}\delta_{m_{3}m_{4}]}^{n_{3}n_{4}]}  \\
        & \qquad \qquad  - \tfrac{1}{6}\Theta_{[m_{1}\dots m_{3}}{}^{[n_{1}}\Theta_{m_{4}]}{}^{n_{2}\dots n_{4}]} \Bigr),
    \end{split} \\
    \begin{split}
        (\Pr{4}{27})_{m_{1}\dots m_{4}}{}^{n_{1}\dots n_{4}} &= \tfrac{1}{8}\Bigl( 3\, \delta_{m_{1}\dots m_{4}}^{n_{1}\dots n_{4}} + \tfrac{15}{2} \Theta_{[m_{1}m_{2}}{}^{[n_{1}n_{2}}\delta_{m_{3}m_{4}]}^{n_{3}n_{4}]}  \\
        & \qquad \qquad  - \tfrac{1}{2}\Theta_{[m_{1}\dots m_{3}}{}^{[n_{1}}\Theta_{m_{4}]}{}^{n_{2}\dots n_{4}]} + \tfrac{1}{7} \Theta_{m_{1}\dots m_{4}}\Theta^{n_{1}\dots n_{4}} \Bigr),
    \end{split}\\
    \begin{split}
        (\Pr{4}{35})_{m_{1}\dots m_{4}}{}^{n_{1}\dots n_{4}} &= \tfrac{1}{8}\Bigl( 4\,\delta_{m_{1}\dots m_{4}}^{n_{1}\dots n_{4}} - 6 \Theta_{[m_{1}m_{2}}{}^{[n_{1}n_{2}}\delta_{m_{3}m_{4}]}^{n_{3}n_{4}]}  \\
        & \qquad \qquad  + \tfrac{2}{3}\Theta_{[m_{1}\dots m_{3}}{}^{[n_{1}}\Theta_{m_{4}]}{}^{n_{2}\dots n_{4}]} - \tfrac{1}{6} \Theta_{m_{1}\dots m_{4}}\Theta^{n_{1}\dots n_{4}} \Bigr).
    \end{split}
\end{align}
Note in particular the helpful relations
\begin{equation}
    \Theta_{mn}{}^{pq}(\lambda_{\rep{7}})_{pq} = -6(\lambda_{\rep{7}})_{mn}, \quad \Theta_{mn}{}^{pq}(\lambda_{\rep{21}})_{pq} = 2(\lambda_{\rep{21}})_{mn} .
\end{equation}

\subsubsection*{Differential operators}

Using the decomposition of differential forms into $\Spin(7)$ representations and taking $f\in \Omega^{0}_{\rep{1}}$, $\alpha \in \Omega^{1}_{\rep{8}}$, $\beta \in \Omega^{2}_{\rep{7}}$, $\gamma\in \Omega^{2}_{\rep{21}}$, $\delta \in \Omega^{3}_{\rep{48}}$, $\mu \in \Omega^{4}_{\rep{27}}$ and $ \nu \in \Omega^{4}_{\rep{35}}$, one can write the exterior derivative as combinations of the following operators
\begin{align}
    &&\dd^{\rep1}_{\rep8}\colon\Omega^{0}_{\rep{1}} &\to \Omega^{1}_{\rep{8}}, &\dd^{\rep1}_{\rep8}f  &= \dd f, &&\\
   && \dd^{\rep8}_{\rep7}\colon\Omega^{1}_{\rep{8}} &\to \Omega^{2}_{\rep{7}}, &  \dd^{\rep8}_{\rep7} \alpha &= \Pr{2}{7} \dd\alpha, &&\\
   && \dd^{\rep8}_{\rep{21}}\colon\Omega^{1}_{\rep{8}} &\to \Omega^{2}_{\rep{21}}, &  \dd^{\rep8}_{\rep{21}} \alpha &= \Pr{2}{21} \dd\alpha, &&\\
   && \dd^{\rep8}_{\rep{35}}\colon\Omega^{1}_{\rep{8}} & \to \Omega^{4}_{\rep{35}}, &  \dd^{\rep8}_{\rep{35}} \alpha &= \Pr{4}{35}\dd *(\alpha\wedge \Theta) ,&&\\
   && \dd^{\rep7}_{\rep{48}}\colon\Omega^{2}_{\rep{7}} & \to \Omega^{3}_{\rep{48}}, &  \dd^{\rep7}_{\rep{48}} \beta &= \Pr{3}{48}\dd\beta , &&\\
   && \dd^{\rep{21}}_{\rep{48}} \colon \Omega^{2}_{\rep{21}}& \to \Omega^{3}_{\rep{48}}, &  \dd^{\rep{21}}_{\rep{48}}\gamma &= \Pr{3}{48}\dd\gamma ,&&\\
   && \dd^{\rep{48}}_{\rep{27}}\colon\Omega^{3}_{\rep{48}} & \to \Omega^{4}_{\rep{27}}, &  \dd^{\rep{48}}_{\rep{27}}\delta &= \Pr{4}{27}\dd\delta ,&&\\
  &&  \dd^{\rep{48}}_{\rep{35}}\colon\Omega^{3}_{\rep{48}} & \to \Omega^{4}_{\rep{35}}, &  \dd^{\rep{48}}_{\rep{35}}\delta &= \Pr{4}{35}\dd\delta.&&
\end{align}
We also impose $(\dd^{\rep p}_{\rep q})^{\dagger} = \dd^{\rep q}_{\rep p}$, where the adjoint is defined by the standard inner product on differential forms. Adapting the arguments made in \cite{bryant2003a}, one can find the following decomposition of the exterior derivative:
\begin{align}
    \dd f &= \dd^{\rep1}_{\rep8}f,\\
    \dd(f\Theta) &= \dd^{\rep1}_{\rep8}f\wedge \Theta, \\[5pt]
    \dd\alpha &= \dd^{\rep8}_{\rep7}\alpha + \dd^{\rep8}_{\rep{21}}\alpha ,\\
    \dd *(\alpha\wedge \Theta) &= -\tfrac{1}{2}\dd^{\rep8}_{\rep1}\alpha\, \Theta + \tfrac{1}{2}(\dd^{\rep8}_{\rep7}\alpha) \cdot \Theta + \dd^{\rep8}_{\rep{35}}\alpha, \\
    \dd(\alpha\wedge \Theta) &= \dd^{\rep8}_{\rep7}\alpha\wedge \Theta + \dd^{\rep8}_{\rep{21}}\alpha\wedge \Theta ,\\
    \dd*\alpha &= -*\dd^{\rep8}_{\rep1}\alpha ,\\[5pt]
    \dd\beta &= -\tfrac{3}{7}*(\dd^{\rep7}_{\rep8}\beta \wedge \Theta) + \dd^{\rep7}_{\rep{48}}\beta ,\\
    \dd(\beta\cdot \Theta) &= -\tfrac{16}{7}\dd^{\rep7}_{\rep8}\beta \wedge \Theta + 4*\dd^{\rep7}_{\rep{48}}\beta ,\\
    \dd * \beta &= * \dd^{\rep7}_{\rep8}\beta ,\\[5pt]
    \dd\gamma &= \tfrac{1}{7}*(\dd^{\rep{21}}_{\rep8}\gamma\wedge \Theta) + \dd^{\rep{21}}_{\rep{48}}\gamma ,\\
    \dd * \gamma &= *\dd^{\rep{21}}_{\rep8}\gamma ,\\[5pt]
    \dd\delta &= \tfrac{1}{8}(\dd^{\rep{\rep{48}}}_{\rep7}\delta)\cdot \Theta + \dd^{\rep{48}}_{\rep{27}}\delta + \dd^{\rep{48}}_{\rep{35}}\delta,\\
    \dd*\delta &= -*\dd^{\rep{48}}_{\rep7}\delta - * \dd^\rep{{48}}_{\rep{21}}\delta ,\\[5pt]
    \dd\mu &= *\dd^{\rep{27}}_{\rep{48}}\mu ,\\[5pt]
    \dd\nu &=  \tfrac{1}{7}\dd^{\rep{35}}_{\rep8}\nu\wedge \Theta - *\dd^{\rep{35}}_{\rep{48}}\nu.
\end{align}
In the above, we have used the notation $\beta \cdot \Theta$ to denote the isomorphism $\Omega^{2}_{\rep{7}}\rightarrow \Omega^{4}_{\rep{7}}$ given by
\begin{equation}
    (\beta\cdot \Theta)_{abcd} = 4 \beta_{[a|}{}^{i}\Theta_{i|bcd]}.
\end{equation}

\subsubsection*{Laplacians}

One can use the relations above to prove various identities for determinants of Laplacians on $\Spin(7)$ manifolds. In particular, one has
\begin{align}
    \det{}'\Delta_{\rep{8}} &= (\det{}'\Delta_{\rep{7}})( \det{}'\Delta_{\rep{1}}), \\
    \det{}'\Delta_{\rep{35}} &= (\det{}' \Delta_{\rep{27}})(\det{}' \Delta_{\rep{8}}), \\
    \det{}'\Delta_{\rep{48}} &= (\det{}'\Delta_{\rep{27}}) (\det{}'\Delta_{\rep{21}}).
\end{align}
As an example, we will prove the first of these relations -- the others follow similarly. Using the decomposition of the exterior derivative into $\dd^{\rep{p}}_{\rep{q}}$, and the fact that $\dd^{2}\alpha=\dd^{2}\beta = 0$, for $\alpha\in \Omega^{1}_{\rep{8}}$ and $\beta \in \Omega^{2}_{\rep{7}}$ we have
\begin{align}
    \Delta_{\rep{1}}f &= \dd^{\rep8}_{\rep1}\dd^{\rep1}_{\rep8}f, \\
    \Delta_{\rep{7}}\beta &= 4\dd^{\rep8}_{\rep7}\dd^{\rep7}_{\rep8}\beta ,\\
    \Delta_{\rep{8}}\alpha &= \dd^{\rep1}_{\rep8}\dd^{\rep8}_{\rep1}\alpha + 4\dd^{\rep7}_{\rep8}\dd^{\rep8}_{\rep7}\alpha .
\end{align}
Then one finds
\begin{equation} 
\begin{split}
    \det{}'\Delta_{\rep{8}} &= \det{}'(\dd^{\rep1}_{\rep8}\dd^{\rep8}_{\rep1} + 4\dd^{\rep7}_{\rep8}\dd^{\rep8}_{\rep7}) \\
    &= \det{}'\left( (\dd^{\rep1}_{\rep8} + 2\dd^{\rep7}_{\rep8})(\dd^{\rep8}_{\rep1} + 2\dd^{\rep8}_{\rep7}) \right) \\
    &= \det{}'\left( (\dd^{\rep8}_{\rep1} + 2\dd^{\rep8}_{\rep7})(\dd^{\rep1}_{\rep8} + 2\dd^{\rep7}_{\rep8}) \right) \\
    &= \det{}' (\dd^{\rep8}_{\rep1}\dd^{\rep1}_{\rep8} + 2\dd^{\rep8}_{\rep1}\dd^{\rep7}_{\rep8} + 2\dd^{\rep8}_{\rep7}\dd^{\rep1}_{\rep8} + 4\dd^{\rep8}_{\rep7}\dd^{\rep7}_{\rep8}) \\
    &= \det{}'(\dd^{\rep8}_{\rep1}\dd^{\rep1}_{\rep8} + 4\dd^{\rep8}_{\rep7}\dd^{\rep7}_{\rep8}) \\
    &= \det{}'(\Delta_{\rep{1}} + \Delta_{\rep{7}}) \\
    &= (\det{}'\Delta_{\rep{1}})(\det{}'\Delta_{\rep{7}}).
    \end{split}
\end{equation}
Note that, in going from the second to the third line, we used the fact that $\det{}'$ is the ($\zeta$-regularised) product of non-zero eigenvalues. In going from the fourth line to the fifth, we used $\dd^{\rep8}_{\rep7}\dd^{\rep1}_{\rep8} = 0$, which is simply the statement that \eqref{eq:Spin(7)_complex} is a complex. The final identity follows from noting that $\Delta_{\rep{1}}\Delta_{\rep{7}}=\Delta_{\rep{7}}\Delta_{\rep{1}}=0$.

\section{Determinants and partition functions}

\subsection{\texorpdfstring{$\zeta$}{zeta}-regularised determinants}\label{app:zeta-dets}

We give a brief outline of $\zeta$-regularised determinants and their properties~\cite{RAY1971145}. We follow the notation of \cite{Pestun:2005rp}.

Given an increasing sequence of positive real numbers $A=\{a_{1}, a_{2},\ldots\}$, we define the $\zeta$-regularised sum of the numbers to be $\zeta_{A}(-1)$ where, for large $\re s$, we define
\begin{equation}
    \zeta_{A}(s) = \sum_{n}a_{n}^{-s} \qquad \re s \gg 0,
\end{equation}
and then extended to the whole of $\bbC$ by analytic continuation. The $\zeta$-regularised product of $A$ is then defined to be
\begin{equation}
    \ee^{-\zeta'_{A}(0)},
\end{equation}
where the prime denotes differentiation with respect to the complex parameter $s$.

Given a vector space $V$ and an operator $A\colon V\rightarrow V$ with only non-negative real eigenvalues, we define the $\zeta$-regularised determinant, denoted $\det{}'A$, to be the $\zeta$-regularised product of its non-zero eigenvalues. That is
\begin{equation}
    \det{}'A \coloneqq \ee^{-\zeta_{A}'(0)}
\end{equation}
where we have used the same symbol for the operator and its sequence of non-zero eigenvalues. For an operator $B\colon V\rightarrow W$, we note the useful identity
\begin{equation}
    |\det{}'B| \coloneqq (\det{}'B^{\dagger}B)^{1/2}.
\end{equation}
Note that since we neglect the zero eigenvalues, we also have 
\begin{equation}
    \det{}'B^{\dagger}B = \det{}'BB^{\dagger}. \label{eq:det_identity_1}
\end{equation}
Given two operators $A,B\colon V\rightarrow V$ that obey $AB=BA=0$, it is simple to show
\begin{equation}
    \det{}'(A+B) = \det{}'A\det{}'B.
\end{equation}
These determinants are useful when looking at Laplacians of differential operators. We will highlight some useful identities for these determinants in the de Rham and Dolbeault complexes. The results in the latter case all naturally generalise to the $\G_{2}\times \G_{2}$ and $\Spin(7)\times \Spin(7)$ complexes we discuss in this paper.

\begin{figure}
    \centering
    \hfill
    \subfigure[]{
        \centering
        \begin{tikzpicture}[scale=1.5]
            \draw[white] (0,0) -- (0.1,0);
            \draw[black] (0,1.5) -- (1.5-2*.08848,1.5);
            \draw[black] (1.5+2*.08848,1.5) -- (3,1.5);
            \draw[black] (1.5,1.5-.08848) -- (1.5,1.5+.08848);
            \draw[black] (1.5+2*.08848,1.5-.08848) -- (1.5+2*.08848,1.5+.08848);
            \draw[black] (1.5-2*.08848,1.5-.08848) -- (1.5-2*.08848,1.5+.08848);
            \draw[black] (1.5-2*.08848,1.5-.08848) -- (1.5+2*.08848,1.5-.08848);
            \draw[black] (1.5-2*.08848,1.5+.08848) -- (1.5+2*.08848,1.5+.08848);
            %
            %
            \node[circle,text=black, inner sep=0pt, minimum size=0pt] at (1,1.7) {$\dd\Omega^{p-1}$};
            \node[circle,text=black, inner sep=0pt, minimum size=0pt] at (2.2,1.7) {$\dd^{\dagger}\Omega^{p+1}$};
        \end{tikzpicture}
        \label{fig:de_Rham_vertex}}
    \hfill
    \subfigure[]{
        \centering
        \begin{tikzpicture}[scale=1.5]
            \draw[black] (-1,-1) -- (1,1);
            \draw[black] (-1,1) -- (1,-1);
            \draw[black] (-0.25,0) -- (0,0.25);
            \draw[black] (0,0.25) -- (0.25,0);
            \draw[black] (0.25,0) -- (0,-0.25);
            \draw[black] (0,-0.25) -- (-0.25,0);
            \node[circle,align=center,text=black,inner sep=0pt,minimum size=0pt] at (-1,0) {$\del^{\dagger}\delb\Omega^{p+1,q-1}$};
            \node[circle,align=center,text=black,inner sep=0pt,minimum size=0pt] at (1,0) {$\del\delb^{\dagger}\Omega^{p-1,q+1}$};
            \node[circle,align=center,text=black,inner sep=0pt,minimum size=0pt] at (0,0.85) {$\del\delb\Omega^{p-1,q-1}$};
            \node[circle,align=center,text=black,inner sep=0pt,minimum size=0pt] at (0,-0.85) {$\del^{\dagger}\delb^{\dagger}\Omega^{p+1,q+1}$};
        \end{tikzpicture}
        \label{fig:Dolbeault_vertex}}
    \hfill
    \caption{A pictorial representation of the Hodge decomposition of differential forms (neglecting harmonic forms). Figure (a) shows the de Rham decomposition of $\Omega^p$ into exact and co-exact pieces which we can view as coming from the left and right of the node respectively. Figure (b) shows the Dolbeault decomposition of $\Omega^{p,q}$. Pictorially, the subspaces can be associated with the squares surrounding the node, corresponding to the direction the double differential maps from.}
\end{figure}
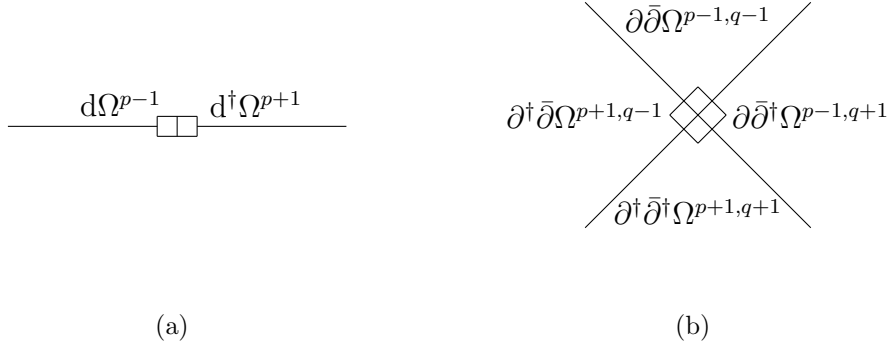

First consider the de Rham Laplacian $\Delta = \dd\dd^{\dagger}+\dd^{\dagger}\dd$. Denoting the space of $p$-forms on an $n$-dimensional manifold $M$ by $\Omega^{p}$, the Hodge decomposition gives
\begin{equation}
    \Omega^{p} = \dd\Omega^{p-1} \oplus \dd^{\dagger}\Omega^{p+1} \oplus H^{p}.
\end{equation}
Figure \ref{fig:de_Rham_vertex} shows this pictorially: we can associate the exact (resp.~co-exact) subspaces with the left (resp.~right) region surrounding the node in the de Rham complex. Note that, by definition, $H^{p}$ is the zero eigenspace of $\Delta$ and so can be neglected when calculating $\det{}'\Delta$. Hence, we can write
\begin{equation}
    \det{}'\Delta^{p} = \det{}'('\!\Delta^{p}) \det{}'(\Delta'^{p}),
\end{equation}
where $'\!\Delta^{p}$ is the restriction of $\Delta^{p}$ to $\dd\Omega^{p-1}$, and $\Delta'^{p}$ is the restriction to $\dd^{\dagger}\Omega^{p+1}$. Observe that one can identify
\begin{equation}
    '\!\Delta = \dd\dd^{\dagger}, \qquad \Delta' = \dd^{\dagger}\dd.
\end{equation}
Due to \eqref{eq:det_identity_1}, we see that
\begin{equation}
    \det{}' ('\!\Delta^{p}) = \det{}'(\Delta'^{p-1}).
\end{equation}
Furthermore, since the Hodge star commutes with the Laplacian, we have
\begin{equation}
    \det{}'\Delta^{p} = \det{}'\Delta^{n-p},
\end{equation}
where $n=\dim M$.

For K\"ahler manifolds, we can refine this further. We have the Laplacians for $\del$, $\delb$ and $\dd$ which are proportional:
\begin{equation}\label{eq:Kahler_laplacians}
    \Delta_{\del} = \Delta_{\delb} = \tfrac{1}{2}\Delta.
\end{equation}
We also have the Hodge decompositions of $\Omega^{p,q}$ with respect to $\del$ and $\delb$:
\begin{align}
    \Omega^{p,q} &= \del\Omega^{p-1,q}\oplus \del^{\dagger}\Omega^{p+1,q}\oplus H^{p,q}_{\del}, \\
    &= \delb \Omega^{p,q-1}\oplus \delb^{\dagger}\Omega^{p,q+1}\oplus H^{p,q}_{\delb}.
\end{align}
By \eqref{eq:Kahler_laplacians}, we have equality of the spaces of harmonic forms, $H^{p,q}_{\del} = H^{p,q}_{\delb}$, which means that we can combine the Hodge decompositions above and write
\begin{equation}\label{eq:Kahler_hodge_decomp}
    \Omega^{p,q} = \del\delb\Omega^{p-1,q-1} \oplus \del\delb^{\dagger}\Omega^{p-1,q+1} \oplus \del^{\dagger}\delb\Omega^{p+1,q-1} \oplus \del^{\dagger}\delb^{\dagger}\Omega^{p+1,q+1} \oplus H^{p,q}.
\end{equation}
Once again, $H^{p,q}$ is the zero eigenspace of $\Delta$ and so can be ignored when computing $\det{}'$. The Laplacian $\Delta$ then decomposes according to its action on the four subspaces in \eqref{eq:Kahler_hodge_decomp}. One can identify these subspaces with the four squares surrounding a vertex in the Hodge diamond, as shown in Figure \ref{fig:Dolbeault_vertex}. Given this decomposition, one finds
\begin{equation}
    \det{}'\Delta = (\det{}'\DeAB) (\det{}'\DeBD) (\det{}'\DeCD)(\det{}'\DeAC),
\end{equation}
where the position of $\scriptstyle\bullet$ denotes the restriction of $\Delta$ to the relevant subspace as labelled in Figure \ref{fig:Dolbeault_vertex}. For example
\begin{equation}
    \overset{\bullet}{\Delta} = \Delta|_{\del\delb\Omega^{p-1,q-1}}.
\end{equation}
Note that we also use this notation for the spaces $\mathcal{A}^{p,q}$, with the replacement $(\partial,\bar\partial)\to(\dd_+,\dd_-)$.
Using \eqref{eq:det_identity_1} and \eqref{eq:Kahler_laplacians} one then finds
\begin{equation}
    \det{}'\overset{\bullet}{\Delta}{}^{p,q} = \det{}'(\Delta\raisebox{1pt}{$\scriptstyle \bullet$})^{p,q-1} = \det{}'\underset{\bullet}{\Delta}^{p-1,q-1} = \det{}' (\raisebox{1pt}{$\scriptstyle \bullet$}\Delta)^{p-1,q}.
\end{equation}
Hence, the value of the determinant depends only on the ``square'' in the Hodge diamond and not the vertex (as is shown in Figure \ref{fig:SU_diagram}).\footnote{These were referred to as the determinants of the ``face'' Laplacians in \cite{Pestun:2005rp}.} One can then use the symmetries of the Hodge diamond to relate the value of determinants on different squares. In particular, for a Calabi--Yau $n$-fold, one can use Hodge duality, complex conjugation, and contraction with the holomorphic $n$-form to see that
\begin{equation}
    \det{}'\Delta^{p,q} = \det{}'\Delta^{q,p} = \det{}'\Delta^{n-p,q} = \det{}'\Delta^{p,n-q} = \det{}'\Delta^{n-p,n-q}.
\end{equation}
For a Calabi--Yau threefold, this leaves us with three independent determinants, as shown in Figure \ref{fig:SU_diagram}.

All of this generalises to the $\G_{2}\times \G_{2}$ and $\Spin(7)\times \Spin(7)$ complexes, where for $\G_{2}$ the maps $\theta_{\pm}$ play the role of Hodge duality and contraction with the holomorphic $n$-form. A small distinction is that in general there is no notion of ``complex conjugation'' and so $\det{}'\hat{\Delta}^{p,q} \neq \det{}'\hat{\Delta}^{q,p}$. However, when one has a genuine $\G_{2}$ or $\Spin(7)$ structure, these determinants are in fact equal, leading to Figures \ref{fig:G2_diagram} and \ref{fig:Spin7_diagram}.

\subsection{Direct calculation of partition function}\label{app:DirectOneLoop}

In Section \ref{sec:G2-quad-action} of the main text, we gave a calculation of the partition function of the target-space theory for the $G_{2}\times G_{2}$ double complex~\eqref{eq:double_complex} via BV quantisation. Here, following~\cite{Pestun:2005rp}, we will show that this calculation agrees with a direct calculation using formal manipulations of the path integral. 

The partition function of the theory is
\begin{equation}
Z=\frac{1}{\Vol(\mathcal{G})}\int\mathcal{D}a\mathcal{D}b\mathcal{D}c\,\ee^{-S_0},
\end{equation}
where the measure is for the fields $b_{11}$, $a_{00}$ and $c_{22}$, $S_0$ is the quadratic target-space action~\eqref{eq:G2_action}, and $\Vol(\mathcal{G})$ is the volume of the gauge group.

\subsubsection*{\texorpdfstring{$b_{11}\wedge \dd_+\dd_- b_{11}$}{c11 d+ d- c11}}

Let us start by focusing on $S_{0}^{a}$, the term in the action that depends on $b_{11}$. Since $S_{0}^{a}$ is a quadratic action for a single real bosonic field with a second-order kinetic operator, the path integral over $b_{11}$ is formally given by
\begin{equation}
\begin{split}
Z^{a}&=\frac{1}{\Vol(\mathcal{G}^{a})}\frac{\Vol(H^{1,1})\Vol(\AomAC^{1,1})\Vol(\AomAB{}^{1,1})\Vol(\AomBD^{1,1})}{\sqrt{\det{}'\dd_{+}\dd_{-}|_{\AomCD^{1,1}}}}\\
&=\frac{1}{\Vol(\mathcal{G}^{a})}\frac{\Vol(H^{1,1})\Vol(\AomAC^{1,1})\Vol(\AomAB{}^{1,1})\Vol(\AomBD^{1,1})}{\sqrt{\det{}'\hat{\Delta}_{C}}},
\end{split}
\end{equation}
where we are again denoting the volume of a (formally infinite-dimensional) space $\Omega$ by $\Vol(\Omega)$. Here the determinant comes from the integral over the component of $b_{11}$ that is orthogonal to gauge transformations (so that the restricted kinetic operator has no kernel). One then needs to account for $b_{11}$ that do come from gauge transformations, which in this case is simply the product of the volumes of $\AomAC^{1,1}$, $\AomAB{}^{1,1}$ and $\AomBD^{1,1}$. One can think of these as $b_{11}$ that are of the form $\dd_{+}
b_{01}+\dd_{-}b_{10}$, where $b_{10}$ and $b_{01}$ can be independent of each other. We then need to compute the (formal) volumes of the spaces. 

First consider $\dd_{-}\colon\AomCD^{1,0}\to\AomAC^{1,1}$. Since $\dd_{-}$ is an invertible map between real vector spaces, the ratio of the volumes is
\begin{equation}
\frac{\Vol(\AomAC^{1,1})}{\Vol(\AomCD^{1,0})}=\sqrt{\det{}'\dd_{-}^{\dagger}\dd_{-}|_{\AomCD^{1,0}}}=\sqrt{\det{}'\hat{\Delta}_{B}},
\end{equation}
where we have observed that the operator $\dd_{-}^{\dagger}\dd_{-}$ acting on $\AomCD^{1,0}$ is simply $\hat{\Delta}_{B'}$. Using the Hodge decomposition, the volume of $\mathcal{A}^{1,0}$ is
\begin{equation}
\Vol(\mathcal{A}^{1,0})=\Vol(\AomCD^{1,0})\Vol(\AomBD^{1,0})\Vol(H^{1,0}),
\end{equation}
where $\Vol(H^{1,0})$ is the space of $\hat{\Delta}$-harmonic $(1,0)$-forms. Finally, using the map $\dd_{+}\colon\AomCD^{0,0}\to\AomBD^{1,0}$, we have
\begin{equation}
\frac{\Vol(\AomBD^{1,0})}{\Vol(\AomCD^{0,0})}=\sqrt{\det{}'\dd_{+}^{\dagger}\dd_{+}|_{\AomCD^{0,0}}}=\sqrt{\det{}'\hat{\Delta}_{A}}.
\end{equation}
Noting that $\Vol(\mathcal{A}^{0,0})=\Vol(\AomCD^{0,0})\Vol(H^{0,0})$, we then have
\begin{equation}
\Vol(\AomAC^{1,1})=\frac{\Vol(H^{0,0})}{\Vol(H^{1,0})}\frac{\sqrt{\det{}'\hat{\Delta}_{B}}}{\sqrt{\det{}'\hat{\Delta}_{A}}}\frac{\Vol(\mathcal{A}^{1,0})}{\Vol(\mathcal{A}^{0,0})}.
\end{equation}
One can find $\Vol(\AomBD^{1,1})$ in a similar fashion:
\begin{equation}
\Vol(\AomBD^{1,1})=\frac{\Vol(H^{0,0})}{\Vol(H^{0,1})}\frac{\sqrt{\det{}'\hat{\Delta}_{B'}}}{\sqrt{\det{}'\hat{\Delta}_{A}}}\frac{\Vol(\mathcal{A}^{0,1})}{\Vol(\mathcal{A}^{0,0})}.
\end{equation}
Finally, we need to calculate $\Vol(\AomAB{}^{1,1})$. Given $\dd_{-}\colon\AomBD^{1,0}\to\AomAB{}^{1,1}$, one has
\begin{equation}
\frac{\Vol(\AomAB{}^{1,1})}{\Vol(\AomBD^{1,0})}=\sqrt{\det{}'\dd_{-}^{\dagger}\dd_{-}|_{\AomBD^{1,0}}}=\sqrt{\det{}'\hat{\Delta}_{A}}.
\end{equation}
Using the above expressions for $\Vol(\AomBD^{1,0})$, one then finds
\begin{equation}
\Vol(\AomAB{}^{1,1})=\frac{1}{\Vol(H^{0,0})}\det{}'\hat{\Delta}_{A}\Vol(\mathcal{A}^{0,0}).
\end{equation}

Putting this together, one has
\begin{equation}
Z^{a}=\frac{1}{\Vol(\mathcal{G}^{a})}\frac{\Vol(H^{1,1})\Vol(H^{0,0})}{\Vol(H^{1,0})\Vol(H^{0,1})}\left(\frac{\det{}'\hat{\Delta}_{B}\det{}'\hat{\Delta}_{B'}}{\det{}'\hat{\Delta}_{C}}\right)^{1/2}\frac{\Vol(\mathcal{A}^{1,0})\Vol(\mathcal{A}^{0,1})}{\Vol(\mathcal{A}^{0,0})}.
\end{equation}
Formally, one identifies $\Vol(\mathcal{G}^{a})$ with $\Vol(\mathcal{A}^{1,0})\Vol(\mathcal{A}^{0,1})/\Vol(\mathcal{A}^{0,0})$, so that the gauge group $\mathcal{G}^{a}$ can be thought of as gauge transformations by $(1,0)$ and $(0,1)$ fields, modulo $(0,0)$-form ghosts. The contribution to the partition function from $S_{0}^{a}$ is then
\begin{equation}
Z^{a}=\frac{\Vol(H^{1,1})\Vol(H^{0,0})}{\Vol(H^{1,0})\Vol(H^{0,1})}\left(\frac{\det{}'\hat{\Delta}_{B}\det{}'\hat{\Delta}_{B'}}{\det{}'\hat{\Delta}_{C}}\right)^{1/2}.
\end{equation}

\subsubsection*{\texorpdfstring{$a_{00}\wedge \dd_+ \dd_- c_{22}$}{c00 d+ d- c22}}

We now compute the contribution of $S_{0}^{b}$, which depends on $a_{00}$ and $c_{22}$. In this case, since $S_{0}^{b}$ is an action for two real bosonic fields with a second-order kinetic operator, the path integral over $a_{00}$ and $c_{22}$ is formally given by
\begin{equation}
\begin{split}
Z^{b}&=\frac{1}{\Vol(\mathcal{G}^{b})}\frac{\Vol(H^{0,0})\Vol(H^{2,2})\Vol(\AomAC^{2,2})\Vol(\AomAB{}^{2,2})\Vol(\AomBD^{2,2})}{\det{}'\dd_{+}\dd_{-}|_{\AomCD^{2,2}}}\\
&=\frac{1}{\Vol(\mathcal{G}^{b})}\frac{\Vol(H^{0,0})\Vol(H^{2,2})\Vol(\AomAC^{2,2})\Vol(\AomAB{}^{2,2})\Vol(\AomBD^{2,2})}{\det{}'\hat{\Delta}_{A}},
\end{split}
\end{equation}
where again the determinant comes from the component of $c_{22}$ that is orthogonal to gauge transformations (so that the restricted kinetic operator has no kernel). Note that $a_{00}$ has no gauge transformations. Again, we need the volumes of the spaces appearing above.

Consider first the map $\dd_{-}\colon\AomCD^{2,1}\to\AomAC^{2,2}$ so that
\begin{equation}
\frac{\Vol(\AomAC^{2,2})}{\Vol(\AomCD^{2,1})}=\sqrt{\det{}'\dd_{-}^{\dagger}\dd_{-}|_{\AomCD^{2,1}}}=\sqrt{\det{}'\hat{\Delta}_{B'}}.
\end{equation}
We also note that
\begin{equation}
\Vol(\mathcal{A}^{2,1})=\Vol(\AomAB{}^{2,1})\Vol(\AomBD^{2,1})\Vol(\AomCD^{2,1})\Vol(\AomAC^{2,1})\Vol(H^{2,1}).
\end{equation}
We now want to write the volume of the various subspaces of $\mathcal{A}^{2,1}$ in terms of lower-degree $\mathcal{A}^{p,q}$. For example, we have
\begin{align}
\Vol(\AomAC^{2,1}) 
 & =\frac{\Vol(H^{1,0})}{\Vol(H^{2,0})\Vol(H^{0,0})}\frac{\det{}'\hat{\Delta}_{A}}{\sqrt{\det{}'\hat{\Delta}_{B}}}\frac{\Vol(\mathcal{A}^{2,0})\Vol(\mathcal{A}^{0,0})}{\Vol(\mathcal{A}^{1,0})}.
\end{align}
Similar calculations for $\Vol(\AomBD^{2,1})$ and $\Vol(\AomAB{}^{2,1})$ give
\begin{align}
\Vol(\AomBD^{2,1}) & =\frac{\Vol(H^{0,1})\Vol(H^{1,0})}{\Vol(H^{0,0})\Vol(H^{1,1})}\frac{\sqrt{\det{}'\hat{\Delta}_{C}}}{\det{}'\hat{\Delta}_{B}}\frac{\Vol(\mathcal{A}^{1,1})\Vol(\mathcal{A}^{0,0})}{\Vol(\mathcal{A}^{1,0})\Vol(\mathcal{A}^{0,1})},\\
\Vol(\AomAB{}^{2,1}) & =\frac{\Vol(H^{0,0})}{\Vol(H^{1,0})}\frac{\det{}'\hat{\Delta}_{B}}{\sqrt{\det{}'\hat{\Delta}_{A}}}\frac{\Vol(\mathcal{A}^{1,0})}{\Vol(\mathcal{A}^{0,0})}.
\end{align}
Using these we have
\begin{align}
\Vol(\AomAC^{2,2}) & =\frac{\Vol(H^{2,0})\Vol(H^{1,1})\Vol(H^{0,0})}{\Vol(H^{2,1})\Vol(H^{1,0})\Vol(H^{0,1})}\frac{\det{}'\hat{\Delta}_{B}}{\sqrt{\det{}'\hat{\Delta}_{A}\det{}'\hat{\Delta}_{C}}}\frac{\Vol(\mathcal{A}^{2,1})\Vol(\mathcal{A}^{1,0})\Vol(\mathcal{A}^{0,1})}{\Vol(\mathcal{A}^{2,0})\Vol(\mathcal{A}^{1,1})\Vol(\mathcal{A}^{0,0})},
\end{align}
from which it is simple to see
\begin{align}
\Vol(\AomBD^{2,2}) & =\frac{\Vol(H^{0,2})\Vol(H^{1,1})\Vol(H^{0,0})}{\Vol(H^{1,2})\Vol(H^{1,0})\Vol(H^{0,1})}\frac{\det{}'\hat{\Delta}_{B'}}{\sqrt{\det{}'\hat{\Delta}_{A} \det{}'\hat{\Delta}_{C}}}\frac{\Vol(\mathcal{A}^{1,2})\Vol(\mathcal{A}^{1,0})\Vol(\mathcal{A}^{0,1})}{\Vol(\mathcal{A}^{0,2})\Vol(\mathcal{A}^{1,1})\Vol(\mathcal{A}^{0,0})}.
\end{align}
Finally we need $\Vol(\AomAB{}^{2,2})$ which is given by
\begin{align}
\Vol(\AomAB{}^{2,2})
 & =\frac{\Vol(H^{1,0})\Vol(H^{0,1})}{\Vol(H^{1,1})\Vol(H^{0,0})}\frac{\det{}'\hat{\Delta}_{C}}{\sqrt{\det{}'\hat{\Delta}_{B}\det{}'\hat{\Delta}_{B'}}}\frac{\Vol(\mathcal{A}^{1,1})\Vol(\mathcal{A}^{0,0})}{\Vol(\mathcal{A}^{1,0})\Vol(\mathcal{A}^{0,1})}.
\end{align}

Putting this all together, we find that the contribution to the partition function is
\begin{equation}
\begin{split}
Z^{b}&=\frac{1}{\Vol(\mathcal{G}^{b})}\frac{\Vol(H^{2,2})\Vol(H^{2,0})\Vol(H^{0,2})\Vol(H^{1,1})\Vol(H^{0,0})^2}{\Vol(H^{2,1})\Vol(H^{1,2})\Vol(H^{1,0})\Vol(H^{0,1})}\frac{\sqrt{\det{}'\hat{\Delta}_{B}\det{}'\hat{\Delta}_{B'}}}{\det{}'^{2}\hat{\Delta}_{A}}\\
&\eqspace{}\times\frac{\Vol(\mathcal{A}^{2,1})\Vol(\mathcal{A}^{1,2})\Vol(\mathcal{A}^{1,0})\Vol(\mathcal{A}^{0,1})}{\Vol(\mathcal{A}^{2,0})\Vol(\mathcal{A}^{0,2})\Vol(\mathcal{A}^{1,1})\Vol(\mathcal{A}^{0,0})}.
\end{split}
\end{equation}
Again, taking $\Vol(\mathcal{G}^{b})$ to cancel the various volumes of the spaces of forms, this simplifies to
\begin{equation}
Z^{b}=\frac{\Vol(H^{2,2})\Vol(H^{2,0})\Vol(H^{0,2})\Vol(H^{1,1})\Vol(H^{0,0})^2}{\Vol(H^{2,1})\Vol(H^{1,2})\Vol(H^{1,0})\Vol(H^{0,1})}\frac{\sqrt{\det{}'\hat{\Delta}_{B}\det{}'\hat{\Delta}_{B'}}}{\det{}'^{2}\hat{\Delta}_{A}}.
\end{equation}

\subsubsection*{Final result}

Combining the contributions from $S_{0}^{a}$ and $S_{0}^{b}$, the partition function is given by
\begin{equation}
Z=\frac{\Vol(H^{2,2})\Vol(H^{2,0})\Vol(H^{0,2})\Vol(H^{1,1})^2\Vol(H^{0,0})^{3}}{\Vol(H^{2,1})\Vol(H^{1,2})\Vol(H^{1,0})^{2}\Vol(H^{0,1})^{2}}\frac{\det{}'\hat{\Delta}_{B}\det{}'\hat{\Delta}_{B}}{\det{}'^{2}\hat{\Delta}_{A}\sqrt{\det{}'\hat{\Delta}_{C}}}.
\end{equation}
Upon taking the cohomologies to be trivial and setting $\det{}' \hat{\Delta}_A = A$, and so on, we have $Z=BB'A^{-2}C^{-1/2}$ in agreement with both the double complex calculation in Section \ref{eq:g2_one_loop} and the BV quantisation calculation in Section \ref{sec:G2-quad-action}.

\section{Review of \texorpdfstring{$O(d,d)\times\bbR^+$}{O(d,d) x R+} generalised geometry}
\label{app:gg}

Generalised geometry is a geometric formalism in which one extends the tangent bundle by a sequence of differential forms to create a vector bundle $T \hookrightarrow E$ that has an enlarged structure group $\GL(d,\bbR) \hookrightarrow \mathcal{G}$. The case we will be interested in is the geometry defined by the vector bundle
\begin{equation}\label{eq:gen_tangent_bundle}
    E = T\oplus T^{*},
\end{equation}
with sections or \emph{generalised vectors} written as $V=v+\lambda$. This bundle has a natural $\Orth{d,d}$ structure which preserves a symmetric bilinear form
\begin{equation}\label{eq:O(d,d)_metric}
    \eta(V,V) = v\lrcorner\lambda.
\end{equation}
One can then take tensor products of $E$ and decompose them according to $O(d,d)$ representations, and sections of such bundles are called generalised tensors.

One can also define a bracket $\bgen{\cdot}{\cdot}$ which gives $E$ the structure of an exact Courant algebroid~\cite{chen2013,roytenberg1999}.\footnote{One also needs a smooth bundle map $\anchor\colon E\rightarrow T$ called the anchor in the definition of the Courant algebroid. We will normally take this to just be the projection onto $T$ in \eqref{eq:gen_tangent_bundle}.} It is called the Courant bracket and is given by
\begin{equation}
    \bgen{v+\lambda}{w+\mu} = \mathcal{L}_{v}w +\mathcal{L}_{v}\mu - \mathcal{L}_{w}\lambda - \tfrac{1}{2}\dd(v\lrcorner\mu - w\lrcorner\lambda),
\end{equation}
where $W=w+\mu$ is another section of $E$.

The Courant bracket is clearly covariant under diffeomorphisms and also under a closed 2-form transformation $b\in \Omega^{2}_{\text{cl}}(M)$ given by
\begin{equation}
    \ee^{b}(v+\lambda) = v+\lambda-v\lrcorner b .
\end{equation}
We therefore have an enlarged automorphism group of the Courant algebroid given by $\Diff \hookrightarrow \GDiff = \Diff \ltimes \Omega^{2}_{\text{cl}}(M)$, whose elements we refer to as generalised diffeomorphisms. These are generated by a local derivative along a generalised vector $V=v+\lambda$ called the Dorfman derivative. The action of the Dorfman derivative is
\begin{equation}\label{eq:dorf}
    \Lgen_{V}W = \mathcal{L}_{v}w + \mathcal{L}_{v}\mu - w\lrcorner\dd\lambda .
\end{equation}
Note that this is not antisymmetric but instead satisfies
\begin{equation}
    \tfrac{1}{2}(\Lgen_{V}W-\Lgen_{W}V) = \bgen{V}{W}, \qquad \tfrac{1}{2}(\Lgen_{V}W+\Lgen_{W}V) = \dd\eta(V,W) .
\end{equation}

We can naturally incorporate the NSNS flux $H$ into the construction by twisting the Dorfman derivative (and hence the Courant bracket) to get the flux twisted derivative
\begin{equation}\label{eq:twisted_dorf}
    \LHgen_{V}W = \mathcal{L}_{v}w + \mathcal{L}_{v}\mu - w\lrcorner \dd \lambda + w\lrcorner (v\lrcorner H).
\end{equation}
An alternative but equivalent way to include the flux is to take \eqref{eq:gen_tangent_bundle} to just be a local definition and allow non-trivial patching by $\Omega^{2}_{\text{cl}}(M)$. That is, for an open subset $\mathcal{U}_{i}\in M$ and $V_{i}=v_{i}+\lambda_{i} \in \Gamma(\mathcal{U}_{i},E)$, $V_{j} = v_{j} + \lambda_{j} \in \Gamma(\mathcal{U}_{j},E)$, there exists a $\Lambda_{ij} \in \Omega^{1}(\mathcal{U}_{i}\cap\mathcal{U}_{j})$ such that, on $\mathcal{U}_{i}\cap \mathcal{U}_{j}$
\begin{equation}\label{eq:flux_patching}
\begin{aligned}
    v_{i} &= v_{j} \\
    \lambda_{i} &= \lambda_{j} - v_{j}\lrcorner\dd\Lambda_{ij}\,\,
\end{aligned}  \qquad \Rightarrow \qquad V_{i} = \ee^{\dd\Lambda_{ij}}V_{j}.
\end{equation}
This patching defines a bundle $E^{H}$. The equivalence of \eqref{eq:twisted_dorf} and \eqref{eq:flux_patching} comes from choosing a global isomorphism $E^{H} \simeq E$. To do so, one must pick a connection $B$ which is locally a 2-form, and patches as\footnote{This non-trivial constraint on triple intersections means $B$ is a connective structure on a gerbe \cite{Hitchin:1999fh}.}
\begin{align}\label{eq:B_patching}
\begin{split}
    B_{i} &= B_{j} + \dd\Lambda_{ij}  \\
    \Lambda_{ij} + \Lambda_{jk} + \Lambda_{ki} &= \dd\Lambda_{ijk}
\end{split}
\begin{split}
    &\text{on }\mathcal{U}_{i}\cap \mathcal{U}_{j}, \\
    &\text{on }\mathcal{U}_{i}\cap \mathcal{U}_{j} \cap \mathcal{U}_{k},
\end{split}
\end{align}
where $\Lambda_{ijk} \in C^{\infty}(\mathcal{U}_{i}\cap \mathcal{U}_{j} \cap \mathcal{U}_{k})$. The flux is determined by this connection via $H=\dd B$ locally. It is easy to see from the patching \eqref{eq:flux_patching} and \eqref{eq:B_patching} that $V\in \Gamma(E)$ if and only if $\ee^{B}V \in \Gamma(E^{H})$. Moreover, it is easy to check that $\Lgen_{\ee^{B}V}\ee^{B}W = \ee^{B}\LHgen_{V}W$. Hence a choice of $B$ defines an isomorphism of algebroids
\begin{equation}\label{eq:equiv_pictures}
    (E^{H},\Lgen) \quad \longleftrightarrow \quad (E,\LHgen).
\end{equation}
It is possible to show that a twist by $H$ and $H'$ are equivalent as algebroids if and only if $H'=H+\dd\alpha$. That is, inequivalent exact Courant algebroids are classified by $[H] \in H^{3}(M)$~\cite{chen2013}.\footnote{In addition, this class must be quantised in string theory.} This equivalence of twisted bundle versus twisted derivative applies to all generalised tensor bundles and we will often move between the two pictures in \eqref{eq:equiv_pictures} and will drop the superscript $H$ to avoid cluttering our notation further.

Since it geometrises the $H$ flux, generalised geometry turns out to be naturally well suited to describe the NSNS sector of string backgrounds. In fact, as was shown in \cite{Coimbra:2011nw}, one can also account for the dilaton by enlarging the structure group further to $\Orth{d,d}\times \bbR^{+}$. 
All tensors should then be appropriately weighted under the $\bbR^{+}$ by including factors of $\det T^{*}$ in the bundles. In particular, we can consider weighted generalised vectors $\tilde{V} \in \Gamma(\tilde E)$ and the induced action of the $\Orth{d,d}$ metric $\eta$
\begin{equation}
    \tilde{E} = E \otimes \det T^{*} \quad \Rightarrow \quad \eta(\tilde{V},\tilde{W}) \in \Gamma\left((\det T^{*})^{2}\right).
\end{equation}

The $\Orth{d,d}$ structure defines a Clifford algebra via
\begin{equation}
    \{\Gamma^{A},\Gamma^{B}\} = \eta^{AB} ,
\end{equation}
where $\eta^{AB}$ are the components of the $\Orth{d,d}$ inner product $\eta$ in some orthonormal frame. One can show that this has a natural representation on the exterior algebra, so that weighted $p$-forms $(\det T^*)^{-1/2}\otimes\ext^{\bullet}T^{*}$ form a spinor representation of $\Spin(d,d)\times \bbR^{+}$. We then call any $\rho\in\Gamma((\det T^*)^{-1/2}\otimes\ext^{\bullet}T^{*})$ a generalised spinor and denote the vector bundle of generalised spinors by $S$. Note that $S$ is reducible as an $\Orth{d,d}\times \bbR^{+}$ representation. There exists a notion of chirality and we can define even/odd spinors to be even/odd polyforms. That is
\begin{equation}
    S = S_{+}\oplus S_{-}, \qquad S_{\pm} = (\det T^*)^{-1/2}\otimes\ext^{\text{ev/odd}}T^{*}.
\end{equation}
More generally we can define the weighted spinor bundles
\begin{equation}
    S^{(p)}_{\pm} = (\det T^*)^{p}\otimes S_{\pm},
\end{equation}
so that $S^{(1/2)}_{\pm}$ corresponds to unweighted polyforms.

There exists a natural $\Orth{d,d}$-invariant pairing on $S$ called the Mukai pairing. Taking $\rho,\mu \in \Gs{S^{(p)}_{\pm}}$, it is given by
\begin{equation}
    \left<\rho, \mu \right> = \sum_{i}\rho_{i} \wedge \sigma(\mu_{d-i}) \in \Gamma((\det T^{*})^{2p}),
\end{equation}
where $\rho_{i}$ means the restriction of the polyform $\rho$ to its degree $i$ component, and $\sigma\colon S\rightarrow S$ is the automorphism defined by\footnote{This convention is different than the one chosen in e.g.~\cite{gualtieri2004}}
\begin{equation}\label{eq:sigma}
    \sigma(\mu_{k}) = (-1)^{k(k+1) / 2}\mu_{k}.
\end{equation}
Note that for $d$ even, $\langle \cdot,\cdot \rangle$ restricts to a pairing on $S^{(1/2)}_{\pm}$, and that for $d=6$ this pairing defines a non-degenerate symplectic structure.

As with conventional geometry, one has a notion of connections, torsion and curvature. A generalised connection is simply a first-order linear differential operator $D$ which acts on a generalised vector in frame indices as
\begin{equation}
    D_A V^B = \partial_A V^B + \Omega_A{}^B{}_C V^C,
\end{equation}
where $\partial_A$ denotes the natural embedding of the ordinary partial derivative in $E$, and the generalised connection one-form $\Omega$ takes values in the adjoint representation of $O(d,d)\times\bbR^+$, so that the action of $D$ has the obvious extension to any generalised tensor with arbitrary conformal weight. The generalised torsion $T$ of such a connection is a generalised tensor defined in terms of the Dorfman derivative~\eqref{eq:dorf} by
\begin{equation}
    T(V)\cdot \alpha = \Dorf^D_V\alpha - \Dorf_V \alpha,
\end{equation}
where $V\in\Gamma(E)$ and $\alpha$ is a generalised tensor. One might also expect there exists a generalised analogue of the Riemannian curvature, however the naive object one would define turns out to not be tensorial, and we find that there is no useful notion of ``generalised curvature'' for an arbitrary generalised connection without specifying additional structure.

\subsection{\texorpdfstring{$\Orth{d}\times \Orth{d}$}{O(d) x O(d)} structures}\label{sec:gen_metrics}

A generalised metric is given by a reduction of $\Orth{d,d}\times \bbR^{+}$ to the maximal compact subgroup $\Orth{d}\times \Orth{d}$~\cite{Coimbra:2011nw,gualtieri2004,Gualtieri:2010fd}. As for many conventional $G$-structures, it is defined by a set of globally non-vanishing tensors $(\Phi,G)$, where $\Phi \in \Gamma(\det\, T^{*})$  -- which specifies the isomorphism between weighted and un-weighted generalised vectors $\tilde{E}\cong E$ --  and $G\colon S^{2}E\rightarrow \bbR$ is a positive-definite inner product on $E$ that is compatible with the $\Orth{d,d}$ metric \eqref{eq:O(d,d)_metric} in the following sense. Using $\eta$ as an isomorphism $E\cong E^{*}$, we can view $G\colon E\rightarrow E$ and then require $G^{2}=1$. Given such a $G$, we get a decomposition of $E$ into eigenbundles of $G$ so that
\begin{equation}\label{eq:gen_metric_decomposition}
    E = C_{+}\oplus C_{-},
\end{equation}
where $C_{\pm}$ are $\eta$-orthogonal subbundles of $E$ such that $\eta|_{C_{\pm}}$ is positive (resp.~negative) definite. The inner product $G$ can then be written
\begin{equation}
    G = \eta|_{C_{+}} - \eta|_{C_{-}}.
\end{equation}
Hence a choice of $G$ is equivalent to a choice of decomposition \eqref{eq:gen_metric_decomposition}.

An alternative definition of a generalised metric $(\Phi,G)$ is given via a choice of conformal split frame of $\tilde{E}$. This is defined to be a local frame $\{\hat{E}^{+}_{a}\} \cup \{\hat{E}^{-}_{\bar{a}}\}$ such that
\begin{align}
    \eta(\hat{E}^{+}_{a},\hat{E}^{+}_{b}) &= \Phi^{2}\delta_{ab}, \\
    \eta(\hat{E}^{-}_{\bar{a}},\hat{E}^{-}_{\bar{b}}) &= -\Phi^{2}\delta_{\bar{a}\bar{b}}, \\
    \eta(\hat{E}^{+}_{a},\hat{E}^{-}_{\bar{b}}) &= 0.
\end{align}
This determines $\Phi$ uniquely and defines $C_{\pm}$ to be the span of $\{\hat{E}^{\pm}\}$.

A generalised metric is equivalent to a choice of conventional metric $g$, $B$-field, and dilaton $\dil$. Indeed, given two independent local orthonormal frames $\hat{e}^+_{a}$, $\hat{e}^-_{a}$ of $T$, the conformal split frame can be written as
\begin{equation}
\begin{split}
    \hat{E}^{+}_{a} &= \ee^{-2\dil}\sqrt{g}\,(\hat{e}^+_{a} + \imath_{\hat{e}^+_{a}}g + \imath_{\hat{e}^+_{a}}B), \label{eq:conf_split_frame} \\
    \hat{E}^{-}_{\bar{a}} &= \ee^{-2\dil}\sqrt{g}\,(\hat{e}^-_{\bar{a}} - \imath_{\hat{e}^-_{\bar{a}}}g + \imath_{\hat{e}^-_{\bar{a}}}B).
\end{split}
\end{equation}
In several applications, it is useful to evaluate  $\Orth{d}\times \Orth{d}$ expressions in which one chooses frames such that $\hat{e}^+_{a}=\hat{e}^-_{a}=\hat{e}_{a}$ are aligned.

A generalised $G$-structure is said to be integrable if there exists a torsion-free generalised connection that is compatible with the structure. An $\Orth{d}\times \Orth{d}$ structure is thus torsion-free if there exists a generalised connection $D$ that satisfies
\begin{equation}
    DG = 0, \qquad D\Phi = 0, \qquad L^{D}_{V} = L_{V},
\end{equation}
where $L_{V}^{D}$ is the Dorfman derivative with all partial derivatives replaced with the connection $D$. We call a connection that satisfies these constraints a generalised Levi-Civita connection. As was shown in \cite{Coimbra:2011nw}, such connections always exist but are not unique. Using a split frame, the torsion-free connection acting on $V = v^{a}\hat{E}^{+}_{a} + v^{\bar{a}}\hat{E}^{-}_{\bar{a}}$ takes the form
\begin{equation}
\begin{aligned}\label{eq:gen-LC}
    D_{a}v^{b} & =\nabla_{a}v^{b}-\tfrac{1}{6}H_{a}{}^{b}{}_{c}v^{c}-\tfrac{2}{d-1}(\delta_{a}{}^{b}\del_{c}\dil-\delta_{ac}\del^{b}\dil)v^{c}+A^+_{a}{}^{b}{}_{c}v^{c},\\
    D_{\bar{a}}v^{b} & =\nabla_{\bar{a}}^{-}v^{b}\equiv\nabla_{\bar{a}}v^{b}-\tfrac{1}{2}H_{\bar{a}}{}^{b}{}_{c}v^{c},\\
    D_{a}v^{\bar{b}} & =\nabla_{a}^{+}v^{\bar{b}}\equiv\nabla_{a}v^{\bar{b}}+\tfrac{1}{2}H_{a}{}^{\bar{b}}{}_{\bar{c}}v^{\bar{c}},\\
    D_{\bar{a}}v^{\bar{b}} & =\nabla_{\bar{a}}v^{\bar{b}}+\tfrac{1}{6}H_{\bar{a}}{}^{\bar{b}}{}_{\bar{c}}v^{\bar{c}}-\tfrac{2}{d-1}(\delta_{\bar{a}}{}^{\bar{b}}\del_{\bar{c}}\dil-\delta_{\bar{a}\bar{c}}\del^{\bar{b}}\dil)v^{\bar{c}}+A^-_{\bar{a}}{}^{\bar{b}}{}_{\bar{c}}v^{\bar{c}},
\end{aligned}
\end{equation}
where $\nabla$ is the Levi-Civita connection for $g$, $H=\dd B$ and $A^{\pm}$ are undetermined tensors satisfying 
\begin{equation}
\begin{aligned}
        A^+_{abc} = - A^+_{acb},\qquad A^+_{[abc]}=0, \qquad A^+_a{}^a{}_b=0,\\
        A^-_{\ba\bb\bc} = - A^-_{\ba\bc\bb},\qquad A^-_{[\ba\bb\bc]}=0, \qquad A^-_{\ba}{}^{\ba}{}_{\bb}=0,
\end{aligned}
\end{equation}
so that they do not contribute to the torsion. The $A^{\pm}$ tensors thus parametrise the failure of the metric-compatibly and vanishing torsion conditions to specify a unique generalised connection.

Thanks to the generalised metric structure, we can use the compatible connection $D$ to define generalised curvatures. The analogue of the Riemann tensor is not unique, i.e., depends on the choice of generalised Levi-Civita, and so it is not a very useful object. However, there exist certain contractions and projections that are uniquely defined. In particular, we can define the generalised Ricci tensor $\GenRic$ and generalised Ricci scalar $\GenS$ via the action of $D$ on either generalised vectors~\cite{Coimbra:2011nw}
\begin{equation}\label{eq:genRicTensor}
    \GenRic_{a\bar b}w_+^a = [D_a, D_{\bar b}]w_+^a,\qquad \GenRic_{\bar a b} w_-^{\bar a} = [D_{\bar a}, D_b]w^{\bar a}_-,
\end{equation}
or spinors
\begin{align}\label{eq:Riccis}
\begin{split}
    \GenRic_{a\bar{b}} \gamma^{a}\epsilon^{+} &= [\gamma^{a}D_{a},D_{\bar{b}}]\epsilon^{+},  \\
    \GenRic_{\bar{a}b}\gamma^{\bar{a}}\epsilon^{-} &= [\gamma^{\bar{a}}D_{\bar{a}},D_{b}]\epsilon^{-} ,
\end{split}
\begin{split}
     -\tfrac{1}{4}\GenS\epsilon^{+} &= (\gamma^{a}D_{a}\gamma^{b}D_{b} - D^{\bar{a}}D_{\bar{a}}) \epsilon^{+}, \\
     -\tfrac{1}{4}\GenS\epsilon^{-} &= (\gamma^{\bar{a}}D_{\bar{a}}\gamma^{\bar{b}}D_{\bar{b}} - D^{a}D_{a})\epsilon^{-}.
\end{split}
\end{align}
Here $\epsilon^{\pm}$ are $S(C_{\pm})$ spinors and the $\gamma^{a}$ are representations of the Clifford algebra induced by the $\Orth{d}$ structure on $C_{\pm}$. Upon explicit evaluation, one finds
\begin{align}
    \GenRic_{ab} &= \Ric_{ab} -\tfrac{1}{4}H_{acd}H_{b}{}^{cd} + 2\nabla_{a}\nabla_{b}\dil +\tfrac{1}{2}\ee^{2\dil}\nabla^{c}(\ee^{-2\dil}H_{cab}) ,\\
    \GenS &= \Scalar + 4\nabla^{2}\dil - 4(\del\dil)^{2} - \tfrac{1}{12}H^{2},
\end{align}
where we have aligned the frames $\hat{e}^+_a = \hat{e}^-_{\bar{a}}$, and $\mathcal{R}_{ab}$ and $\mathcal{R}$ are the conventional Ricci tensor and scalar for $g$. The right-hand side of these are simply the equations of motion in the absence of RR fluxes, and hence both $\GenRic_{ab}$ and $\GenS$ vanish on on-shell. In particular, since a background which is supersymmetric and solves the Bianchi identity $\dd H=0$ automatically solves the equations of motion, both $\GenRic_{ab}$ and $\GenS$ must vanish for supersymmetric backgrounds. This crucial result is used many times in the main text.


\subsection{Generalised Calabi--Yau}\label{app:gcy}

A generalised Calabi--Yau structure is a reduction of the structure group to $\SU(n,n)$ where $d=2n$. It is defined by a nowhere-vanishing complex pure spinor $\Psi$. Given such a spinor, one can define the null space $L_{\Psi}$
\begin{equation}
    L_{\Psi} = \{V \in \Gamma(E) \;|\; \slashed{V}\rho = V^{A} \Gamma_{A}\Psi =0\}.
\end{equation}

A generalised Calabi--Yau structure is then given by a $\Psi$ satisfying
\begin{equation}
    \dim_{\bbC}L_{\Psi} = d, \qquad \langle \Psi,\bar{\Psi} \rangle \neq 0.
\end{equation}
A spinor satisfying the first condition is said to be pure, and the associated null space is said to be maximally isotropic. The generalised Calabi--Yau structure is integrable (i.e.~there exists a torsion-free compatible connection) if and only if
\begin{equation}
    \dd\Psi = 0.
\end{equation}

Hitchin showed that for $d=6$ these structures can be described via a variational problem~\cite{hitchin2003}. Indeed, consider a real chiral spinor $\rho$ which is stable in the sense of \cite{hitchin2001stable}. Since $\langle \cdot,\cdot \rangle$ defines an $\Orth{d,d}$ invariant symplectic structure, there is an associated moment map $\mu\colon S_{\pm}\rightarrow \mathfrak{g}^{*}$ given by
\begin{equation}
    \mu(\rho)(a) = \tfrac{1}{2}\langle a\cdot \rho , \rho \rangle \qquad \forall \, a \in \Orth{d,d}.
\end{equation}
Then we can consider the following map which is an invariant quartic homogeneous function in $\rho$
\begin{equation}
    q\colon S_{\pm} \rightarrow (\det T^{*})^{2}, \qquad q(\rho) = \tr\left(\mu(\rho)^{2}\right).
\end{equation}
It turns out that $\rho$ defines an $\SU(n,n)$ structure if and only if $q(\rho)<0$, which is an open condition on $\rho$. Such a $\rho$ is known as stable. Note that $q(\rho)\in \Gamma((\det T^{*})^{2})$ which has a canonical orientation and hence a well-defined notion of a negative section. The real spinor $\rho$ then becomes the real part of the complex pure spinor $\Psi$, with the imaginary part $\hat{\rho}$ given by the first variation of the functional
\begin{equation}\label{eq:GCY_Hitchin_functional}
    H(\rho) = \int_{M} \sqrt{-\frac{q(\rho)}{3}} \qquad \Rightarrow \qquad \delta H = \int_{M} \langle \delta \rho , \hat{\rho} \rangle.
\end{equation}
Note that $H$ is a homogeneous functional of degree 2 in $\rho$. Denoting the space of stable spinors of definite chirality by $U$, one can show that there is an integrable complex structure $\mathcal{J}$ on $\rho \in \Gamma(U)$ and that the second variation of the functional $H$ is given by
\begin{equation}
    \delta^{2}H(\delta_{1}\rho,\delta_{2}\rho) = \int_{M} \langle\delta_{1}\rho , \mathcal{J}\delta_{2}\rho \rangle.
\end{equation}

Now suppose we fix some $\rho \in \Gamma(U)$ such that $\dd\rho = 0$ and only allow variations within the cohomology class of $\rho$. That is, we take $\delta \rho = \dd b$ for some real polyform $b$. Then by \eqref{eq:GCY_Hitchin_functional} we have
\begin{equation}
    \delta H (\dd b) = \int_{M} \langle\dd b, \hat{\rho} \rangle = \int_{M} \langle b, \dd \hat{\rho} \rangle =0 \qquad \Rightarrow \qquad \dd\hat{\rho} = 0.
\end{equation}
Therefore, stationary points of $H$ within a fixed cohomology class $[\rho]$ correspond to $\SU(3,3)$ structures with $\dd\rho=\dd\hat{\rho} = 0$, that is $\dd\Psi = \dd(\rho + \ii \hat{\rho}) = 0$. Hence, stationary points correspond to integrable $\SU(3,3)$ structures.

\subsection{The generalised Hitchin functional for integrable \texorpdfstring{$\gxg$}{G2 x G2} structures}\label{app:gen-hitchin-g2}

Turning now to the generalised geometry of a seven-dimensional manifold, in the main text we describe $\gxg$ structures in terms of a pair of $C_{\pm}$ spinors. However, following \cite{Jeschek2005,Witt:2004vr}, one can also define them through a $\Spin(7,7)\times\bbR^+$ globally defined nowhere-vanishing real chiral spinor $\rho \in S_{\pm}$ that is stable in the sense of \cite{hitchin2001stable}. By a simple dimension count, one has that the spinor lives in an open orbit of $\Spin(7,7)\times \bbR^{+}$.
 
One can define an operator $\square_{\rho}$ which maps spinors of one chirality to the other given by
\begin{equation}
    \square_{\rho}\colon S_{\pm} \to S_{\mp}, \qquad \square_{\rho}(\alpha) = \ee^{B}*\sigma(\ee^{-B}\alpha),
\end{equation}
where $\alpha$ is a generalised spinor, $*$ is the Hodge operator associated to the Riemannian metric $g$ and $\sigma$ was given in~\eqref{eq:sigma}. If we work in the flux twisted differential picture instead, then we can just write $\square_{\rho}(\alpha) = \mathop*\sigma(\alpha)$. It is possible to show that this is an $\Orth{7}\times \Orth{7}$ covariant map and hence a generalised $\gxg$ structure can be equivalently described by either a stable $\rho \in \Gamma(S_{\pm})$ or a stable $\square_{\rho}\rho\in \Gamma(S_{\mp})$. The $\gxg$ structure is then said to be integrable (there exists a compatible torsion-free generalised connection) if and only if
\begin{equation}
    \dd\rho = \dd\square_{\rho}\rho = 0,
\end{equation}
which is the analogue of $\dd\Psi=\dd(\rho+\ii\hat\rho)=0$ for an $SU(3,3)$ structure. For concreteness, we will take $\rho \in \Gamma(S_{-})$, and so $\square_{\rho}\rho \in \Gamma(S_{+})$.

To match the description of generalised $\gxg$ structures given in the main text around \eqref{sec:gen-g2g2} where we consider the spinors $\epsilon_{\pm} \in S(C_{\pm})$, we note that there is also a natural isomorphism between these bundles and the bundle of $O(d,d)\times\bbR^+$ spinors $S$ as
\begin{equation}
    S \simeq S(C_{+})\otimes S(C_{-}),
\end{equation}
and under this isomorphism we associate
\begin{equation}
    \ee^{-2\dil}\ee^{B}(\epsilon_{+}\otimes \epsilon_{-}) = \rho + \square_{\rho}\rho.
\end{equation}

In general, a $\G_{2}\times \G_{2}$ structure defines a local $\SU(3)$ structure on the manifold. We can then write $\rho, \square_{\rho}\rho$ explicitly in terms of the local $\SU(3)$ structure of the manifold. While the results of this paper will hold in general, we will mostly be interested in the case where the generalised structure is induced from a genuine $\G_{2}$ structure. In that case, we can write
\begin{align}
    \rho &=  \ee^{-2\dil}\ee^{B}(-\gphi + \vol), \\
    \square_{\rho}\rho &=  \ee^{-2\dil}\ee^{B}(1-\mathop*\gphi).
\end{align}

We can write the generalised $\gxg$ structure more explicitly in terms of the local $\SU(3)$ structure defined by the $\epsilon_{\pm}$. This $\SU(3)$ structure locally defines a 1-form $\alpha$, a 2-form $\omega$, and two 3-forms $\psi_{\pm}$ which can be viewed as the real and imaginary parts of a holomorphic 3-form on some 6-dimensional $\mathcal{D}\subset T$ that is orthogonal to $\alpha$. There is also a scalar $\cos a$, where $a$ is the angle between $\epsilon_{+}$ and $\epsilon_{-}$ as 8-dimensional real vectors. Without loss of generality, we can take $\rho \in \Gamma(S_{-})$ and can write
\begin{align}
    \rho &= \ee^{-2\dil}\ee^{B}\left(s\alpha - c(\psi_{+} + \omega\wedge \alpha) - s\psi_{-} - s\tfrac{1}{2}\omega^{2}\wedge \alpha + c\vol_{g}\right), \\
    \square_{\rho}\rho &=  \ee^{-2\dil}\ee^{B}\left(c + s\omega - c(\psi_{-}\wedge \alpha + \tfrac{1}{2}\omega^{2}) + s\psi_{+}\wedge \alpha - s\tfrac{1}{6}\omega^{3}\right),
\end{align}
where $s$ and $c$ are shorthand for $\sin a$ and $\cos a$, and $\vol_{g}$ is the volume form defined by the metric. While individually the tensors in these expressions are defined only where $s\neq 0$, the precise combinations that appear can be written as bilinears of $\epsilon_{\pm}$ and so are globally defined. When $s=0$, the spinors $\epsilon_{\pm}$ become parallel and the $\SU(3)$ stabiliser degenerates to a $\G_{2}$ defined by some 3-form $\gphi$. At these points one finds
\begin{align}
    \rho &=  \ee^{-2\dil}\ee^{B}(-\gphi + \vol), \\
    \square_{\rho}\rho &=  \ee^{-2\dil}\ee^{B}(1-\mathop*\gphi).
\end{align}

As for $\SU(3,3)$ structures, one can understand integrable generalised $\gxg$ structures via a variational approach~\cite{Witt:2004vr}. Since $\rho \in \Gamma(S_{-})$ must be in an open orbit of $\Spin(7,7)\times \bbR^{+}$, we can consider a function
\begin{equation}
    q(\rho) = \left< \rho, \square_{\rho}\rho \right> \in \Gamma(\det T^{*}) ,
\end{equation}
where we think of this as defined on $U\subset S_{-}$, the space of stable $\rho$. As shown in \cite{Witt:2004vr}, this is a homogeneous function of degree 2 in $\rho$ and the first variation is given by
\begin{equation}
    \delta q(\delta \rho) = \left<\delta \rho, \square_{\rho}\rho \right>.
\end{equation}
Integrating $q$ over $M$, one obtains the Hitchin functional for $\gxg$ structures:
\begin{equation}\label{eq:gen-Hitchin-g2}
    H(\rho) = \int_M  \left< \rho, \square_{\rho}\rho \right> .
\end{equation}
If we assume that $\dd\rho = 0$ and vary only within a cohomology class $\delta \rho = \dd b$, we find that the extrema of the Hitchin functional are given by
\begin{equation}
    \delta H(\delta \rho) = \int_{M} \left<\dd b,\square_{\rho}\rho \right> = \int_{M}\left< b,\dd\square_{\rho}\rho\right> = 0 \qquad \Rightarrow \qquad \dd\square_{\rho}\rho = 0.
\end{equation}
Hence the functional extremises on integrable $\gxg$ structures.


\input{main.bbl}

\end{document}

%% file: main.bbl
\providecommand{\href}[2]{#2}\begingroup\raggedright\endgroup